\documentclass[aps,showpacs,showkeys,nofootinbib,twocolumn]{revtex4}
\usepackage{epsfig,amssymb,bm,graphics,color}
\newcommand*{\ee}{e^+e^-}

\newcommand*{\CEP}{\phi_{CEP}}

\begin{document} {\normalsize }

\title{Multi-photon regime of non-linear Breit-Wheeler and Compton processes\\
in short linearly and circularly polarized laser pulses}
%
%%%%%%%%%%%%%%%%%%%%%%%%%%%%%%%%%%%%%%%%%%%%%%%%%%%
 %%%
 \author{A.~I.~Titov}
\email[]{atitov@theor.jinr.ru}
 \affiliation{
 Bogoliubov Laboratory of Theoretical Physics, JINR, Dubna 141980, Russia}
 \author{A.~Otto, and B.~K\"ampfer}
%\email[]{kaempfer@hzdr.de}
\affiliation{
Helmholtz-Zentrum  Dresden-Rossendorf, 01314 Dresden, Germany,\\%
Institut f\"ur Theoretische Physik, TU~Dresden, 01062 Dresden, Germany}
 \begin{abstract}
 Non-linear Breit-Wheeler $e^+e^-$ pair production and its crossing
 channel - the non-linear Compton process - in the multi-photon regime
are analyzed for linearly and circularly polarized short laser pulses.
The contribution of multi-photon processes to total cross sections
is determined and we also show
that (i) the azimuthal angular distributions of outgoing electrons
in these processes differ on a qualitative level, and
(ii) they depend on the polarization properties of the pulses.
A finite carrier envelope phase $\CEP$ leads to a non-trivial non-monotonic behavior of
 the azimuthal angle distributions of the considered processes. That effect
can be used for the  $\CEP$ determination.
\end{abstract}

\maketitle

The non-linear Breit-Wheeler (BW) process refers to the decay of a probe photon
$\gamma'$ (energy $\omega'$, four-momentum $k'$) into an
electron-positron ($e^+ e^-$) pair while traversing a laser pulse
(characterized by central frequency $\omega$, wave four-vector $k$ and
polarization four-vector $a$ with $a \cdot k = 0$, where the dot stands
for the scalar product), symbolically $\gamma' \to e_L^- + e_L^+$. Here,
the label ``$L$" points to the laser-dressed $e^\pm$ wave functions.
The non-linear Compton (C) scattering,
symbolically $e_L\to \gamma' +  e_L'$, as related (by crossing symmetry on
amplitude level) to the BW process, is also a subject of intensive study.

In the plane-wave approximation, one can further distinguish monochromatic
laser beams with an infinitely long duration
(the infinite-pulse approximation [IPA]) or a pulse of finite duration
(the finite-pulse approximation [FPA]). In the latter case,
the bandwidth effects cause a distribution of frequencies around the  central
one, as evidenced by the power spectrum of the pulse. The IPA case
has been analyzed in depth by the Ritus group~\cite{RitusGroup} some time ago
and summarized in a well-known review paper~\cite{Ritus-79}.
For completeness, we mention the pioneering papers by Reiss~\cite{Reissold}
and  the review papers~\cite{Mourou1,Piazza1,Narozhny2015v},
see also recent publications~\cite{Mourou2,Piazza2,Ilderton:2018,Seipt:2017ckc}.
%\marginpar{Refs.}
It turned out that the polarization
of the laser has a noticeable impact on the pair-production probability due to
the different angular momenta of the emerging $e_L^\pm$.
(For a complete treatment of polarization effects,
cf.\ \cite{Ivanov:2004vh}.)

Note that the kinematics of non-linear C and BW processes are vastly different. In the
weak-field regime, $\xi \ll 1$ (with $\xi$ the reduced field intensity \cite{LL}
defined in our context in Eq.~(\ref{S244}) below)
where the series expansion in powers of the
fine-structure constant $\alpha$ applies, the physical regions over the Mandelstam
$s-t-u$ plane are fairly different \cite{LL,Harvey}.
This is related to the fact that BW is a threshold
process requiring $s > 4 m^2$
(where $s$ and $m$ are the square of the total energy
in the center of mass system (c.m.s.) and the electron mass, respectively),
while C has a Thomson limit and is allowed
at $s > m^2$, i.e.\  it is always above the threshold. These differences continue
into the multi-photon and non-perturbative regimes.
In the case of the BW process, it is natural to use the dimensionless
threshold variable $\zeta=4m^2/s$~\cite{TitovPEPAN} instead of $s$.
The region $\zeta>1$ automatically
selects the sub-threshold multi-photon regime.
The  SLAC experiment E-144~\cite{E-144}
has tested this multi-photon regime at $\xi \leq 0.35$
by the trident process $e^- + L \to e^- + e^+ + e^-$
which combines the subprocesses of non-linear Compton backscattering
$e^- + L \to e^- + \gamma'$ and non-linear BW.
The envisaged LUXE
experiment \cite{LUXE} will probe the non-linear BW process at $\xi > 1$. The
photons $\gamma'$ are thereby delivered by bremsstrahlung of the XFEL driver-beam (L)
with energies ${\cal O}(10$ GeV). The LUXE laser, when less tightly focused,
therefore also provides access to the multi-photon regime and can be used
for benchmarking issues. Note also the interesting goal of LUXE as access to
``measuring the boiling point of the vacuum of quantum
electrodynamics" \cite{Hartin:2018sha}.
The search for a possible manifestation of multi-photon regime
in Compton scattering in analogy with BW
seems to be important and is the subject of our present work.

Since high and ultrahigh laser intensities are customarily achieved by
chirped pulse amplification, one is forced to take into account
the effect of the pulse duration, i.e.\ one has to analyze the FPA case.
%{\color{red}
A number of important phenomena
associated with non-linear C and BW processes
within finite pulses are discussed in the above-mentioned review papers.
Some further entries in the fairly extended literature are given by
%we can also add original studies for C and BW processes in
Refs.~\cite{Mackenroth-2011,Boca-2009,Boca-2012,KrajewskaPRA85,HeinzlPRA81,MackenrothPRL105,TitovEPJD,SKPRA88,Dinu}
%%%%%
and
\cite{A3,Nousch,KrajewskaPRA86,TitovPRA,CEPTitov,Piazza3,Piazza4,BW4,BW5},
respectively.
%}

 Below, we focus on intense optical lasers with reduced field intensity
 $\xi^2\leq0.1$
 as well as short and very short laser pulses
 with a small or very small number of field oscillations.
 Considering both the total cross sections
 and the azimuthal angle distributions of
 the outgoing  particles, we identify observables
 sensitive to the initial pulse polarization, taking into
 account a finite carrier envelope phase $\CEP$~\cite{CEP_1,CEP_2}.
 The manifestation of the carrier envelope phase,
  sometimes called ``the absolute phase of the laser pulse" \cite{BrabecKrausz},
  is a peculiar quantum-mechanical phenomenon. It
  manifests itself in many diverse
  processes involving laser pulses, such as
  the $\ee$ pair production~\cite{KrajewskaPRA86}, non-linear Compton
  scattering~\cite{Li2018,Li2018lqs},
  electron tunneling in a coupled double-quantum-dot system~\cite{CDQDS},
  and strong-field ionization~\cite{SFI}.
  In  the considered C and BW processes, this phenomenon is related to
  a non-trivial interference of
  the outgoing or probe photons with the background laser field.

 The aim of the present study is to give
 a comparative sequential analysis of
 two elementary quantum processes in
 a coherent electromagnetic (e.m.) field
 (i.e.\ laser pulses with finite pulse duration)
 with linear and circular polarizations
 in the multi-photon regime for non-asymptotic kinematics,
 attainable at existing laser facilities, thus
 highlighting the most important observables.

 Our paper is organized as follows.
 In Sect.~II we discuss the laser field model.
 The non-linear Breit-Wheeler process and
 the non-linear Compton scattering are analyzed in Sects.~III and~IV, respectively.
 Our summary is given in Sect.~V.
 In the Appendix, for convenience and easy reference,
 we provide some useful expressions of the  infinite-pulse approximation.

\section{The laser field model}

 We first assume that the external electric field (laser pulse) is
 determined by the electromagnetic (e.m.) four-potential
 in the radiation gauge $A^{i}=(0,{\mathbf A}^{i})$
 as ${\mathbf E}^{i}=-\partial {\mathbf A}^{i}/\partial t$,
where the label $i=0$ or $1$ corresponds
 to the linear ($lin$) or circular ($circ$) polarizations, respectively:
 \begin{eqnarray}
  \mathbf{A}^{(i)}(\phi) &=& f(\phi) \left[ \mathbf{a}_1\cos(\phi+\CEP)\right.\nonumber\\
&&+
\left. \delta^{i1}\,  \mathbf{a}_2\sin(\phi+\CEP)\right],
  \label{I1}
 \end{eqnarray}
 where $\delta^{ij}$ is the Kronecker delta. Thus, $\mathbf{A}^{(0)}\equiv\mathbf{A}^{(lin)}$ and $\mathbf{A}^{(1)}\equiv\mathbf{A}^{(circ)}$.
The quantity
$\phi=k\cdot x$ is the invariant phase with wave four-vector
$k=(\omega, \mathbf{k})$, obeying the null-field property
$k^2 = k \cdot k=0$
%\footnote{Effects of an ambient medium,
%e.g.\ a plasma, can be accommodated in a modified dispersion
%relation, $k^2 \ne 0$ \cite{Mackenroth:2018rtp}, as customary
%done in many phenomenological QCD approaches,
%e.g.\ in \cite{Kampfer:1999ff}.}
implying $\omega = \vert\mathbf{k}\vert$,
 $ \mathbf{a}_{(1,2)} \equiv \mathbf{a}_{(x,y)}$;
 $|\mathbf{a}_x|^2=|\mathbf{a}_y|^2 = a^2$, $\mathbf{a}_x \mathbf{a}_y=0$;
transversality means $\mathbf{k} \mathbf{a}_{x,y}=0$ in the present gauge;
in addition, $A_z = 0$.
For the sake of definiteness,
the envelope function $f(\phi)$ is chosen as hyperbolic secant:
 \begin{eqnarray}
 f(\phi)=\frac{1}{\cosh\frac{\phi}{\Delta}}~.
 \label{I3}
 \end{eqnarray}
 The dimensionless quantity $\Delta$ is related to the pulse duration
 $2\Delta=2\pi N$, where $N$ characterizes the number of cycles in
 the pulse.
 It is related to the time duration of the pulse $\tau=2N/\omega$.
$N < 1$ means sub-cycle pulses
(for the dependence of some observables on the envelope shape, see
\cite{TitovPRA} for example).

 The quantity $\CEP$ is the carrier envelope phase (CEP).
 The interplay of CEP and the azimuthal angle of the outgoing electron
 is important and will be the subject of detailed considerations below.

 For convenience, we recollect the relation between
$\xi^2 = e^2 a^2 / m^2$
 and the averaged laser pulse intensity $\langle I_L \rangle$
 for circularly/linearly polarized pulses~\cite{TitovPEPAN}
 \begin{eqnarray}
  \xi^2= \frac{N}{N_0^{(i)}}
  \frac{5.62\cdot 10^{-19}}{\omega^2_{[{\rm eV}^2]}}
  \langle I_L \rangle_{\left[\frac{\rm W}{\rm  cm^2}\right]} .
  \label{S244}
\end{eqnarray}
The normalization factors $N^{(i)}_0$ are related to the
average density of the e.m.\ field $\langle \cal{E} \rangle$
and are expressed through the envelope functions as
\begin{eqnarray}
N^{(lin)}_0&=&\frac{1}{2 \pi}\int\limits_{-\infty}^{\infty}
d\phi \left(f^2(\phi)+{f'}^2(\phi)\right)\,\cos^2\phi , \nonumber\\
N^{(circ)}_0&=&\frac{1}{2 \pi}\int\limits_{-\infty}^{\infty}
d\phi \left(f^2(\phi)+{f'}^2(\phi)\right)
\label{II2}
\end{eqnarray}
with the asymptotic values at $\Delta/\pi \gg1$:
$N^{(circ)}_0\simeq\Delta/\pi$,
and $N^{(lin)}_0\simeq\Delta/2\pi$.

We use natural units with
 $c=\hbar=1$, $e^2/4\pi = \alpha \approx 1/137.036$.

 \section{Non-linear Breit-Wheeler process}

\subsection{Cross sections}

 As mentioned above,
 we consider essentially multi-photon events, where
 a finite number of laser photons are involved in
 the $\ee$-pair production. This allows for sub-threshold
 $\ee$-pair production with $s<s_{\rm thr}$ or $\zeta>1$.
 In the following, we analyze the dependence of
 cross sections on $\zeta$ and on the e.m.\ field intensity which is described
 by the reduced field intensity parameter
 $\xi^2$. We will also analyze the differential
 cross sections as a function of the azimuthal angle
 of the outgoing electron (or positron) for different values $\CEP$.
 In all cases, we provide a direct comparison of
 results for linear and circular polarizations.

 The differential cross sections read
\begin{eqnarray}
\frac{d \sigma^{(i)}}{d\phi_{e} }
=\frac{\alpha^2\,\zeta}{4m^2\xi^2 N^{(i)}_0}
\,\int\limits_\zeta^{\infty}\,d\ell \,v(\ell) \,
\int\limits_{-1}^{1} d\cos\theta_{e}
\, w^{(i)}{(\ell)}~,
\label{II1}
\end{eqnarray}
where $\ell$ is here an auxiliary continuous variable.
The azimuthal angle of the outgoing electron, $\phi_{e}$, is defined as
 $\cos\phi_{e}={\mathbf a_x}{\mathbf p}_{e}/a |{\mathbf p}_{e}|$.
 It is related to the azimuthal angle of the positron by
 $\phi_{e^+}=\phi_{e} + \pi$.
 Furthermore,  $\theta_{e}$ is the polar angle
 of the outgoing electron, $v$ is the electron (positron) velocity
 in the center of mass system (c.m.s.).

 The lower limit of the integral over the variable $\ell$
 is the threshold parameter $\zeta$.
%=s_{\rm thr}/s\equiv4m^2/s$.
 The region of $\zeta<1$
 corresponds to the above-threshold $\ee$-pair production, while
 the region of $\zeta>1$ is for the sub-threshold pair production
enabled by multi-photon and bandwidth effects.
 We keep our notation of \cite{CEPTitov} and
 denote by $k(\omega,{\mathbf k})$,
 $k'(\omega',{\mathbf k}')$,
 $p(E,{\mathbf p})$ and $p'(E',{\mathbf p}')$ the four-momenta of the
 background (laser) field (\ref{I1}), the
 incoming probe photon, the outgoing positron and the outgoing electron,
 respectively.  The important variables $s$, $v$ and $u$
 are determined by $s={2k\cdot k' }= 2(\omega'\omega -{\mathbf k}'{\mathbf k})$
(with $\mathbf k' \mathbf k = - \omega' \omega$ for head-on geometry),
 $v^2=(\ell s-4m^2)/\ell s$,
 $u\equiv(k'\cdot k)^2/\left(4(k\cdot p)(k\cdot p')\right)
 =1/(1-v^2\cos^2\theta_e)$.

\subsubsection{Linear polarization}

 We evaluate the partial probabilities $w^{(lin)} (\ell)$ in Eq.~(\ref{II1})
 by a direct trace calculation of the square
 of the transition matrix element, similar to the case of
 IPA~\cite{RitusGroup,Ritus-79}, and express them through
 the generalized basis functions $\widetilde A_m(\ell)$, which are analog
 of the IPA basis functions $A_m(n)$ (for circular polarization see~\cite{TitovPRA}).
 Thus, we have
 \begin{eqnarray}
 &&\frac12\, w^{(lin)}(\ell) =
(2u_{\ell}+1)|\widetilde A_0(\ell)|^2\nonumber\\
 &&\qquad-2z{u_{\ell}}\zeta\cos\phi_e
 {\rm Re}[\widetilde A_0(\ell)\widetilde A_1^*(\ell)]\nonumber\\
 &+& \xi^2 {\rm Re}[\widetilde A_0(\ell)\widetilde A_2^*(\ell)]
 +\xi^2(2u-1)|\widetilde A_1(\ell)|^2~,
 \label{II3}
 \end{eqnarray}
 where $u_{\ell}=u_{\rm max}=\ell/\zeta$. The
 generalized basis functions $\widetilde A_m(\ell)$ read
 \begin{eqnarray}
 \widetilde A_m(\ell)
 =\frac{1}{2\pi}\int\limits_{-\infty}^{\infty}d\phi
 \cos^m(\phi+\tilde\phi)\,{\rm e}^{i\ell\phi
 -i{\cal P}^{(lin)}(\phi)}
 \label{II4}
 \end{eqnarray}
with
\begin{eqnarray}
{\cal P}^{(lin)}(\phi) &=&
\tilde\alpha(\phi)- \tilde\beta(\phi)~,\label{II55}\\
\tilde\alpha(\phi)&=&\alpha\int\limits_{-\infty}^{\phi}d\phi'
f(\phi')\cos(\phi'+\CEP)~,\label{II5}\\
%\alpha &=&z\cos\phi_e \qquad\qquad \beta=u\ell\xi^2/u_\ell \label{II5}\\
\tilde\beta(\phi)&=&4\beta\int\limits_{-\infty}^{\phi}d\phi'
f^2(\phi')\cos^2(\phi' + \CEP)~, \nonumber\\
\alpha&=&z\cos\phi_{e},\qquad\beta=\frac{u\ell\xi^2}{4u_\ell}~.
\label{II6}
\end{eqnarray}

The integrand of the function  $\widetilde A_0(\ell)$ in Eq.~(\ref{II4})
does not contain the envelope function $f(\phi)$ and therefore it is divergent.
One can regularize it by using the prescription of~\cite{Boca-2009} which leads to
\begin{eqnarray}
\widetilde A_0(\ell)&=&
\frac{1}{2\pi\ell}
\int\limits_{-\infty}^{\infty} d\phi f(\phi)\,
{\rm e}^{i\ell\phi -i{\cal P}^{(lin)}(\phi)}
\cos(\phi +\CEP)
\nonumber\\
&\times&\left({\alpha}
-{4\beta}f(\phi)\cos(\phi + \CEP)
\right).
\label{II7}
\end{eqnarray}
This equation results in the identity
\begin{eqnarray}
\ell\widetilde A_0(\ell) -  {\alpha}\widetilde A_1(\ell)
+ 4\beta\widetilde A_2(\ell) =0
\label{II8}
\end{eqnarray}
and allows to express the partial probability in
the form
\begin{eqnarray}
\frac12\,w(\ell)^{(lin)} &=&
|\widetilde A_0(\ell)|^2 + \xi^2(2u-1)\nonumber\\
 &\times &\left(|\widetilde A_1(\ell)|^2 -
 {\rm Re}[\widetilde A_0(\ell)\widetilde A^*_2(\ell)]\right)~,
 \label{II9}
 \end{eqnarray}
 which resembles the expression
 for the probability in the IPA case, i.e.\ a monochromatic background field~\cite{Ritus-79}
\begin{eqnarray}
\frac12\, w^{(lin)}_n = A_0^2
 + \xi^2(2u-1)\left( A_1^2
 - A_0A_2\right)~,
 \label{II99}
 \end{eqnarray}
 by replacing the basis functions $ \widetilde A_m(\ell)\to
 A_m\equiv A_m(n\alpha\beta)$ determined as
 \begin{equation}\label{II10}
 A_m(n\alpha\beta)=\frac{1}{2\pi}\int\limits_{-\pi}^{\pi} d\phi
 \cos^m(\phi)\,\,{\rm e}^{in\phi - i\alpha\sin\phi + i\beta\sin2\phi}
 \end{equation}
 with $\beta=\beta/(1+\xi^2/2)$ and $z=z/(1+\xi^2/2)$,
 as well as with obvious substitutions
 $\CEP=0$, $\ell\to n$,
 $\int\limits_\zeta d\ell\to \sum\limits_{n=n_{\rm min}}$,
see \cite{Ritus-79} for details.

\subsubsection{Circular polarization}

As mentioned above, the non-linear BW process in a
circularly polarized short laser pulse was considered in some detail in
Refs.~\cite{TitovPEPAN,TitovPRA,CEPTitov}.
For convenience and easy reference we reproduce the  main
expressions of these studies.
 The partial probabilities are expressed through
 basis functions $Y_{\ell},\,X_{\ell}$:
\begin{eqnarray}
\label{II12}
&& w^{(circ)}(\ell) = 2 |\widetilde Y_\ell(z)|^2+\xi^2(2u-1)\nonumber\\
&& \hspace*{-3mm}
\times \left(|Y_{\ell-1}(z)|^2 + |Y_{\ell+1}(z)|^2 -2
{\rm Re}\,(\widetilde Y_\ell(z)X^*_\ell(z))\right)
\end{eqnarray}
with
\begin{eqnarray}
Y_\ell(z)&=&\frac{1}{2\pi} {\rm e}^{-i\ell(\phi_e-\CEP)}\int\limits_{-\infty}^{\infty}\,
d\phi\,{f}(\phi)
\,{\rm e}^{i\ell\phi-i{\cal P}^{(circ)}(\phi)} ~,\nonumber\\
X_\ell(z)&=&\frac{1}{2\pi}{\rm e}^{-i\ell(\phi_e-\CEP)} \int\limits_{-\infty}^{\infty}\,
d\phi\,{f^2}(\phi)
\,{\rm e}^{i{\ell} \phi-i{\cal P}^{(circ)}(\phi)}~,\nonumber\\
%%%
\widetilde Y_\ell(z)&=&\frac{z}{2l} \left(Y_{\ell+1}(z) +
Y_{\ell-1}(z)\right) - \xi^2\frac{u}{u_l}\,X_\ell(z)~,
\label{YX1}
\end{eqnarray}
 where
 \begin{eqnarray}
{\cal P}^{(circ)}(\phi)&=&z\int\limits_{-\infty}^{\phi}\,d\phi'\,
\cos(\phi'-\phi_e+\CEP)f(\phi')\nonumber\\
&-&
\xi^2\frac{\ell u}{u_\ell}\int\limits_{-\infty}^\phi\,d\phi'\,f^2(\phi')~.
\label{YX2}
\end{eqnarray}

Equation (\ref{II12}) recovers the known expression for
the partial probability in the IPA case \cite{Ritus-79}
\begin{eqnarray}
w^{(circ)}_n&=&
 2 J^2_n(z') +\xi^2(2u-1)\times\nonumber\\
&&\left(J^2_{n-1}(z')+ J^2_{n+1}(z')-2J^2_n(z')\right)
\label{II13}
\end{eqnarray}
by the substitutions $\ell\to n$, $|\widetilde Y_\ell(z)|^2\to J^2_n(z')$,
$|Y_{\ell\pm1}(z)|^2\to J^2_{n\pm1}(z')$,
${\rm Re}\,(\widetilde Y_\ell(z)X^*_\ell(z))\to J^2_n(z')$,
and
$
z'=(2n\xi)/(1+\xi^2)^{1/2}\sqrt{{u}/{u_n}\left(1-{u}/{u_n}\right)}
$
with $u_n=n(k\cdot k')/2m^2_*$.\\

\subsection{Numerical results}

The dependence of the cross sections on the dynamic variables
$\xi^2,\,\zeta,\,u$, and angle $\phi_{e'}$
for circular and linear
polarizations is determined by the properties of the basis functions
$Y_{\ell},\,X_{\ell}$ and $\widetilde A_m(\ell)$, respectively.
This dependence is different for the two polarizations
and manifests itself in both the total and differential cross sections.
\begin{figure}[th]
 \includegraphics[width=0.48\columnwidth]{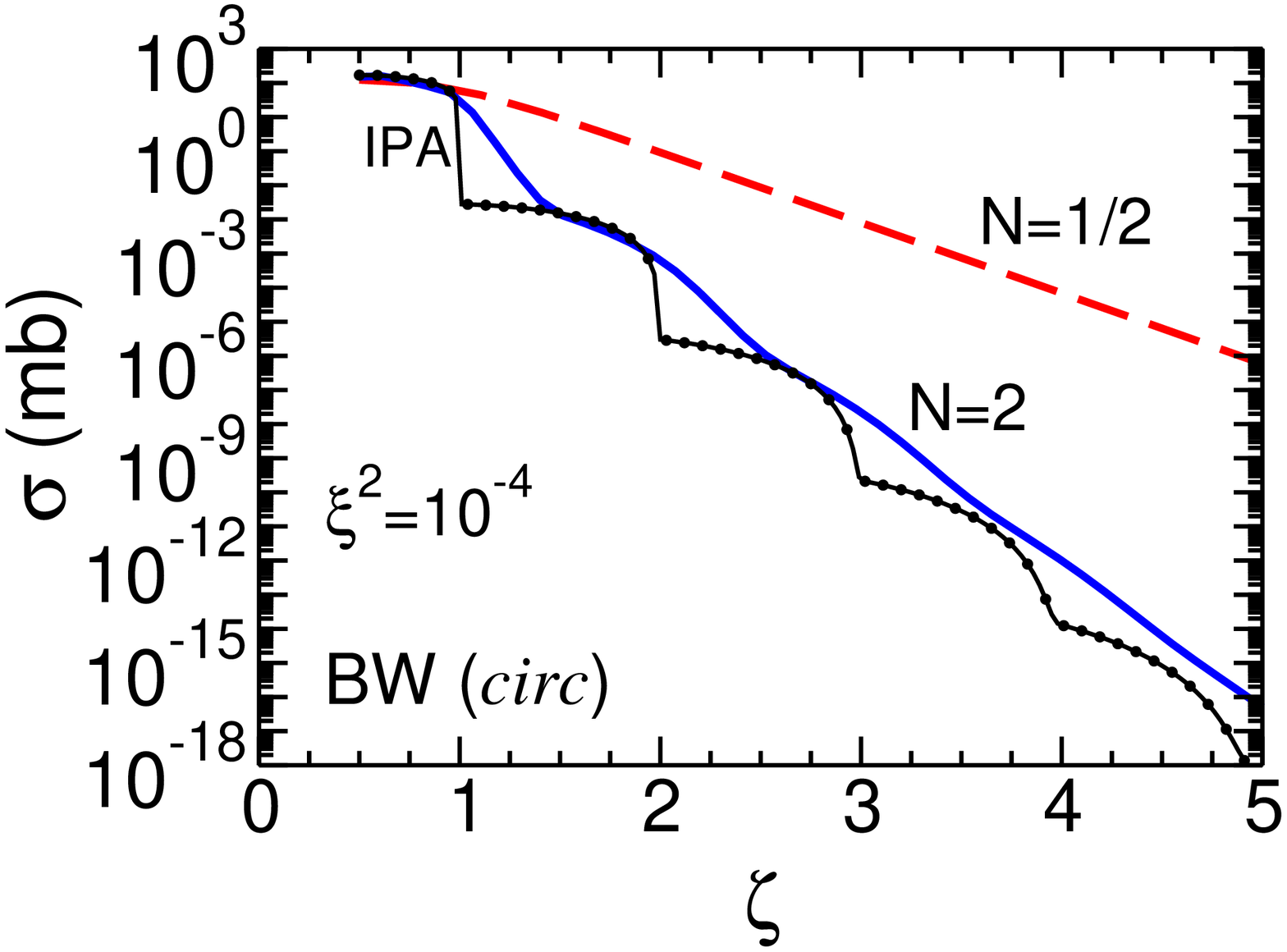}\hfill
 \includegraphics[width=0.48\columnwidth]{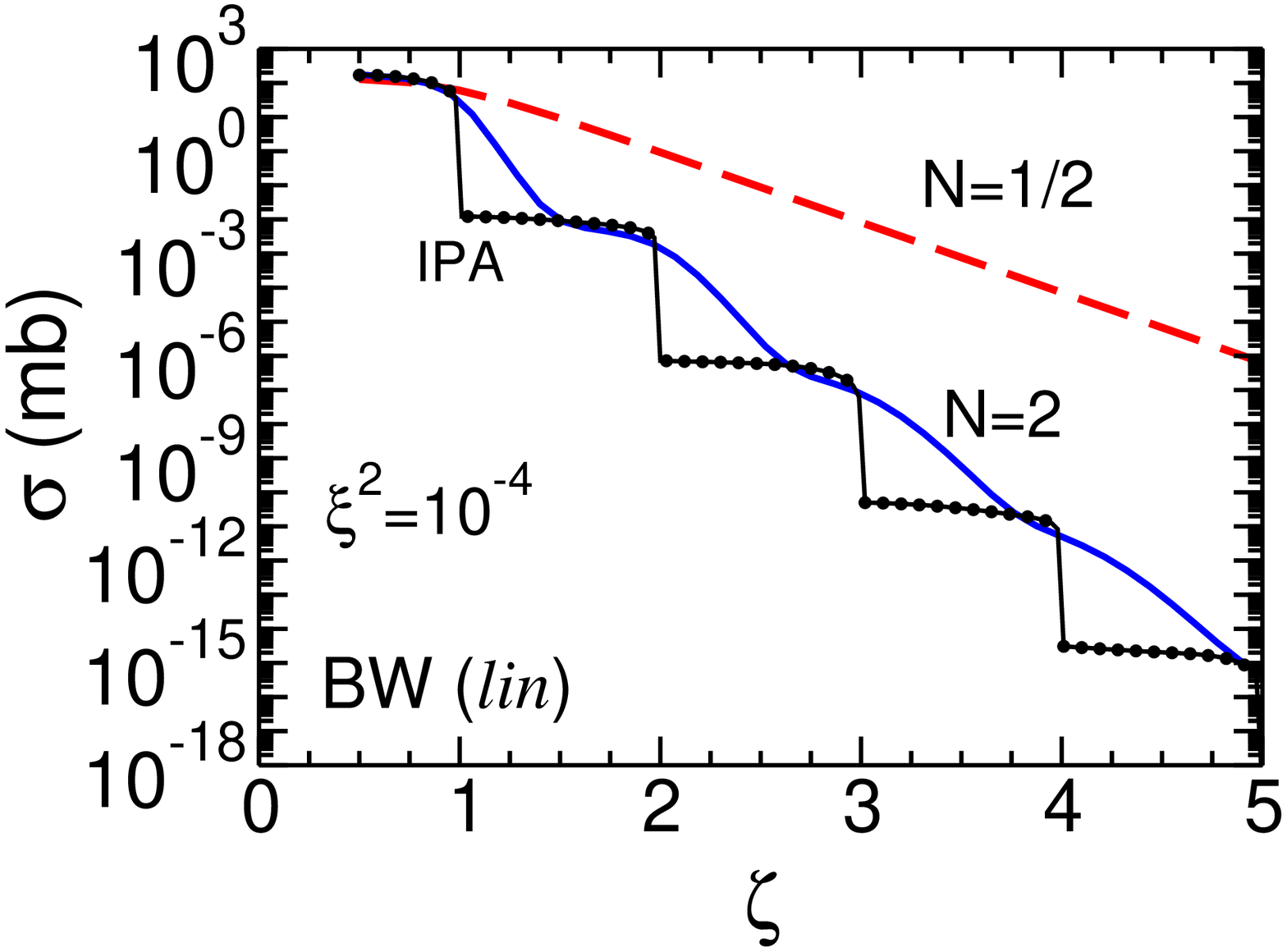}\\
 ~\\
 \includegraphics[width=0.48\columnwidth]{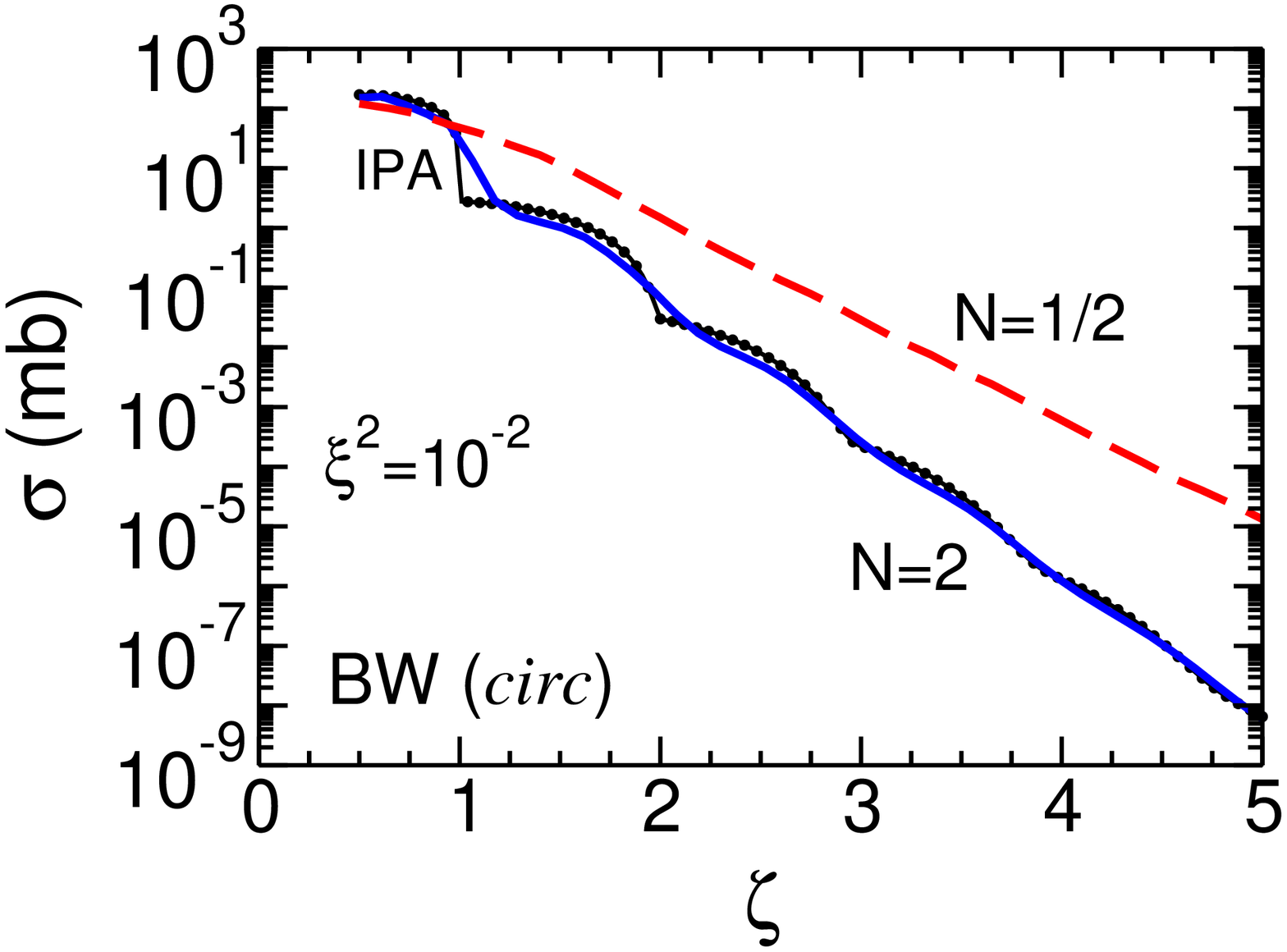}\hfill
 \includegraphics[width=0.48\columnwidth]{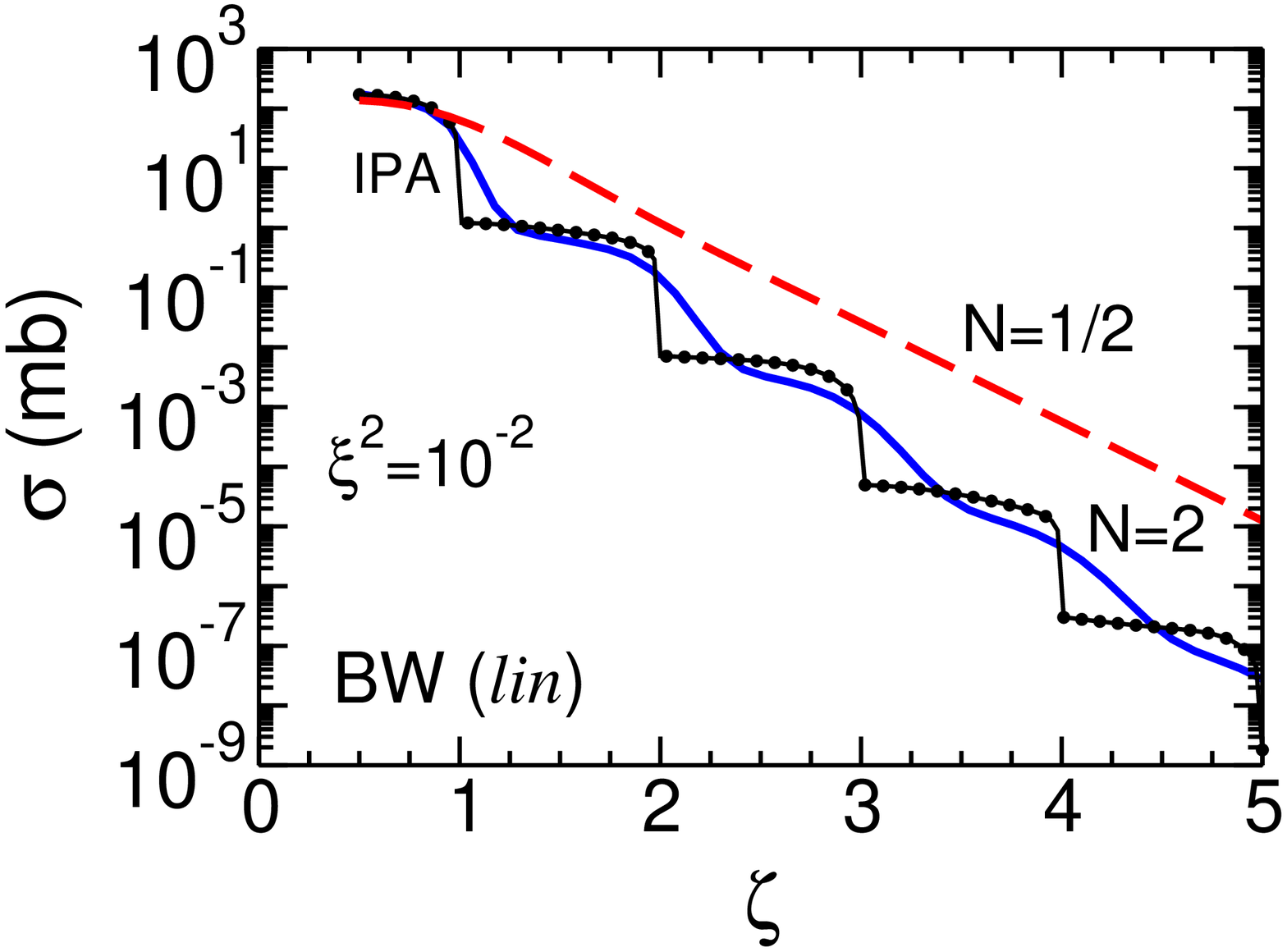}
 ~\\
  \includegraphics[width=0.48\columnwidth]{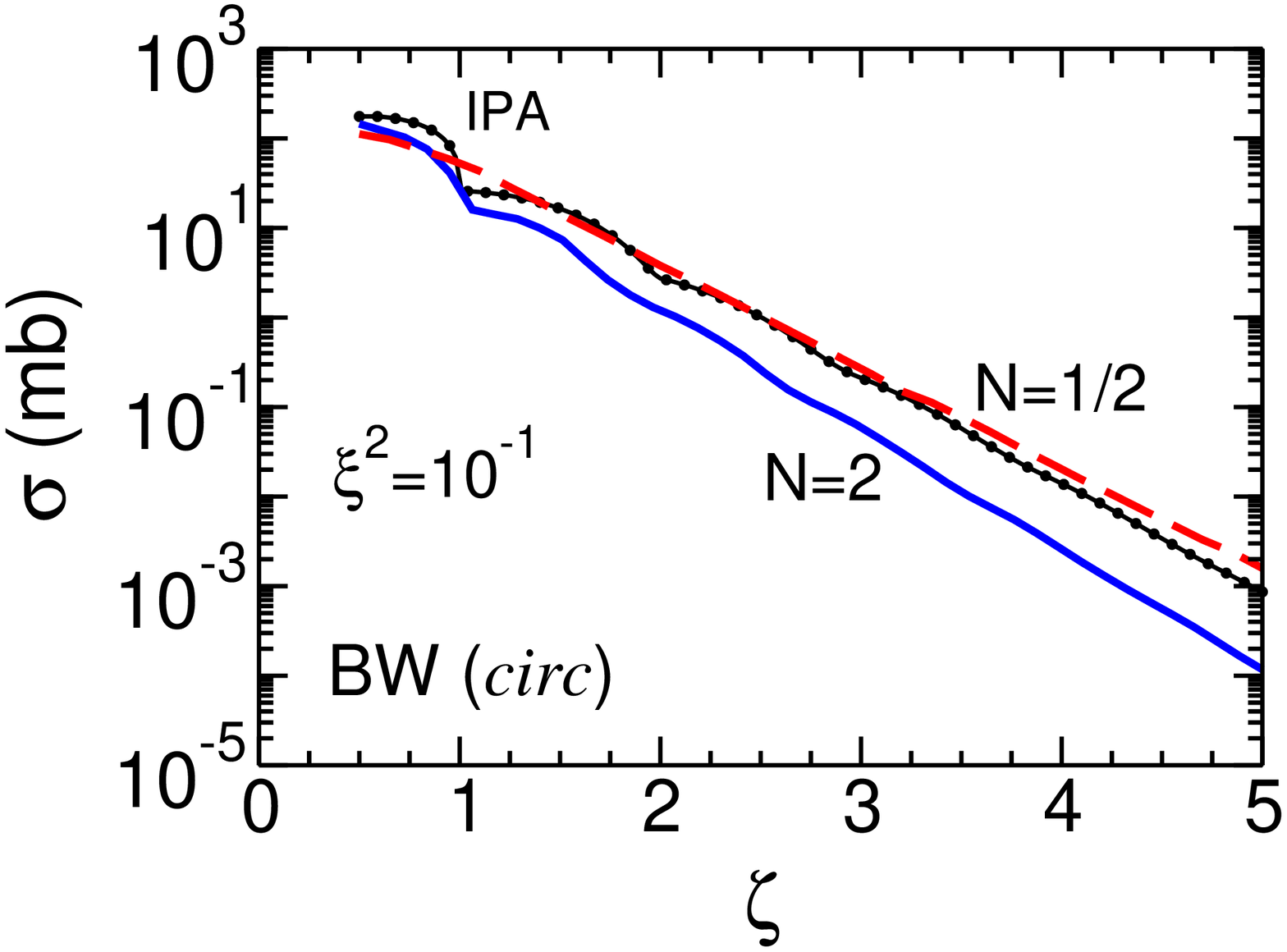}\hfill
  \includegraphics[width=0.48\columnwidth]{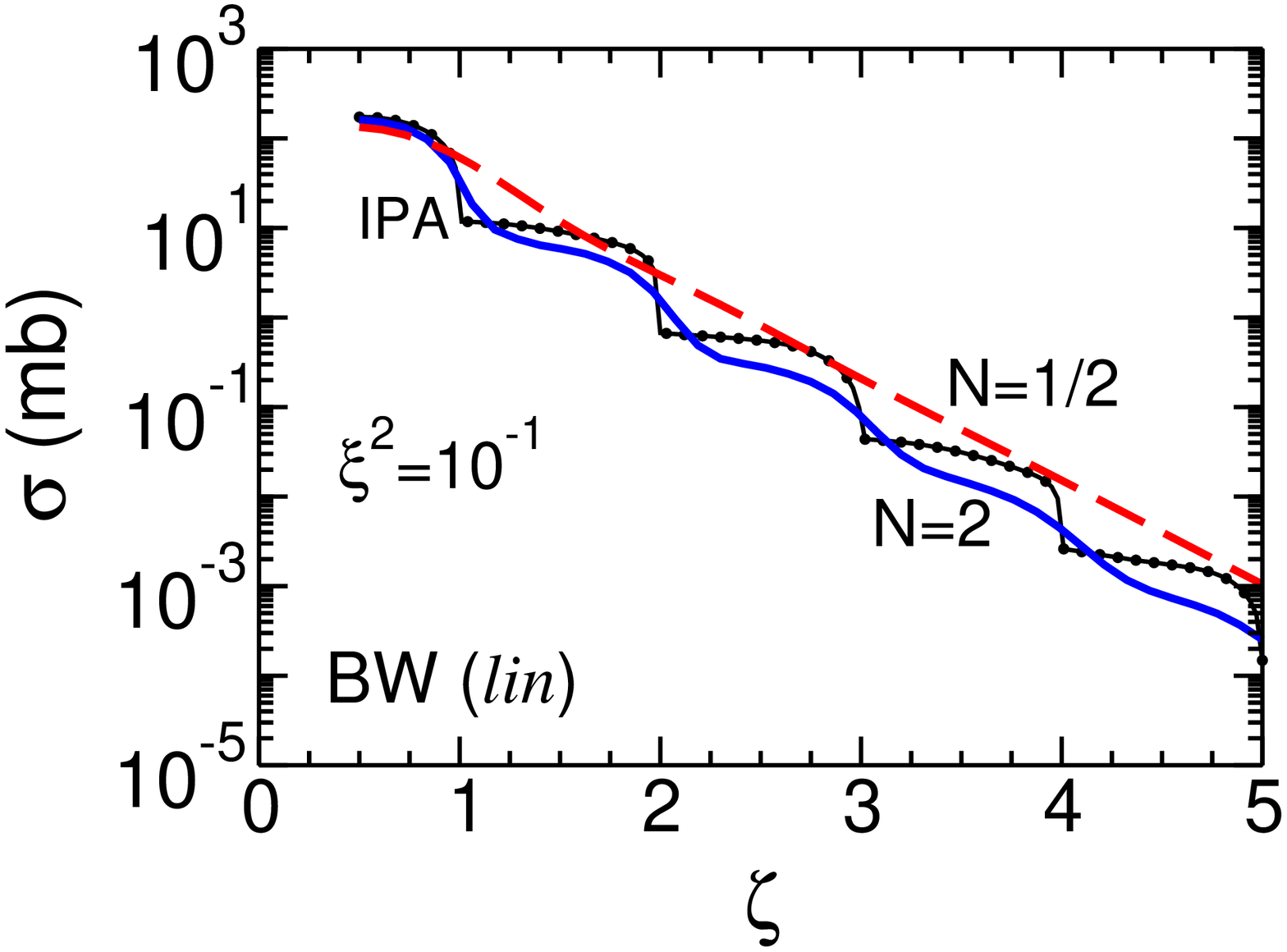}\\
 \caption{\small{(Color online)
The total cross sections of the non-linear Breit-Wheeler $\ee$-pair production
as a function of the sub-threshold parameter $\zeta$ for
$\xi^2=10^{-4}$ (top), $10^{-2}$ (middle) and $10^{-1}$ (bottom).
The left and right columns correspond to the $circ$ and
$lin$ polarizations, respectively. The red dashed and blue solid
curves are for ultra-short and short pulses with number of cycles $N=1/2$
and $2$, respectively. The thin solid curves marked by dots are
for the IPA case.
 \label{Fig:01}}}
 \end{figure}

\subsubsection{Total cross sections}

 The total cross sections as a function of the
 sub-threshold parameter $\zeta$ are exhibited in Fig.~\ref{Fig:01}
 in the upper, middle and lower panels
for $\xi^2=10^{-4},\,10^{-2}$ and $10^{-1}$, respectively.
 The left and right panels in Fig.~\ref{Fig:01}
 correspond to the circular and
 linear polarizations, respectively.
 The red dashed and blue thick solid curves
 correspond to ultra-short and short pulses with the
 number of oscillations in a pulse  $N=1/2$
 and $2$, respectively. The thin solid curves marked by dots
 are for the IPA case, see (\ref{II14}),
 i.e. a monochromatic laser background field.
%%%%%%%%%%%%%%%%%%%%%%%%%%%%%%%%%%%%%%%%%%%%%%%%%%%%%%%%%%%%%%%
 In both circular and linear polarizations, the theoretical model
 yields a step-like behavior for IPA, where each new step
 with $\zeta$
 closest to its integer valuer $n_\zeta$ corresponds
 to opening a channel with the number of
 participating photons exceeding $n_\zeta$, in accordance with Eq.~(\ref{II14}).
 %{\sl In case of linear polarization
 %such step-like behavior is more sharp and
 %in the case of circular polarization step-like behavior
 %vanishes with increasing values of $\xi^2$ and $\zeta$.}
%\marginpar{???}
 %%%%%%%%%%%%%%%%%%%%%%%%%%%%%%%%%%%%%%%%%%%%%%%%%%%%%%%%%%%%%%%%%%%%%5

For finite pulses with $N\gtrsim 2$, one can see a flattening of the step-like
 behavior, and qualitatively they are similar
 to each other for both polarizations.
 In the case of a sub-cycle pulse with $N=1/2$, the cross sections are greatly
 enhanced and they are completely smooth.
No qualitative difference of linear and circular polarizations is recognizable.

\begin{figure}[tb]
 \includegraphics[width=0.48\columnwidth]{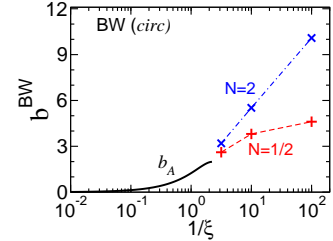}
  \caption{\small{(Color online)
Comparison of the slopes of the exponential dependence of the cross
sections $\sigma^{(circ)}(\zeta)$ as a function of $1/\xi$.
The symbols $+$ and $\times$ are for $N=1/2$ and $2$, respectively;
the thick solid curve corresponds to the asymptotic value (\ref{A2}).
 \label{Fig:0001}}}
 \end{figure}

 The cross sections decrease almost exponentially
 with increasing $\zeta$:
 $\sigma^{(i)} \propto \exp[-b^{(i)}\,\zeta]$, where the slopes
$b^{(i)}$
 depend on the pulse duration and field intensity $\xi^2$.
 Thus, it increases with increasing pulse duration (or $N$)
 and increasing $\xi$.
 Such an exponential behavior resembles the probability of the
non-linear Breit-Wheeler
 process in the asymptotic limit
 $\xi^2\gg1$ and $\zeta\gtrsim 2\xi^3$~\cite{Ritus-79,TitovPRA}
 \begin{eqnarray}
 \sigma^{(\rm BW)}_A\sim\frac{1}{\xi}\exp[-b_{\rm A}\zeta]~,
 \label{A1}
 \end{eqnarray}
 where
 \begin{equation}
 b_{\rm A}= \frac{4}{3\xi}\,\left(1-\frac{1}{15\xi^2}\right)~.
 \label{A2}
 \end{equation}
 Despite the fact that the values of $\xi$ and $\zeta$
 used in the present numerical calculations are far from their asymptotic values,
 in Fig.~\ref{Fig:0001} we nevertheless  compare for completeness
the slopes obtained in our numerical
 calculation with their asymptotic values.
 The symbols $+$ and $\times$ are for the pulses with $N=1/2$ and $2$, respectively,
 and the thick solid curve corresponds to the asymptotic value given by Eq.~(\ref{A2}).
 One can see a clear tendency for the numerical values
 of the slopes to converge their asymptotic limit.

 In the asymptotic limit, the probability of
 $\ee$ creation for the circular polarization is greater by a
 factor of $\left(2\pi\zeta/3\xi \right)^{3/2}$ compared to the
 case of linear polarization~\cite{Ritus-79}.
 Our numerical calculation for $\xi<1$ is far from the asymptotic regime
 and results in the same order of magnitude
 of the $\ee$ production cross sections for circular and linear polarizations.

Finally, we note that the total cross sections shown in Fig.~\ref{Fig:01} as a function
of the multi-photon (threshold) parameter $\zeta$ are determined by an interplay
of two effects: (i) the multi-photon dynamics itself and (ii) the pulse shape effect.
The relative contributions of these effects vary depending on the field intensity.
They manifest themselves most vividly at low field intensities,
where they can be separated to some extent.
The first, dynamic aspect is determined
by the properties of the basis functions (\ref{II4}), (\ref{YX1})
and leads to a strong decrease
of the cross sections as the number of photons involved in the process
(or the parameter $\zeta$) increases.
The second effect is the modulation of the high-momentum
 components which are determined by a Fourier image of the envelope
 function $f(\phi)$ in (\ref{I1}). So, in the case of low intensities
 and (ultra-) short pulses, the partial probabilities as a function of
 $\ell$ are proportional to the square of
$f(\ell)=\int d\phi f(\phi) exp(-i\ell\phi)/2\pi$.
 The dominant contribution to the cross section~(\ref{II1}) comes from
 $\ell=\zeta$ and, therefore, the shape of the cross section resembles
 $f^2(\ell)$ and decreases with increasing $\zeta$.
 These effects -- for circular polarization -- are discussed in some
detail in~\cite{TitovPRA}.
 When the pulse width increases, the shape effect becomes negligible and
 the cross section is determined by the dynamic aspect, i.e.\ the
 properties of the basis functions, which
 approach the IPA basis functions for large widths (or large values of $N$).
 At intermediate field intensity $\xi$, there is an interplay of these two effects
 and it seems to be difficult to separate them.
 It turns out that in some cases the cross sections for finite pulses
 may be  smaller than the IPA prediction.

\subsubsection{Azimuthal angle distributions}

The dependence of the differential cross sections
on the azimuthal angle $\phi_e$  of the outgoing electron  for
linear and circular polarizations is mainly
determined by the phase factors
$\exp[-i(\ell\phi-{\cal P}^{(i)}(\phi))]$ in Eqs.~(\ref{II4}) and
(\ref{YX1}), respectively.
%\marginpar{eq. numbers correct?}
The case of circular polarization is considered in detail
in \cite{CEPTitov}, and therefore, below we discuss this case briefly
only for completeness.

\begin{figure}[tb]
 \includegraphics[width=0.48\columnwidth]{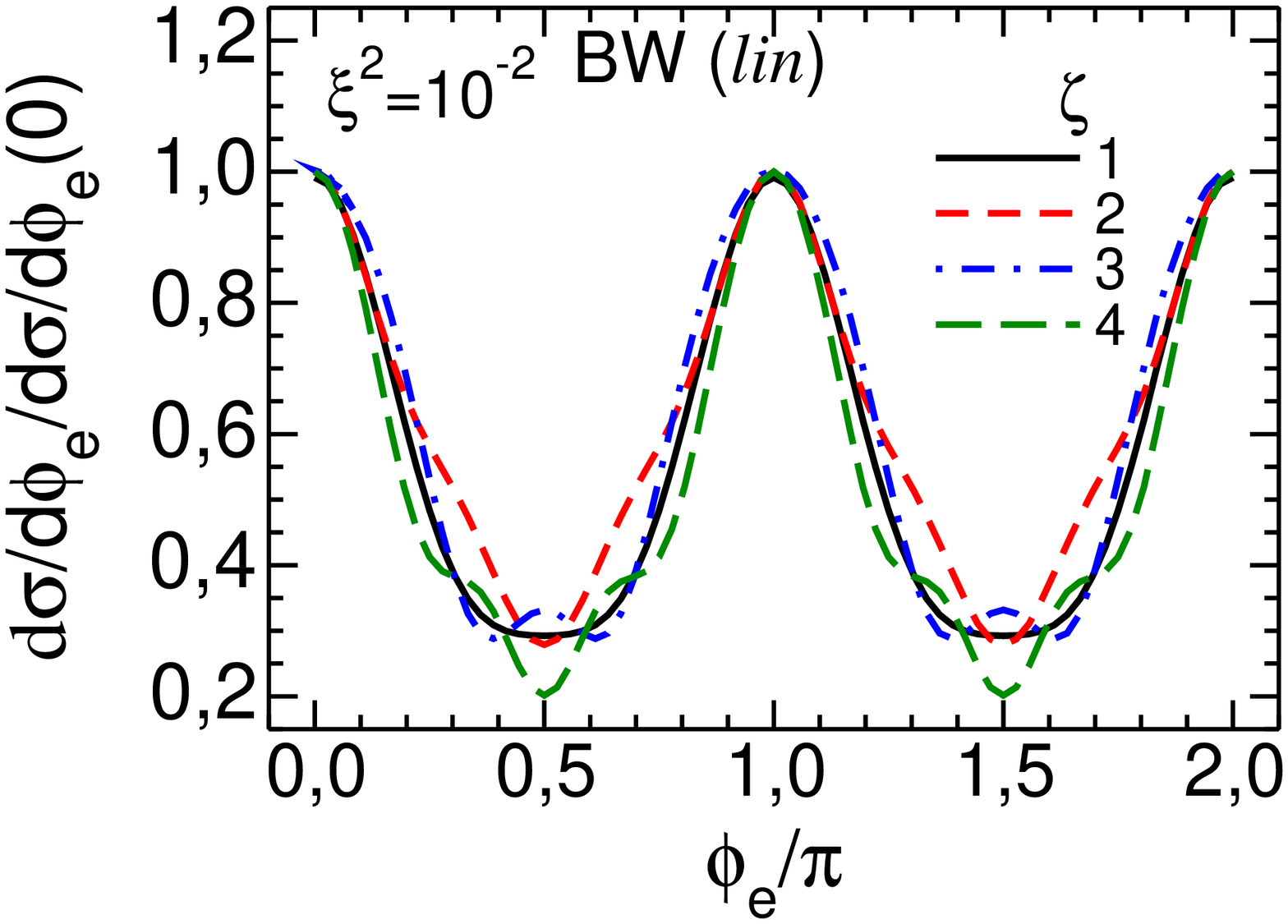}\hfill
 \includegraphics[width=0.48\columnwidth]{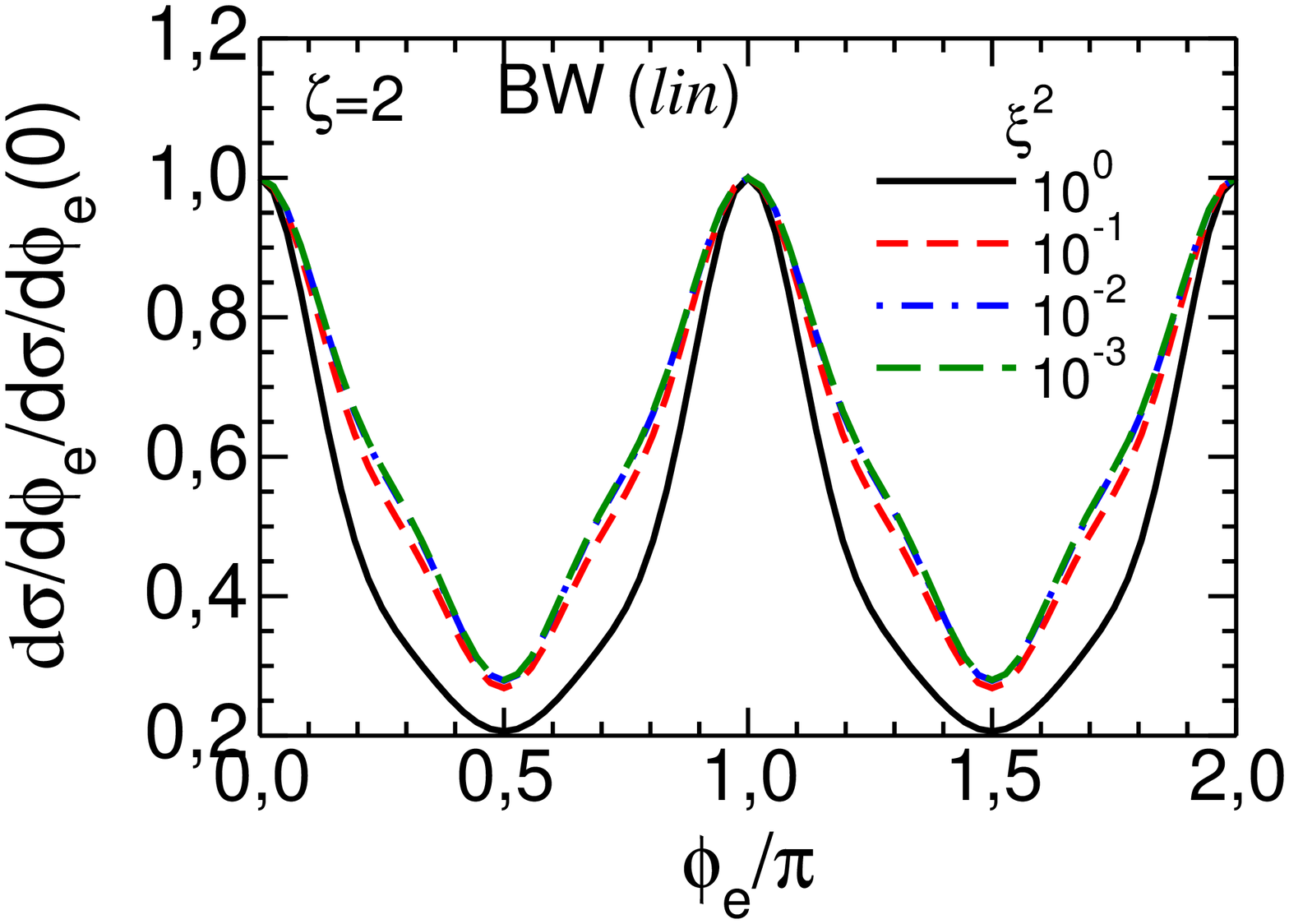}
 \caption{\small{(Color online)
The differential cross sections $d \sigma^{(lin)} /d \phi_{e}$
normalized to their maximum value at $\phi_{e}=0$.
The left and right panels show the variation with
$\xi^2$ and $\zeta$, respectively, according to the legends.
For $lin$ polarization and IPA.
 \label{Fig:03}}}
 \end{figure}

First of all note that, in the case of a circularly polarized
pulse, the differential cross section in IPA
does not depend on $\phi_e$: it is constant and
equal to $\sigma^{(circ)}_{\rm tot}/2\pi$.
In the case of linear polarization and IPA, the dependence
of $d\sigma^{(lin)}/d\phi_e$ on $\phi_e$ exhibits a non-monotonic behavior
with maxima and minima. The shape of
azimuthal angle distributions is shown in Fig.~\ref{Fig:03},
where the corresponding cross sections, normalized
to their respective maximum value at $\phi_{e}=0$, are exhibited.
The left and right panels correspond to different
values of e.m.\ field strength $\xi^2$ and sub-threshold parameter $\zeta$,
respectively.

Qualitatively, the reason of such a behavior is the following.
%For simplicity, consider case of small $\xi^2,\,z\ll 1$.
The main dependence of the basis functions $A_m(n\alpha\beta)$ in~(\ref{II10})
on $\alpha=z\cos\phi_e$ is determined by the oscillating factor
in the exponent
\begin{eqnarray}\label{II144}
&& A(n) \sim \frac{1}{2\pi}\int\limits_{-\pi}^{\pi} d\phi
 {\rm e}^{in\phi - iz\cos\phi_e\sin(\phi) + \beta\sin(2\phi)}
 \nonumber\\
&& \propto  \sum\limits_{k}
 \frac{(-z\cos\phi_{e'})^k}{2^k\,k!}
 \int\limits_{-\pi}^{\pi} d\phi
 {\rm e}^{in\phi}
 \left({\rm e}^{i\phi} - {\rm e}^{-i\phi} \right)^k .
 \end{eqnarray}
The dominant contribution to the latter sum comes from the term with $k=n$,
where $n=n_{\rm min}= Int (\zeta)\geq 1$.
Here, $Int (\zeta) \equiv \left \lfloor {\zeta}\right \rfloor$
is the floor function which returns the maximum integer
less than or equal to $\zeta$.
Since  the differential cross section
is a quadratic form of $A(n)$, one can estimate
\begin{eqnarray}\label{II15}
\frac{d\sigma^{(lin)}}{d\phi_{e}}\propto \cos^{2n_{\rm min}}\phi_{e}~,
\end{eqnarray}
which leads to
maxima at the points $\phi_{e}=0,\,\pi,$ and $2\pi$ and minima
at $\phi_{e}=\pi/2$ and $3\pi/2$.
This is in qualitative
agreement with full numerical results shown in Fig.~\ref{Fig:03}.

\begin{figure}[t]
 \includegraphics[width=0.48\columnwidth]{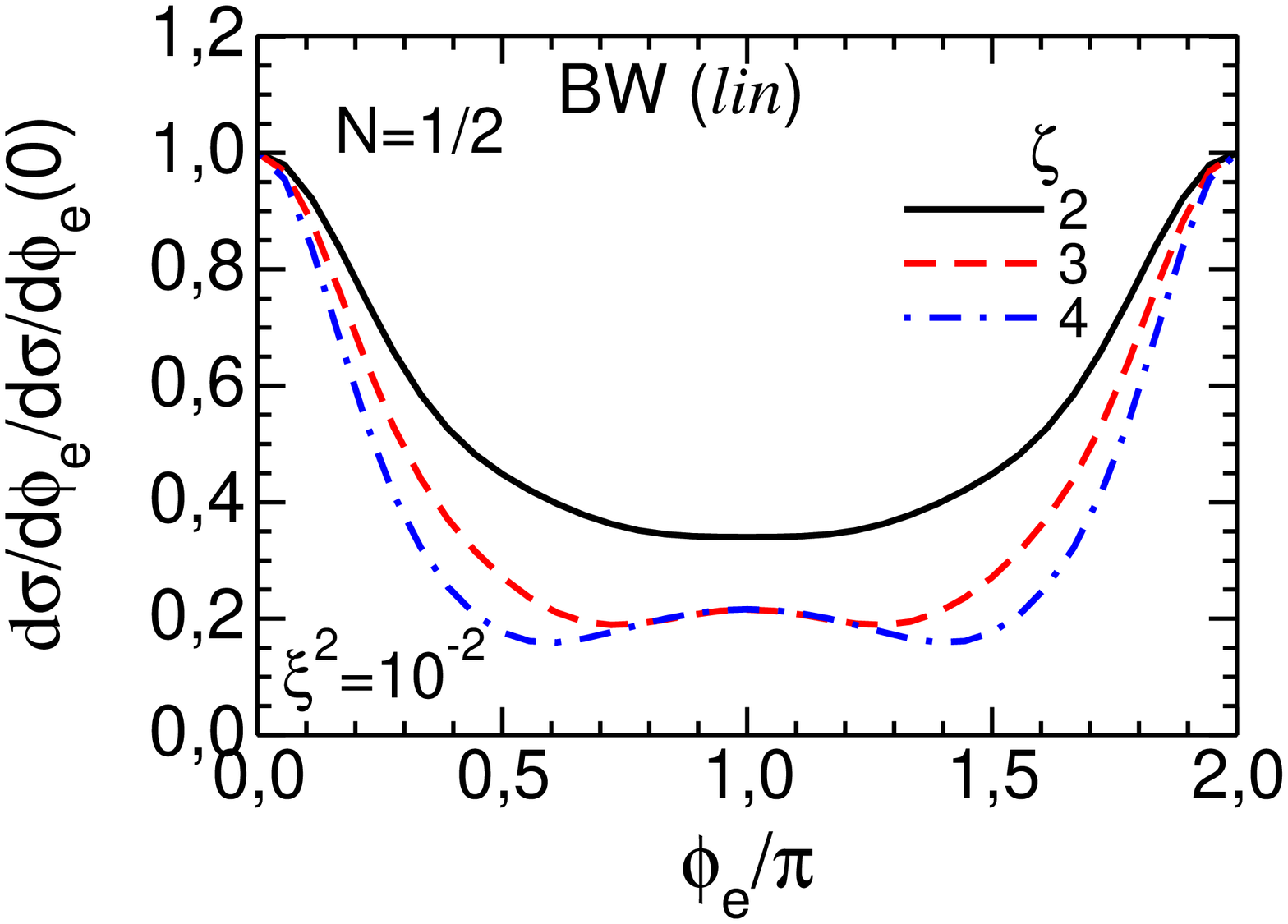}\hfill
 \includegraphics[width=0.48\columnwidth]{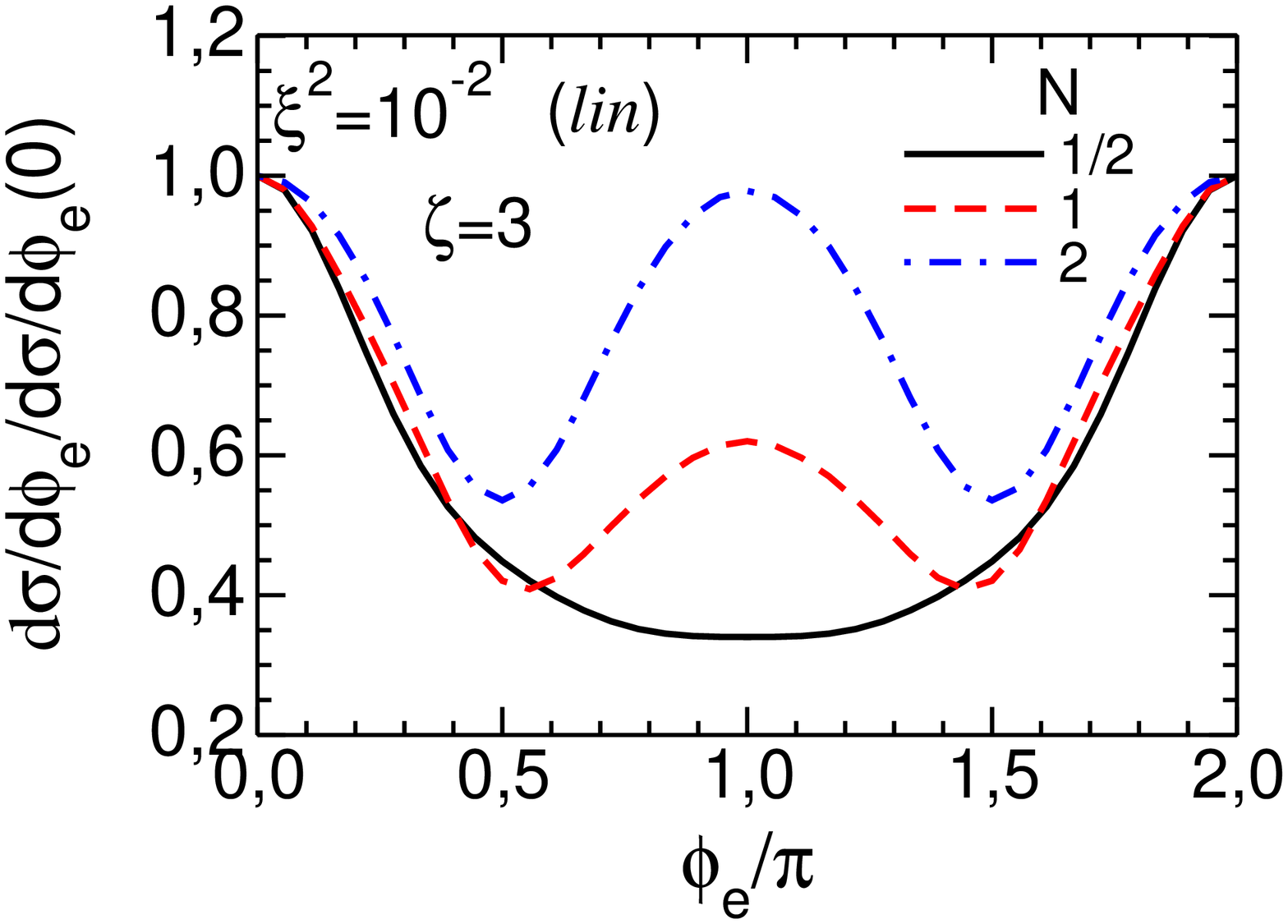}
 \caption{\small{(Color online)
The differential cross sections $d\sigma^{(lin)} / d\phi_{e}$
normalized to the respective maximum value at $\phi_e=0$ for $\xi^2=10^{-2}$.
The left and right panels are for different
values of $\zeta$ for $N=1/2$ and for different values of $N$ for
$\zeta=3$, respectively.
For $lin$ polarization and FPA and $\CEP=0$.
 \label{Fig:04}}}
 \end{figure}

 The differential cross sections for FPA
 and $\CEP=0$ are exhibited in Fig.~\ref{Fig:04}.
 Again, in order to emphasize the shape of the distribution we present
 azimuthal angle distributions of the outgoing electron
 ${d\sigma^{(lin)}}/{d\phi_{e}}$
 normalized to their respective maximum value at $\phi_e=0$.

The results for a sub-cycle pulse with $N=1/2$ for different values of $\zeta$
are exhibited in the left panel.
One can see a strong enhancement of the cross sections at $\phi_e=0,\,2\pi$
and some decrease of them when $\phi_e$ tends to $\pi$.
Qualitatively, this is due to the fact that the oscillating factor in
the basis functions $A(\ell)$ (cf. Eq.~(\ref{II10})) is proportional to
the exponent
\begin{eqnarray}\label{II16}
 {\rm e}^{i(l- z\cos\phi_e)\phi}~,
 \end{eqnarray}
with lowest oscillations and maximum contributions to the
differential cross sections just at $\phi_e=0,\,2\pi$.

The right panel of Fig.~\ref{Fig:04} illustrates the evolution
of the differential cross sections with increasing pulse duration (or $N$).
The result for $N=2$ is close to that for
the IPA case (cf. Fig.~\ref{Fig:03}).

\begin{figure}[hb]
 \vspace{3mm}
 \includegraphics[width=0.48\columnwidth]{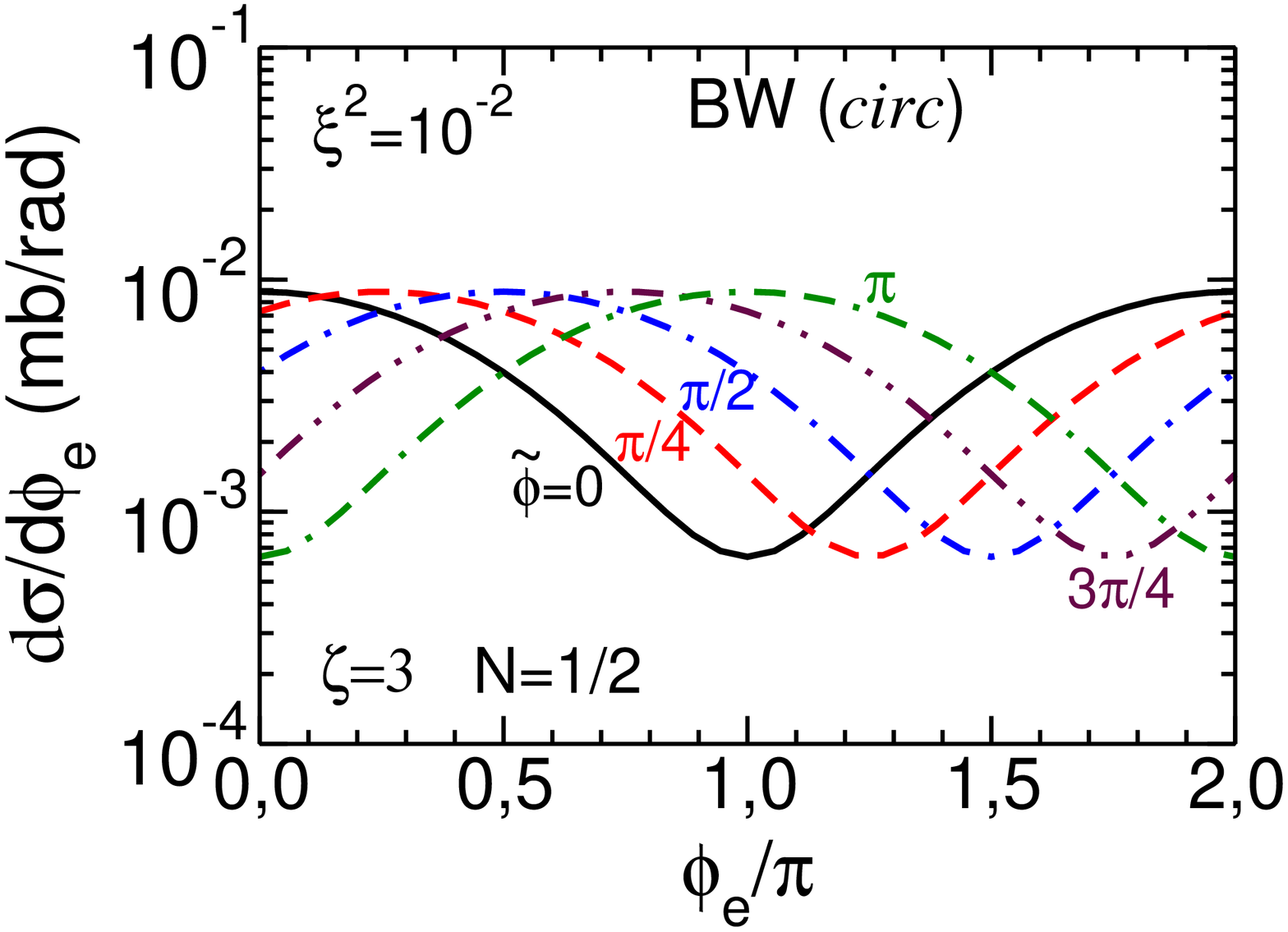}\hfill
 \includegraphics[width=0.48\columnwidth]{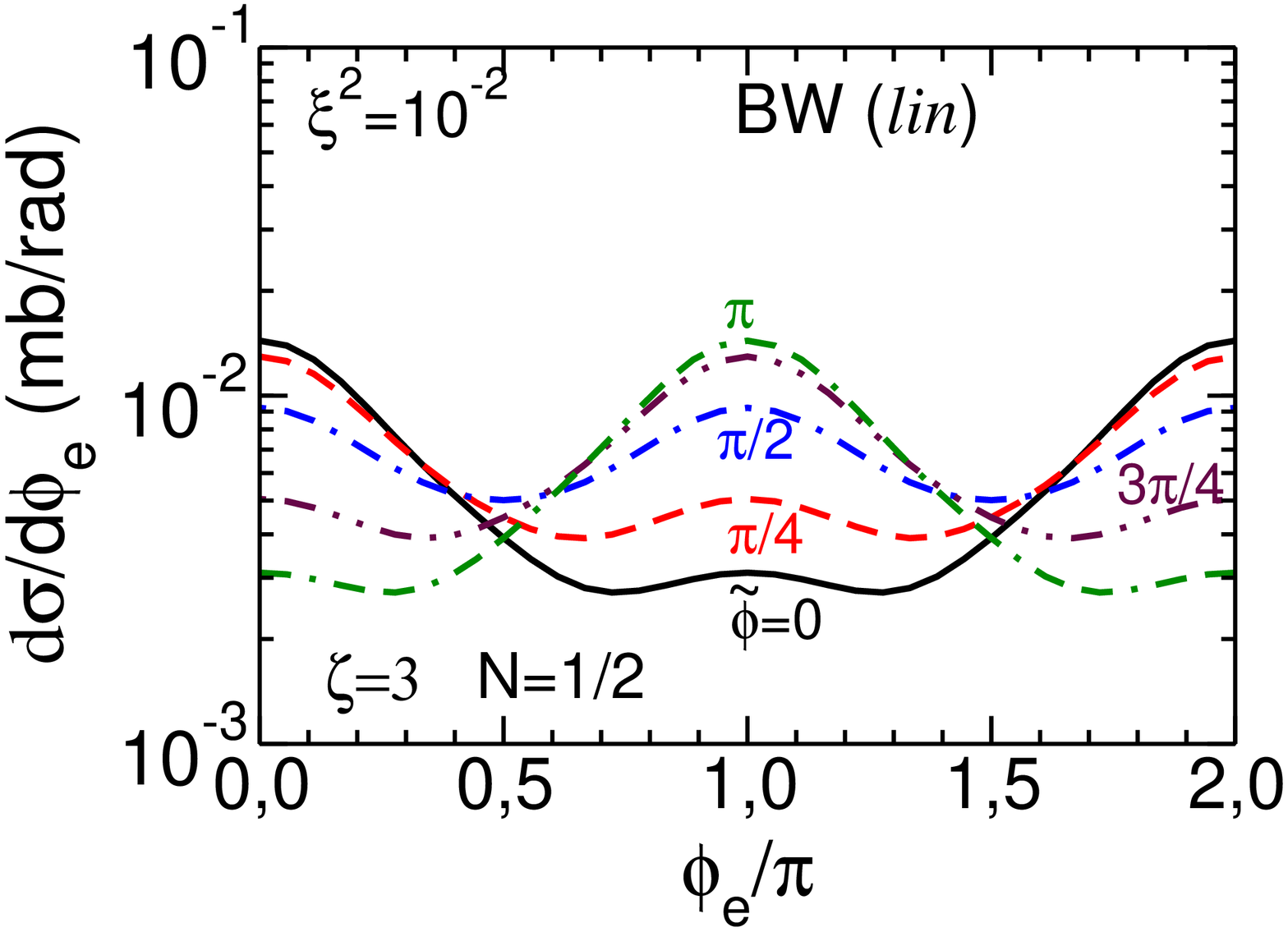}
 ~\\
 \includegraphics[width=0.48\columnwidth]{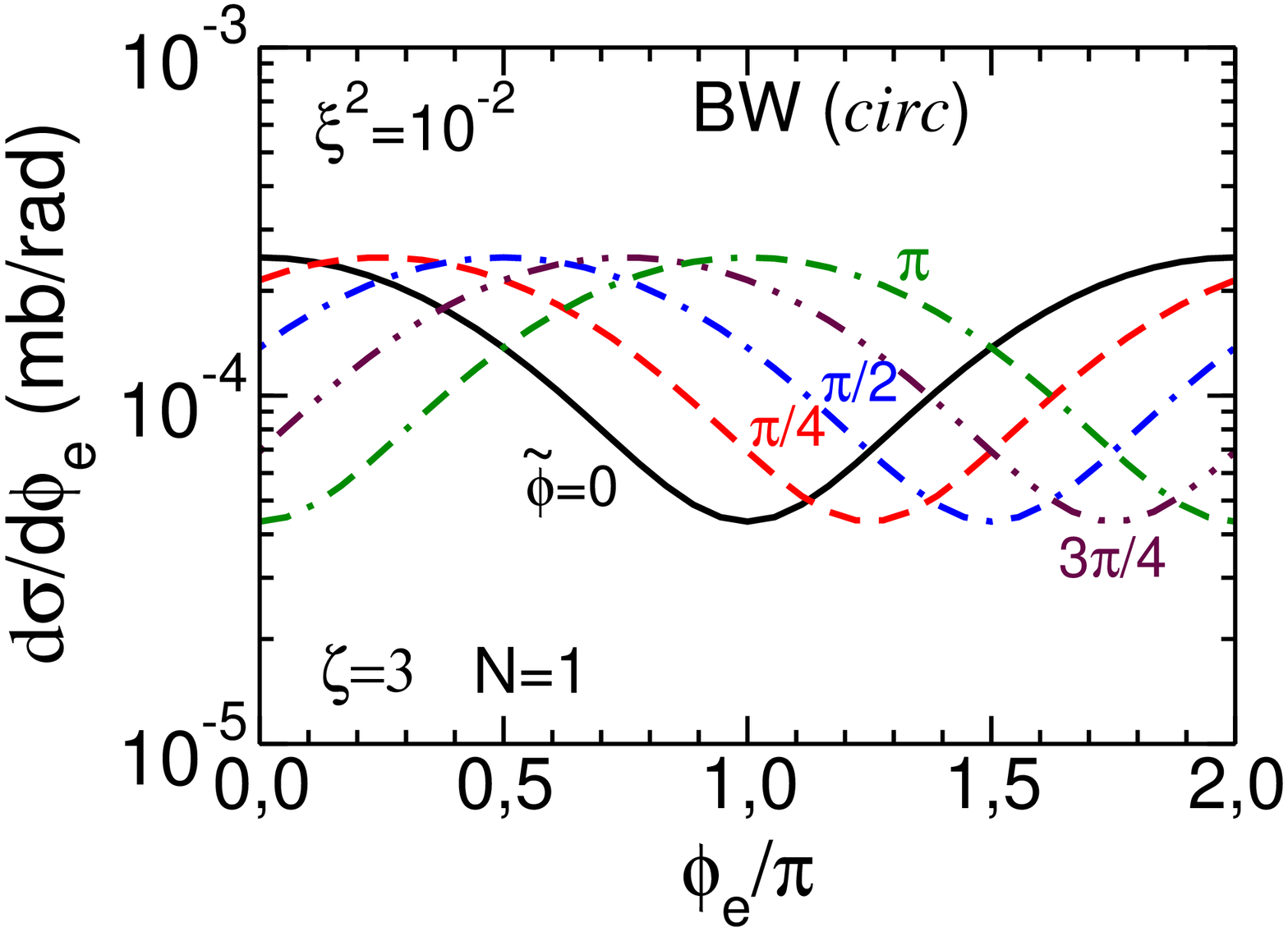}\hfill
 \includegraphics[width=0.48\columnwidth]{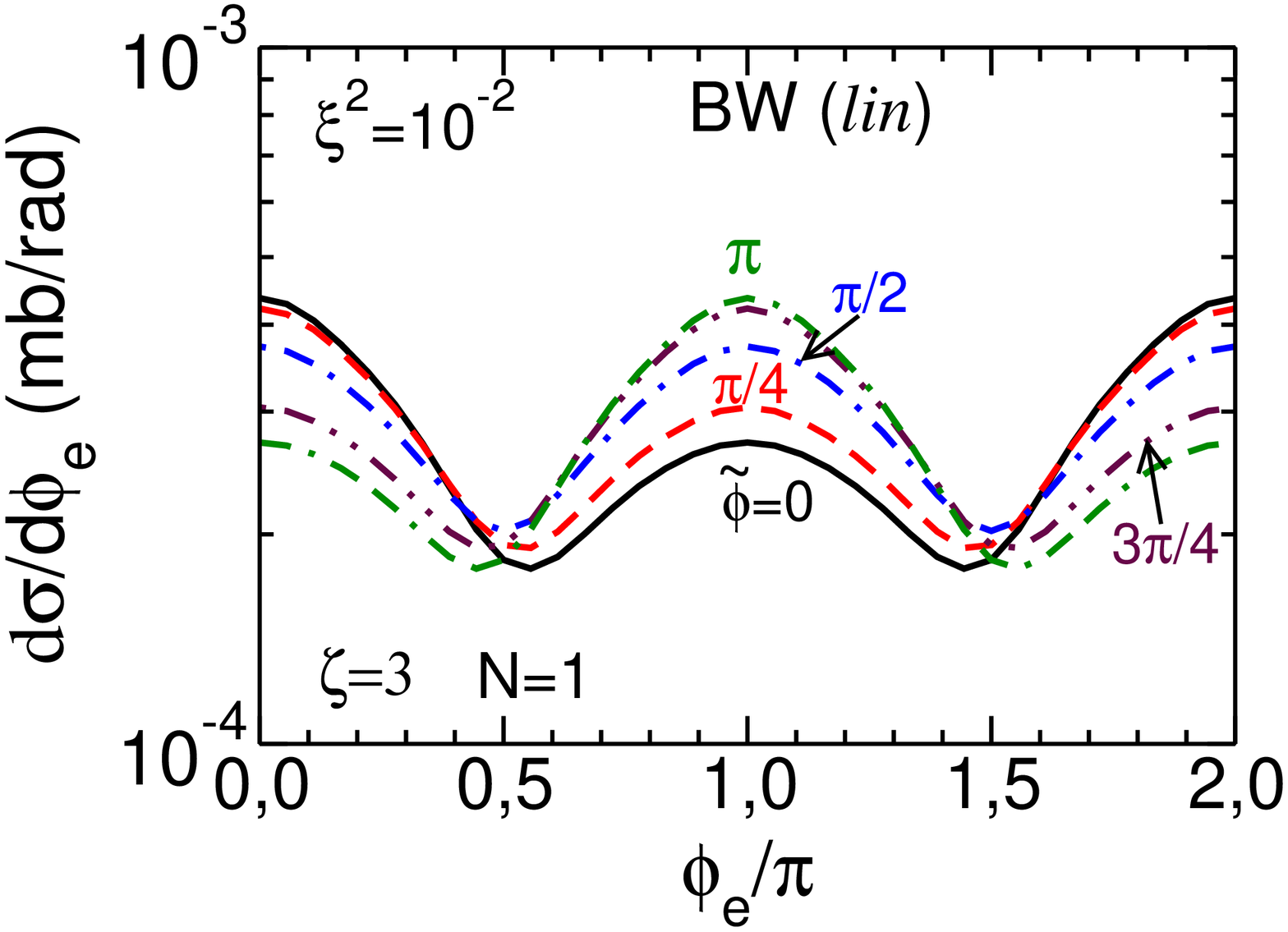}
 ~\\
 \includegraphics[width=0.48\columnwidth]{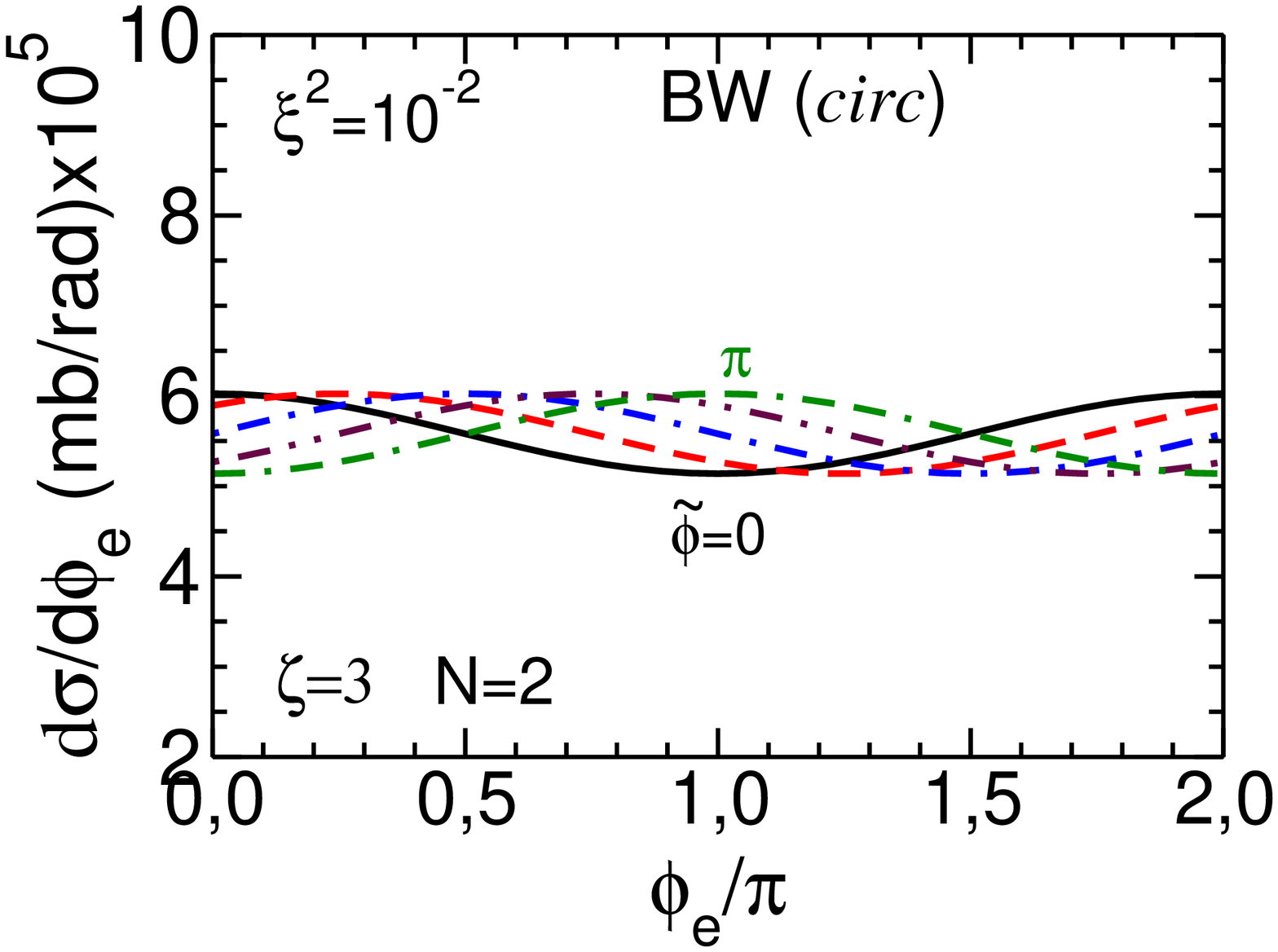}\hfill
 \includegraphics[width=0.48\columnwidth]{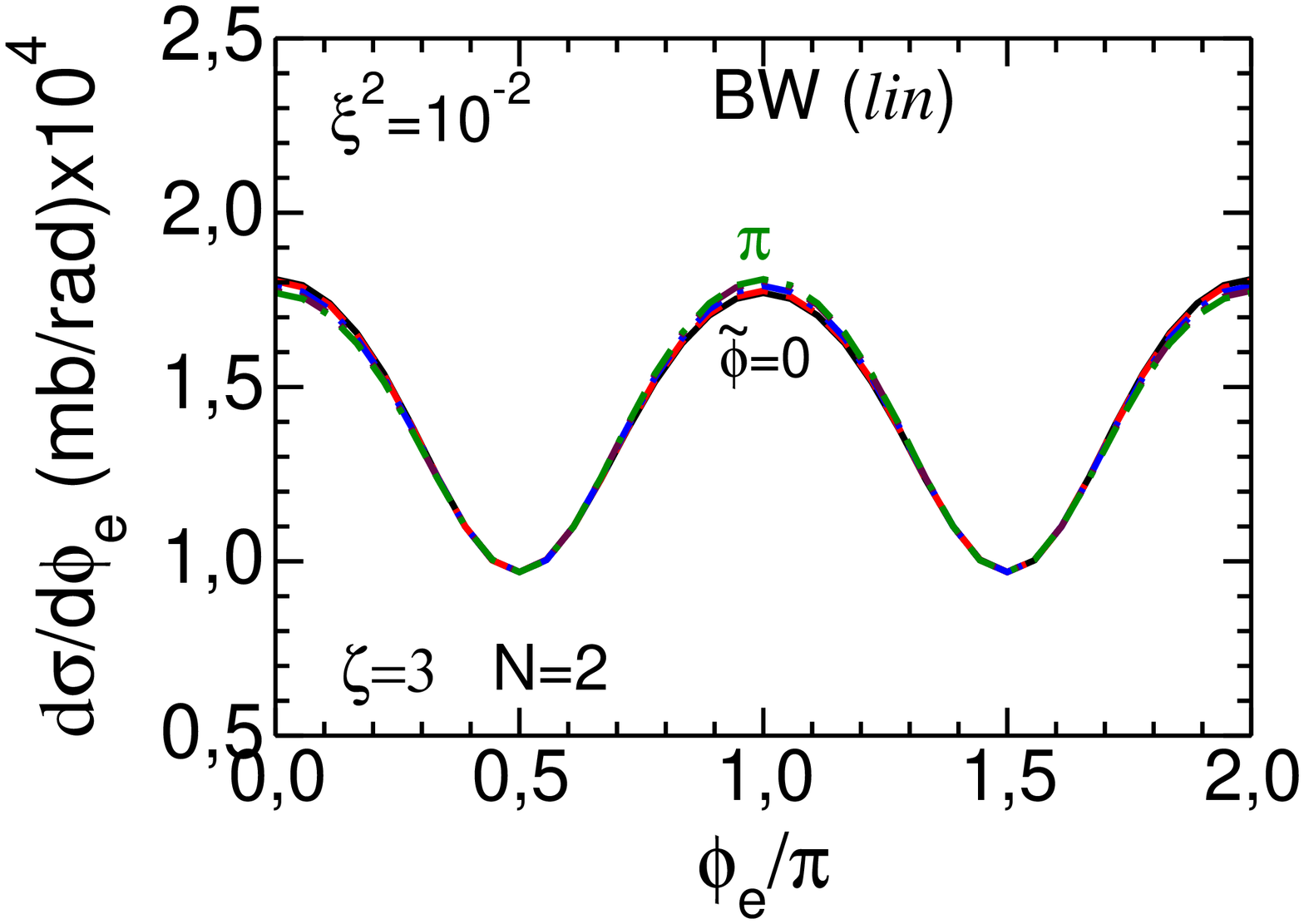}
 \caption{\small{(Color online)
 The differential cross sections $d \sigma / d \phi_e$
 as a function of $\phi_e$ for different values of $\tilde \phi \equiv\CEP$
 for $\zeta=3$ at $\xi^2=0.01$.
The left and right panels are for $circ$ and
$lin$ polarizations, respectively.
The top, middle and bottom panels are for $N=0.5$, 1 and 2,
respectively.
 \label{Fig:05}}}
 \end{figure}

The interplay of the azimuthal angle $\phi_{e}$ of the outgoing electron
and the carrier envelope phase $\CEP$ is interesting and important.
Thus, an analysis of the corresponding angular distribution can serve
as a method for the $\CEP$ determination~\cite{CEPTitov}.
Our results for circular and linear polarizations
are exhibited in Fig.~\ref{Fig:05} in the left and right panels,
respectively. The
calculations are performed for different pulse durations at
fixed $\xi^2=10^{-2}$ and $\zeta=3$.

 In the case of the circular polarization,
 one can see a clear bump-like structure in the cross sections.
 The bump position coincides with the corresponding value of
 the carrier phase. In ultrashort pulses,  the relative prominence of the bumps
 can reach an order of magnitude.
%\marginpar{???}
 The reason for such a behavior is
 that the basis functions $Y_\ell$ and $X_\ell$ are determined by the integral
 over $d\phi$ with a rapidly oscillating exponential function $\exp[i\Psi]$ with
\begin{eqnarray}
\Psi= \ell \phi
 & - & z \cos(\phi_e-\CEP)\int\limits_{-\infty}^{\phi} d\phi'\, f(\phi')\cos\phi'
 \nonumber\\
 & - & z \sin(\phi_e-\CEP)\int\limits_{-\infty}^{\phi} d\phi'\, f(\phi')\sin\phi'.
 \label{UU8}
 \end{eqnarray}
For short pulses, the envelope function $f(\phi)$ is
essentially non-zero
in the neighborhood of $\phi=0$, and the symmetry property $f(\phi)=f(-\phi)$ holds.
The integral on the sign-changing function $\sin(\phi)f(\phi)$
is much smaller than the integral on the sign-constant function $\cos(\phi)f(\phi)$
in the region of small $|\phi|$, where $f(\phi)$ is finite.
 This results in the inequality
 \begin{eqnarray}
 \int\limits_{-\infty}^{\phi} d\phi'\, f(\phi')\cos\phi'
 \gg
 \int\limits_{-\infty}^{\phi} d\phi'\, f(\phi')\sin\phi'~,
 \label{UU81}
\end{eqnarray}
 which leads to the fact
 that the main contribution to the probability comes from the region
 $\phi_e\simeq\CEP$, which is confirmed by the result of our full
 calculation shown in Fig.~\ref{Fig:05} (left panels).
 The effect of the carrier envelope phase decreases with increasing pulse
 duration, and for $N\ge 2$ it becomes small, where the relative height of the bumps
 is about 1.2 and the cross sections become close to the constant value
  $\sigma^{(circ)}/2\pi$.

  The azimuthal angle distributions in the case of linear polarization
  are exhibited in the right panels of Fig.~\ref{Fig:05}.
  The effect of a finite value of $\CEP$ is most pronounced for sub-cycle pulses
  (cf.\ right top panel in~Fig.~\ref{Fig:05}):
%\marginpar{Fig. 6?}
Similar to IPA,
  the cross sections have a bump at $\phi_e=\pi$.
  But now, the  height of the bump depends on  $\CEP$.
  It has a maximum at $\CEP=\pi$ and becomes negligibly small when $\CEP=0$.
%\marginpar{which $\phi$?}
  This is explained by the fact that the oscillation factor~(\ref{II16}) is modified
  as
\begin{eqnarray}\label{II20}
 {\rm e}^{i(l- z\cos\phi_e\cos\CEP)\phi}~,
 \end{eqnarray}
 where $\phi=k\cdot x$ is invariant phase.
 This factor results in minimal oscillations
 of the basis functions $\widetilde A_{\ell}$
 or maximum values of cross sections for the combinations
$\phi_e=0$, $\CEP=0$, and $\phi_e=\pi$, $\CEP=\pi$.
Similar to the circular polarization, the impact of varying $\CEP$ decreases with
increasing pulse duration (or $N$),
and at $N\gtrsim2$ it becomes insignificant,
making the differential distribution very close to the IPA prediction.

\section{Non-linear Compton scattering}

Compton scattering, symbolically
 $e^-+ L \to e^{-}{}' + \gamma' $, is considered here
 as the spontaneous emission
 of a photon by an electron in an external e.m.\ field (\ref{I1}).
 As mentioned in the introduction,
 the non-linear Compton process in the Furry picture
 is described as $e_L^- \to e_L^- {'} + \gamma'$.
 Below we use standard notations and definitions for the Compton process:
 %slightly modified version of Refs.~\cite{TitovEPJD,CEPTitov} for
 %keeping the main notation and definition unchanged.
 the four-momenta of the
 incoming electron, background (laser) field (\ref{I1}),
 outgoing electron and photon are $p(E,{\mathbf p})$, $k(\omega,{\mathbf k})$,
 $p'(E',{\mathbf p}')$, $k'(\omega',{\mathbf k}')$,
 respectively. The variables $\cos\theta'$ and $\phi_{e'}$
 are the polar and azimuthal angles of the outgoing photon
 and the outgoing electron, respectively. We consider the interaction of an
 initial electron with energy $E=4$~MeV with an optical laser beam with
 frequency $\omega=1.55$~eV in head-on collision geometry.

\subsection{Basic equations}

 %%%%%%%%%%%%%%%%%%%%%%%%%%%%%%%%%%%%%%%%%%%%%%%%%%%%%%%%%%%%%%%%%%%%%%%%%%%%%%%%%%
 For a finite pulse,
 similar to the BW process, the truncated differential cross section
of non-linear Compton scattering is determined by the integral
 of the partial contributions,
\begin{eqnarray}
 \frac{d\sigma^{(i)}(\kappa)}{d\cos\theta'}=
 \int\limits_{\kappa}^{\infty}d\ell
% \int\limits_{-1}^{1}d\cos\theta'
 \frac{d\sigma^{(i)}_{\ell}}{d\cos\theta'}~,
 \label{III1}
\end{eqnarray}
where the auxiliary continues variable $\ell$ appears (similarly to the
variable $\ell$ in the BW process) in the Fourier
integral of the corresponding transition matrix elements~\cite{TitovEPJD}.
The product  $\ell \omega$ has the meaning
of an energy fraction of the laser beam involved
in the  non-linear C process.
Without loss of generality, the lower limit of the integral
$\kappa\equiv\ell_{\rm min}$ is chosen as a dynamical parameter. Its
physical meaning will be discussed below.
The notion ``truncated differential cross section"
implies the non-zero lower limit of the integral~(\ref{III1}).
Instead of the conventional internal Ritus variable
$u=(k\cdot k')/(k\cdot p')$ with $u_{\rm min}=0$ and
$u_{\rm max}=u_\ell=2\ell(k\cdot p)/m^2$
we use the variable $\cos\theta'$ with constant limits of integration,
which is useful for the subsequent qualitative
analysis and convenient for the numerical calculation of multi-dimensional
integrals for total cross sections with rapidly oscillating integrands.

The differential cross section in the integrand is
\begin{eqnarray}
 \frac{d\sigma^{(i)}_{\ell}}{d\cos\theta'}
 =\frac{\alpha^2}{\xi^2(p\cdot k )N_0^{(i)}}\,
 F(\ell,\cos\theta')
 \int\limits_{0}^{2\pi}d\phi_{e'}
 \,w^{(i)}(\ell)~
 \label{III22}
\end{eqnarray}
with
 \begin{eqnarray}
 F(\ell,\cos\theta')=\frac{{\omega'}^2}{\ell\omega(E+|\mathbf {p}|)}~,
 \end{eqnarray}
 where the flux factors $N^{(i)}_0$ are defined in Eq.~(\ref{II2}).
 The frequency $\omega'$ of the emitted photon
 is related to the variable $\ell$ and the polar angle $\theta'$
 of the direction of the momentum $\mathbf{k}'$ thought the
 conservation laws as
\begin{eqnarray}
\omega' \equiv  \omega'_\ell=\frac{\ell\,\omega (E+|\mathbf {p}|)}{E + |\mathbf {p}| \cos\theta'
 +\ell \omega(1-\cos\theta') }~.
 \label{III3}
\end{eqnarray}

The partial probabilities $w^{(i)}$
% for the linear $i=0(L)$ and the circular $i=1(C)$ polarizations
read
\begin{eqnarray}
\frac12\, w^{(lin)}(\ell)&=&-|\widetilde A_0(\ell)|^2
 + \xi^2\left(1+\frac{u^2}{2(1+u)}\right)\nonumber\\
 &\times &\left(|\widetilde A_1(\ell)|^2 -
 {\rm Re}\widetilde A_0(\ell)\widetilde A_2^*(\ell)\right)~,
 \label{III6}
 \end{eqnarray}
\begin{eqnarray}
&& w^{(circ)}(\ell)=
 -2 |\widetilde Y_\ell(z)|^2+\xi^2\left( 1 +\frac{u^2}{2(1+u)}\right)\,\times
 \nonumber\\
 && \left(|Y_{\ell-1}(z)|^2+ |Y_{\ell+1}(z)|^2
 -2{\rm Re}\,(\widetilde Y_\ell(z)X^*_\ell(z))\right).
 \label{III7}
\end{eqnarray}
 The basis functions $\widetilde A(\ell)$, and  $Y_{\ell},\,X_{\ell}$ and
 $\widetilde Y_\ell(z)$
 are defined in Eqs.~(\ref{II4}) and~(\ref{YX1}), respectively.
%\marginpar{eq. numbers}
 For the dynamical variable $z$ we use the standard definition
 $z=2\ell\xi\left(({u}/{u_\ell})(1-{u}/{u_\ell})\right)^{1/2}$.
 The partial probabilities $w^{(i)}$
 resemble the corresponding expressions
 in the IPA (cf. Eqs.~(\ref{III8}) and (\ref{III9})).

Evaluating Eq.~(\ref{III1}), we express first
the variable $u$ in the partial probabilities
through $\cos\theta'$ in c.m.s. ($\cos\theta'_c$)
as $u=\omega_c(1-\cos\theta'_c)/(E_c +\omega_c\cos\theta'_c)$
with $\omega_c=\ell(p\cdot k)/2\sqrt{s}$, $E_c=(s+m^2)/2\sqrt{s}$
and then express $\cos\theta'_c$ via $\cos\theta'$ in the lab.\ system as
$\cos\theta'_c= (v-\cos\theta')/(v\cos\theta' - 1)$ with
$v=-(|{\mathbf p}| -\ell\omega)/(E + \ell\omega)$.

Often, analysis of the non-linear C process is constrained
entirely to the energy ($\omega'$)
and angular ($\theta'$ and $\phi'$) distributions of the outgoing photon.
Our approach here allows for an easy access to the polar photon angle
$\theta'$ and the azimuthal  final-state electron angle $\phi_{e'}$.
Such a mixed phase space distribution is to be contrasted with analyses
which focus entirely on the kinematics of the outgoing electron~\cite{DSeipt}.\\

 Now, we would like to stress that the differential cross section
 (\ref{III1}) with (\ref{III22})
 for $\ell>1$ has a sharp maximum in the backward hemisphere in the vicinity of
 $\theta'=\theta'_0\simeq175^o$,
 similar to that in the case of IPA (cf. Fig.~\ref{Fig:001} below).
 %{\color{red}
 A sharp increase of the cross section in the backward hemisphere
 is a well-known property of Compton scattering~\cite{LL}.
 In the case of non-linear Compton scattering, when a finite number
 of background field photon modes contribute,
 a sharp maximum appears~\cite{Harvey}, the position of which
depends on initial-state  kinematics
 and is less sensitive to the field intensity,
 at least for $\xi^2<10$ (cf.\ \ref{A.2}).

In a practical
 study of the Compton scattering, one can choose
 another angle $\theta'<\theta'_0$
 (in accordance with the experimental set-up), remembering that (i) the cross section
 in this case will be smaller, and (ii) qualitatively
 the main results do not depend on the choice of this angle.

 Discussing the physical meaning of the dynamical parameter $\kappa$
 it is convenient to consider the ratio $R(\ell,\theta')=\omega'_\ell/\omega'_1$, where
 $\omega'_1$ is the frequency of the photon  emitted
 in the interaction of an initial electron with a single photon of the pulse
 at the same angle
 \begin{eqnarray}
 \omega'_1=\frac{\omega (E+|\mathbf {p}|)}
 {E + |\mathbf {p}| \cos\theta' + \omega(1-\cos\theta') }~.
 \label{III10}
\end{eqnarray}
 \begin{figure}[tb]
 \vspace*{5mm}
 \includegraphics[width=0.48\columnwidth]{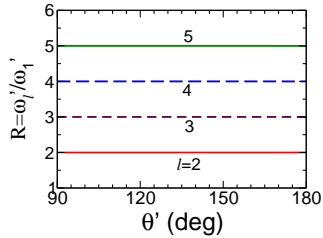}\hfill
 \caption{\small{(Color online)
The ratio $R={\omega'_\ell}/{\widetilde\omega'_1}$ as a function of
$\theta'$ for different values of $\ell$.
 \label{Fig:99}}}
 \end{figure}
 The ratio $R$ as a function of $\theta'$ in the backward hemisphere for the chosen kinematics is exhibited in Fig.~\ref{Fig:99} for different values of $\ell$.
It can be seen that $R$ is practically independent of $\theta'$. Thus,
the relation between $R$, $\ell$ and $\theta'$ reads
\begin{eqnarray}
 R=
 \frac{\ell}{1+\delta(\ell-1)}~,
 \label{III11}
 \end{eqnarray}
 where
 \begin{eqnarray}
 \delta=\varepsilon
 \frac{1-\cos\theta'}{1+v_e\cos\theta'+\varepsilon(1-\cos\theta')}
 \label{III122}
 \end{eqnarray}
 with
 $\varepsilon=\omega/E$ and $v_e=|{\mathbf p}|/E$.
 For the chosen kinematics,
 $\varepsilon\simeq3.85 \times 10^{-7}$ and
 $\delta$ varies from $3.8$$ \times 10^{-7}$ to  $9.5\times 10^{-5}$
 when $\theta'$ varies from $\pi/2$ to $\pi$, respectively.
 Therefore, with great accuracy, the lower limit of the integral in (\ref{III1})
 may be chosen as $\kappa=\ell_{\rm min}=R=\omega'/\omega'_1$ in a wide
 interval of $\theta'$ in the backward hemisphere.
 As mentioned above, the product $\ell \omega$ has the meaning of
 the pulse energy involved into the process.
 The multi-photon dynamics refers to $\ell>1$.
 Since in our definition $\ell_{\rm min}=\kappa$,
 the cross section $\sigma(\kappa)$ with
 $\kappa>1$ corresponds to the multi-photon Compton regime.
 This is the physical meaning of $ \kappa> 1 $
 as a value reflecting the onset of multi-photon
 dynamics.

 In order to isolate multi-photon events,
 one has to install a detector at fixed polar angle $\theta'$ and
 register only such photons with the frequencies higher than $\omega'_{\kappa}$
 (or $\omega'\ge\omega'_{\kappa}$) with $\kappa>1$.
 A preferred angle of detection
 is in the backward hemisphere,
 where the cross section is significantly larger.

Recall that the cross sections~(\ref{III1})
are integrated over the azimuthal angle
$\phi_{e'}$. This means that a selection of
multi-photon events would be performed as a sum of events
in the interval $0\geq \phi_{e'}\geq \pi$
with an appropriate binning (where we assume the symmetry of the azimuthal angle
distribution with respect to the substitution $\phi_{e'}\to 2\pi - \phi_{e'}$).
Technically, a corresponding measurement seems to be quite feasible,
especially since our subsequent study implies
exactly the same analysis of azimuthal angle distributions.
However, bearing in mind that  the frequencies of outgoing photons
$\omega'_\kappa$ and $\omega'_1$, respectively, are independent of $\phi_{e'}$, the
conclusion about the physical meaning of
$\kappa>1$ as a suitable starting point of
switching on the corresponding multi-photon regime remains appropriate.
On the other hand, the favorable intervals for the study of the azimuthal angle
distribution may be understood from our analysis
of the corresponding azimuthal angle distributions given  below in subsection
{\it 2}.
For  IPA and $circ$ polarization, the problem is simplified
since the azimuthal angle distribution is isotropic in this case.

\subsection{Numerical results}

The dependences
of the non-linear C process on the variables $\ell,\,z\,\,u$ and
the field strength $\xi^2$ for $lin$ and $circ$
polarizations are realized through the basis functions
$\widetilde A_m(\ell)$ and
$Y_{\ell},\,X_{\ell}$, respectively
and, in general, are different for these two cases,
both for the truncated total cross sections and for the differential distributions.

\begin{figure}[tb]
 \includegraphics[width=0.48\columnwidth]{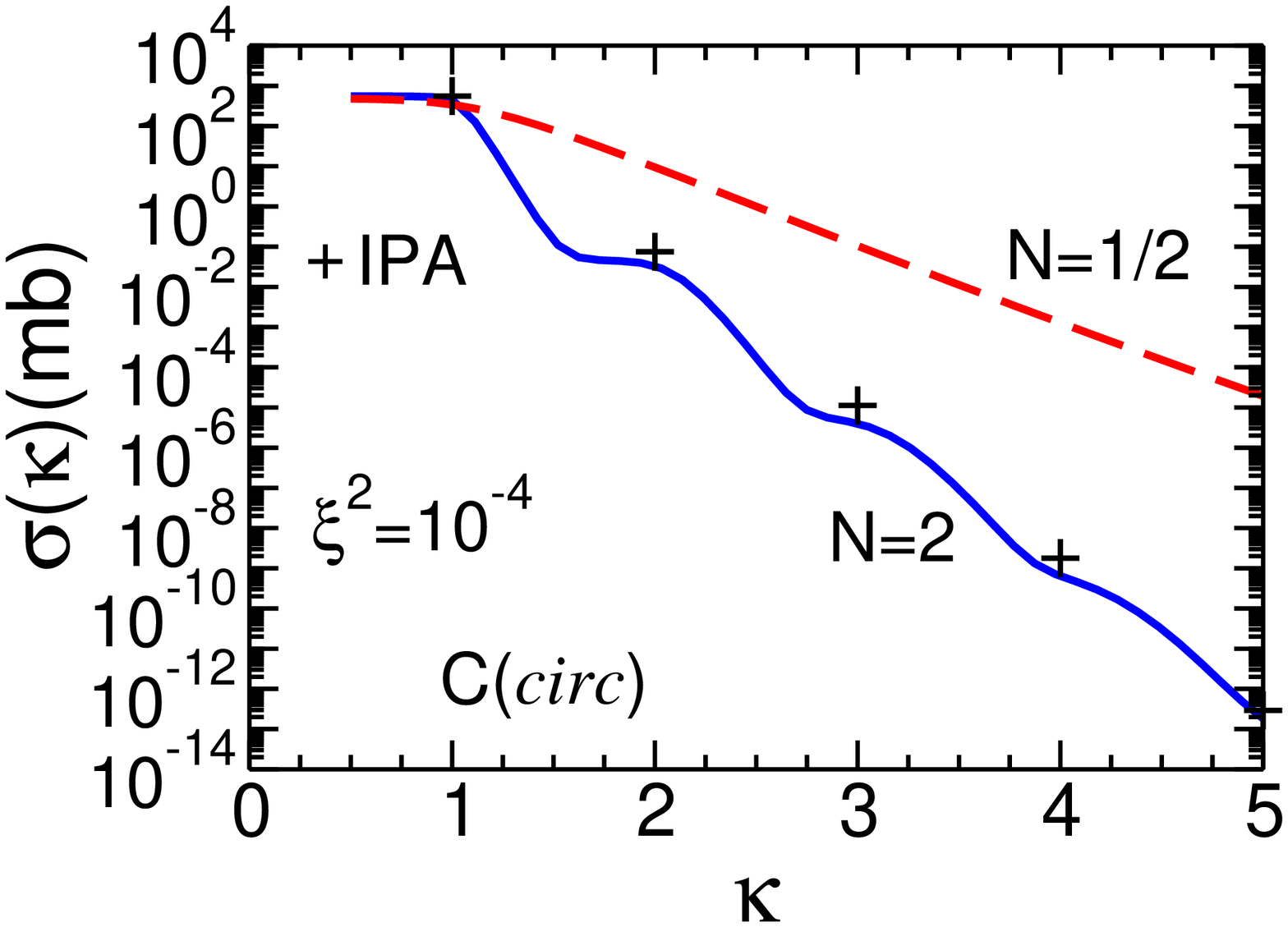}\hfill
 \includegraphics[width=0.48\columnwidth]{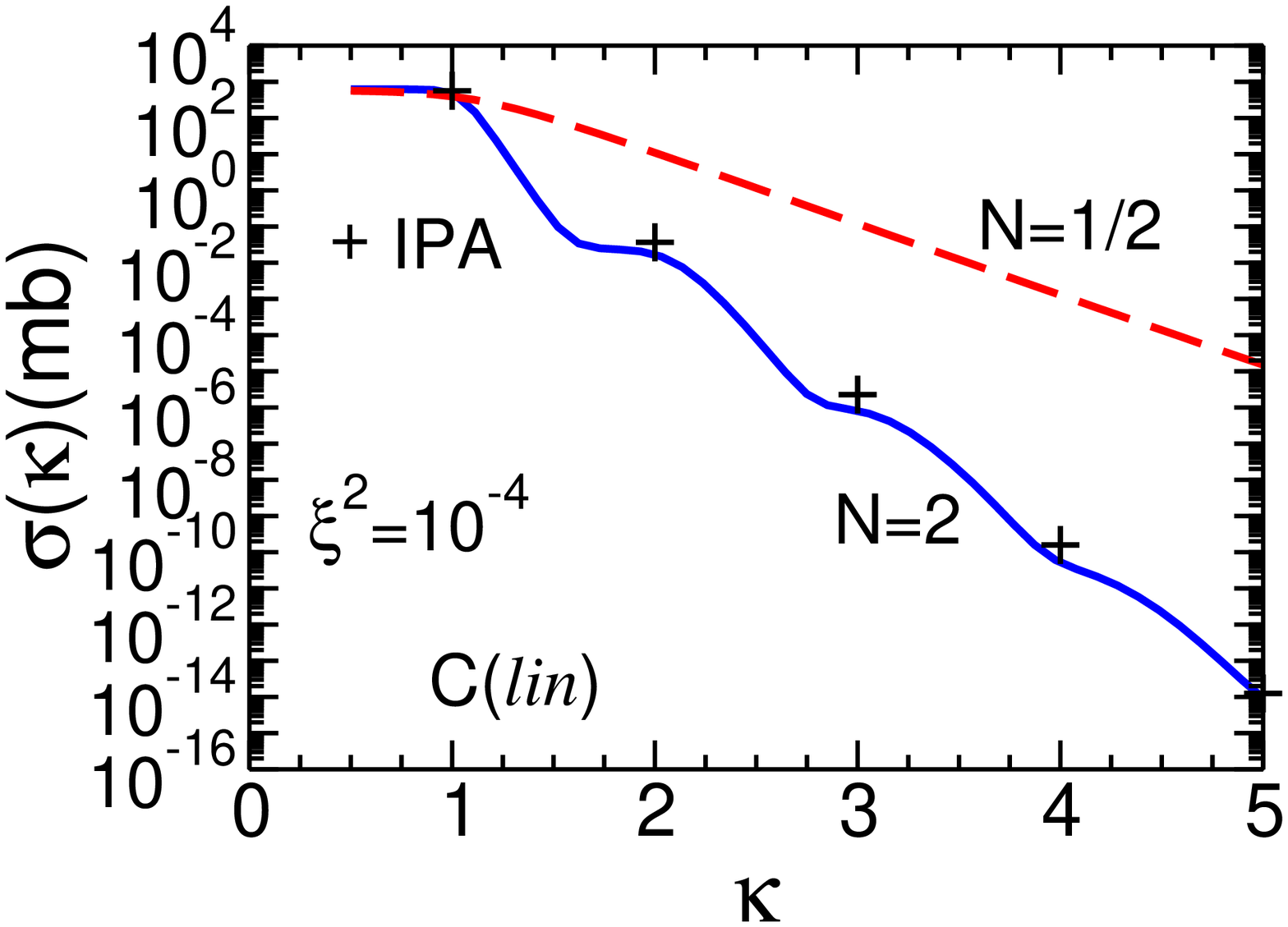}\\
 ~\\
  \includegraphics[width=0.48\columnwidth]{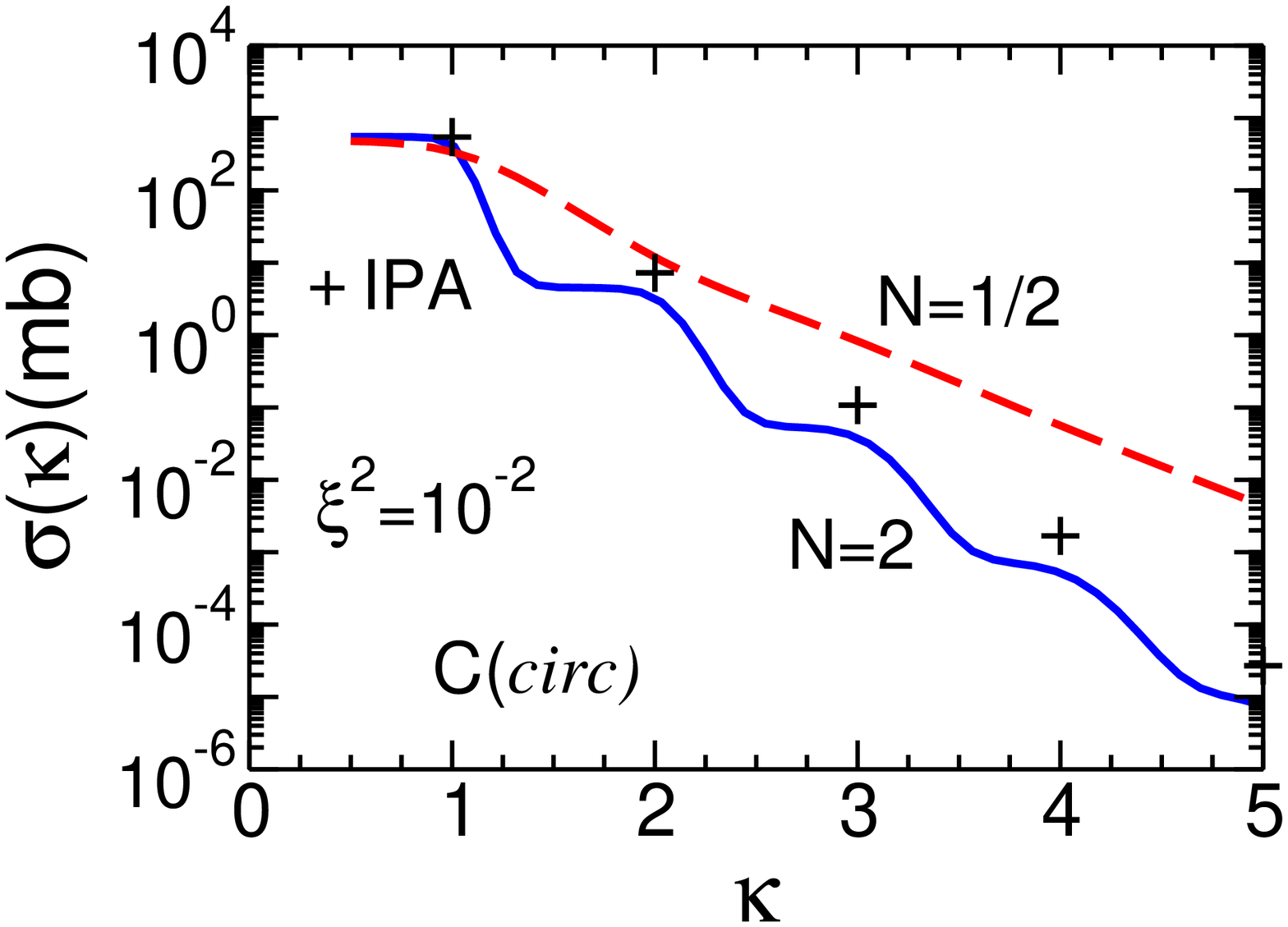}\hfill
 \includegraphics[width=0.48\columnwidth]{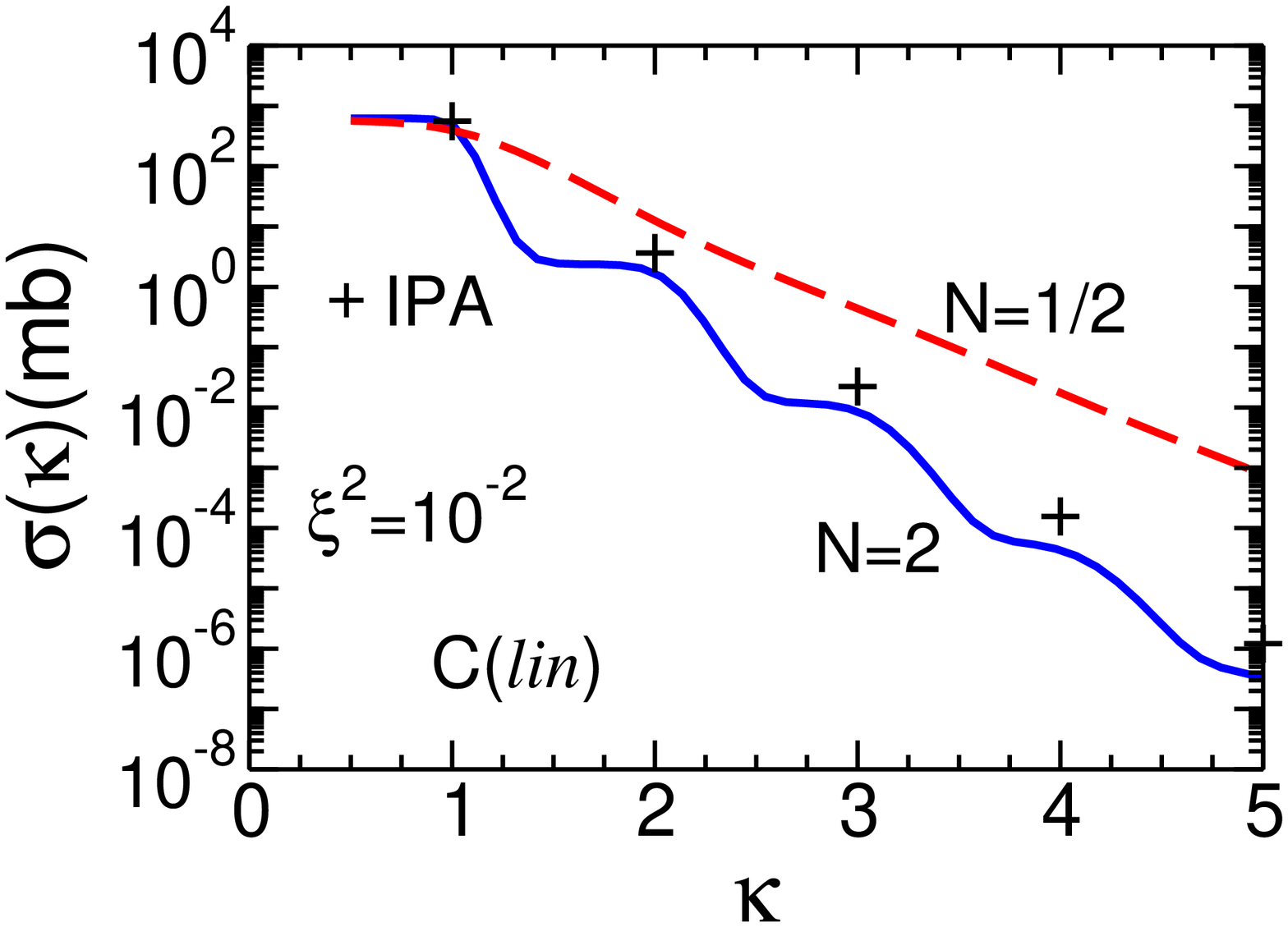}\\
 ~\\
  \includegraphics[width=0.48\columnwidth]{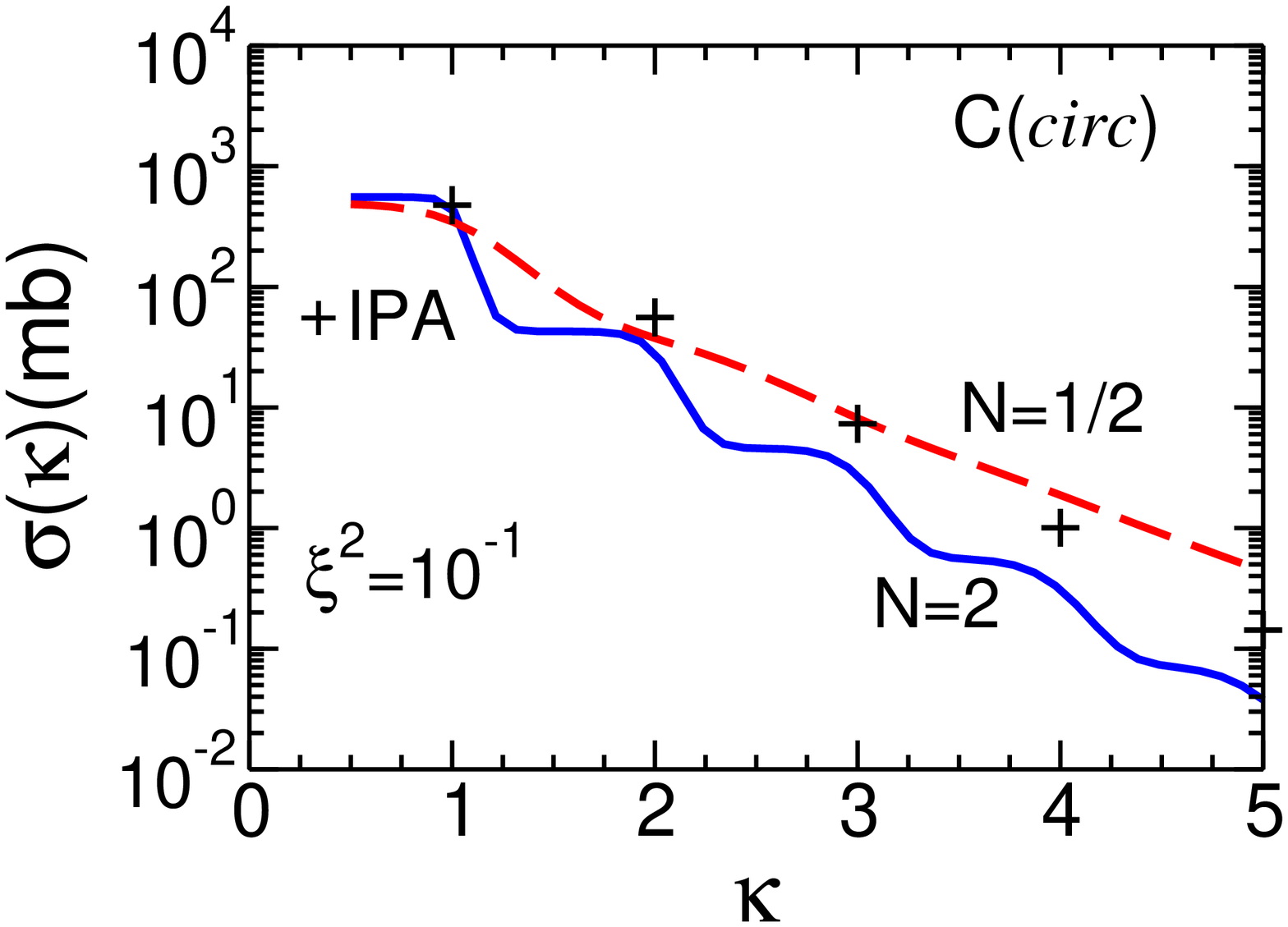}\hfill
 \includegraphics[width=0.48\columnwidth]{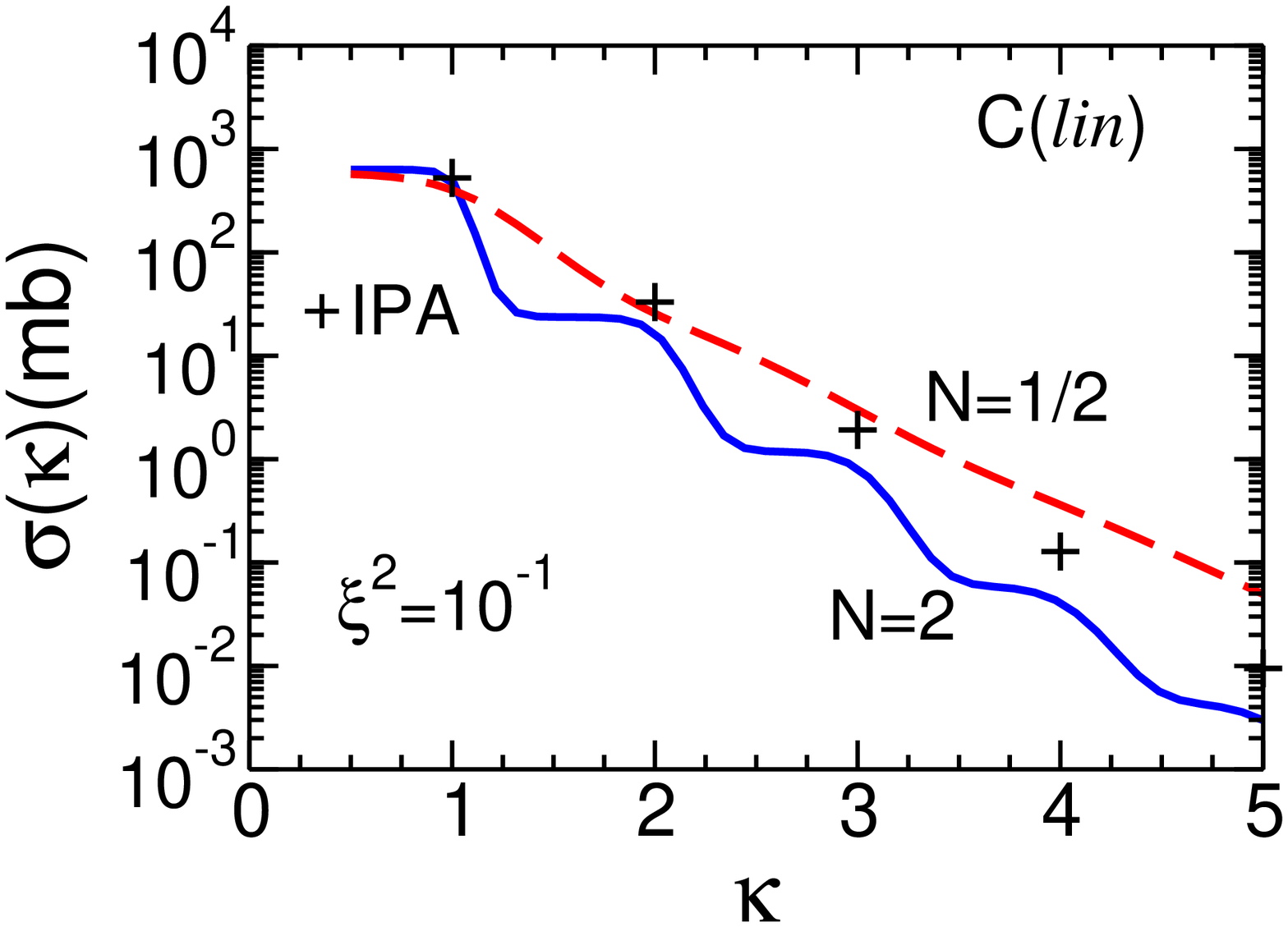}\\
 \caption{\small{(Color online)
The truncated cross sections $\sigma (\kappa)$ as a function of the
parameter $\kappa$ at
$\xi^2=10^{-4}$ (top), $10^{-2}$ (middle) and $10^{-1}$ (bottom).
The left and right columns correspond to the $circ$ and
$lin$ polarizations, respectively. The dashed and thick solid
curves are for ultra-short and short pulses with the number of cycles $N=1/2$
and $2$, respectively. The crosses are for the IPA. The minimal values
of corresponding harmonics are chosen as integer parts of $\kappa$.
 \label{Fig:011}}}
 \end{figure}

\subsubsection{Truncated total cross section}

 The truncated total cross sections integrated over $d\cos\theta'$
\footnote{
The notion of a ``truncated cross section" is to emphasize
that we are employing a non-zero lower limit of
 the integration, $\ell_{min}=\kappa$. The phrase ``partially integrated cross
 section" is also suitable for the distinction to the total cross section
 integrated over the full out-phase space.}
 \begin{eqnarray}
 \sigma^{(i)}(\kappa)=
 \int\limits_{\kappa}^{\infty}d\ell
 \int\limits_{-1}^{1}d\cos\theta'
 \int\limits_{0}^{2\pi}d\phi_{e'}
 \frac{d\sigma^{(i)}_{\ell}}{d\phi_{e'}d\cos\theta'}~,
 \label{III1_}
\end{eqnarray}
 as a function of the
 threshold parameter $\kappa$ at e.m.\ field strength parameters
 $\xi^2=10^{-4},\,10^{-2}$ and $10^{-1}$ are shown in Fig.~\ref{Fig:011}
 in the upper, middle and lower panels, respectively.
 The left and right panels correspond to the circular and
 linear polarizations, respectively.
 Remembering that the multi-photon regime arises at $\kappa>1$,
 for completeness we extend our consideration to smaller values $\kappa \geq0.5$
 and $n_{\rm min} \geq 1$.
 The dashed and thick solid curves
 correspond to the ultra-short and short pulses with the
 number of oscillations $N=1/2$ and $2$, respectively.
 The crosses are for the IPA. The values of corresponding
 minimum number of harmonics
 $n_{\rm min}$ in Eq.~(\ref{III14}) are chosen as integer parts
 of $\kappa$.
 %One can see a sizable difference between predictions for
 %linear and circular polarizations.
 %with the number of participating photons exceeding $n_\kappa$.
 At $\kappa \approx 1$, where only one photon
 from the pulse participates, the results  of all calculations are very
 close to each other. The cross sections
 decrease for increasing $\kappa$. In the case of $N=2$, the shape
 of the curves resembles the step-like behavior. For the sub-cycle pulse
 with $N=1/2$ the cross sections decrease exponentially with $\kappa$,
 $\sigma \propto \exp[-b^{(i)}\kappa]$.
 The slopes $b^{(i)}$ depend on the field intensity and, in general,
 on the polarization.
 In the case of linear polarization, the slope $b^{(lin)}$ is larger.

 \begin{figure}[tb]
% \vspace*{5mm}
 \includegraphics[width=0.48\columnwidth]{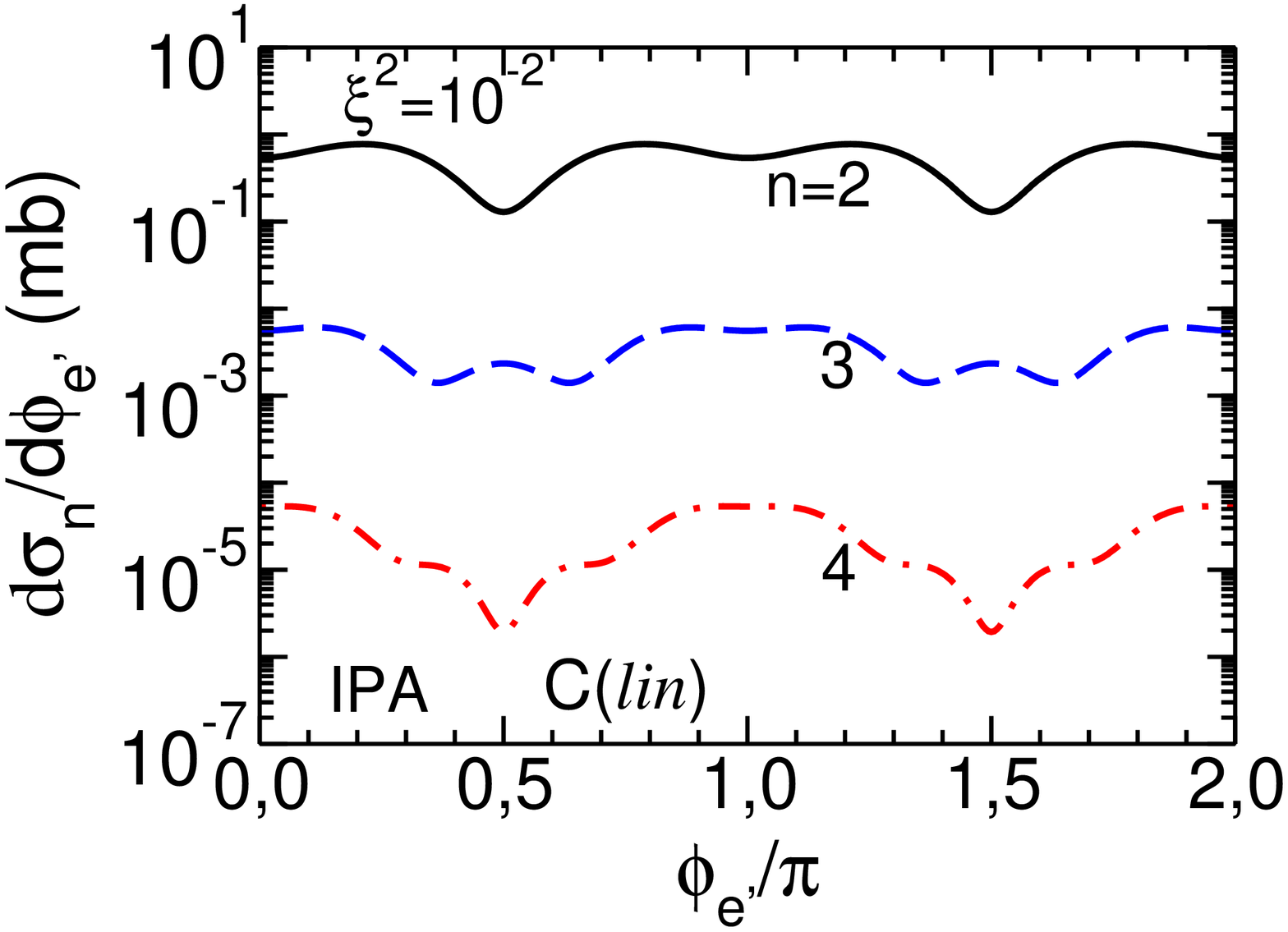}\hfill
 \includegraphics[width=0.48\columnwidth]{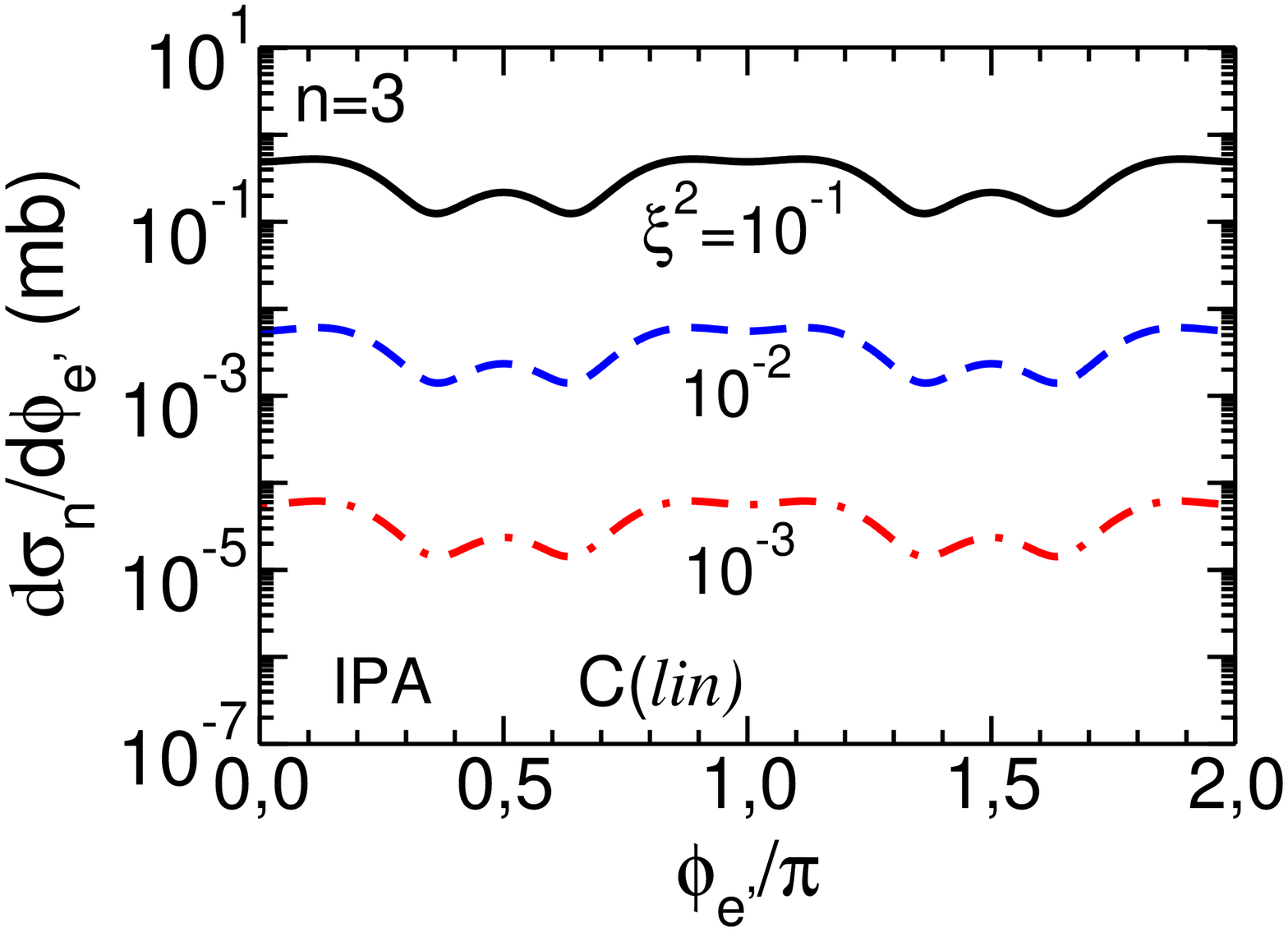}
 \caption{\small{(Color online)
The differential cross sections ${d\sigma^{(lin)}_n}/{d\phi_{e'}}$ in IPA.
The left and right panels correspond to calculations for various
values of harmonic number $n$
at $\xi^2=10^{-2}$ and for various values of $\xi^2$ at $n=3$,
respectively.
 \label{Fig:013}}}
 \end{figure}

Note that the exponential decrease of the total cross sections as
a function of $\kappa$ has the same reason as in the case of the
non-linear BW process
(cf.\ Fig.~\ref{Fig:01}), where the cross sections are functions
of $\zeta$. Now, the threshold parameter $\kappa$ is an analog of the parameter $\zeta$
and the previous discussion at the end of Sect.~III.B applies to the
non-linear Compton scattering, considered here.

\subsubsection{Azimuthal angle distributions}

Analog to the non-linear BW process, the
shape of the differential cross sections as a function of the azimuthal
angle $\phi_{e'}$
of non-linear C scattering for linear and circular polarizations is
determined mainly by the phase factors
${\cal P}^{(i)}$ in Eqs.~(\ref{II55}) and (\ref{YX2}),
%\marginpar{correct numbers?}
respectively. Again, for a monochromatic
circularly polarized beam (IPA case), the differential cross section
does not depend on $\phi_{e'}$.

 \begin{figure}[tb]
% \vspace*{5mm}
 \includegraphics[width=0.48\columnwidth]{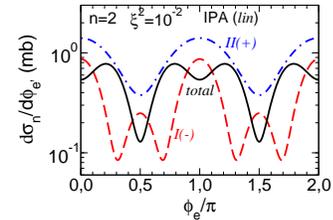}
 \caption{\small{(Color online)
The differential cross sections of the non-linear C process
in IPA for the harmonic $n=2$ and $\xi^2=10^{-2}$ together with the
separate contributions of the first ($I(-)$) and second $(II(+))$ terms
in the partial probability (\ref{III8}). The thick solid curve
is for the coherent sum of both contributions.
 \label{Fig:0133}}}
 \end{figure}

In the case of linear polarization, the azimuthal angle
distribution (in IPA) shown in Fig.~\ref{Fig:013} exhibits
non-monotonic behavior. For example, in the case of $n = 2$ and
$\xi^2 = 10^{-2}$, one sees
some local minima  at $\phi_{e'}=0,\,\pi$
and $2\pi$, some deepening at  $\phi_{e'}=\pi/2,\,3\pi/2$
and bumps at $\phi_{e'}=\pi/4,\,3\pi/4,\,5\pi/4$ and
$7\pi/4$. The left and right panels of Fig.~\ref{Fig:013}
correspond to calculations for different harmonic
numbers $n$ at $\xi^2=10^{-2}$
and different values of $\xi^2$ at $n=3$, respectively.
All cross sections are symmetric under the transformation
$d\sigma_n (\phi_{e'}) \to d\sigma_n(2\pi-\phi_{e'})$.

The reason of the non-monotonic shape of distributions
is the destructive interference of the first and second
terms in the partial probability
in Eq.~(\ref{III8}).
 Thus, Fig.~\ref{Fig:0133} exhibits
the differential cross section of  the non-linear C process at
$n=2$ and $\xi^2=10^{-2}$ together with separate contributions
of the first ($I$) and second ($II$) terms
in the partial probability in Eq.~(\ref{III8}).
One can see that, at points at $\phi_{e'}=0,\,\pi$ and $2\pi$, the absolute
values of $(I)$ and ($II$) are close to each other and, being opposite
in sign, they mutually compensate each other.
At the points  $\phi_{e'}=\pi/4,\,3\pi/4$, $5\pi/4$ and $7\pi/4$,
the contribution of the first term is negligible which leads to
certain local maxima. At the points $\phi_{e'}=\pi/2$ and $3\pi/2$, one can also
observe some local compensation of the two terms
 which leads to the corresponding deepening in the distribution.
 However, the depth of these local minima
 is less than in the first case and, generally,
 depends on the values of $n$ and $\xi^2$.

\begin{figure}[tb]
 \includegraphics[width=0.48\columnwidth]{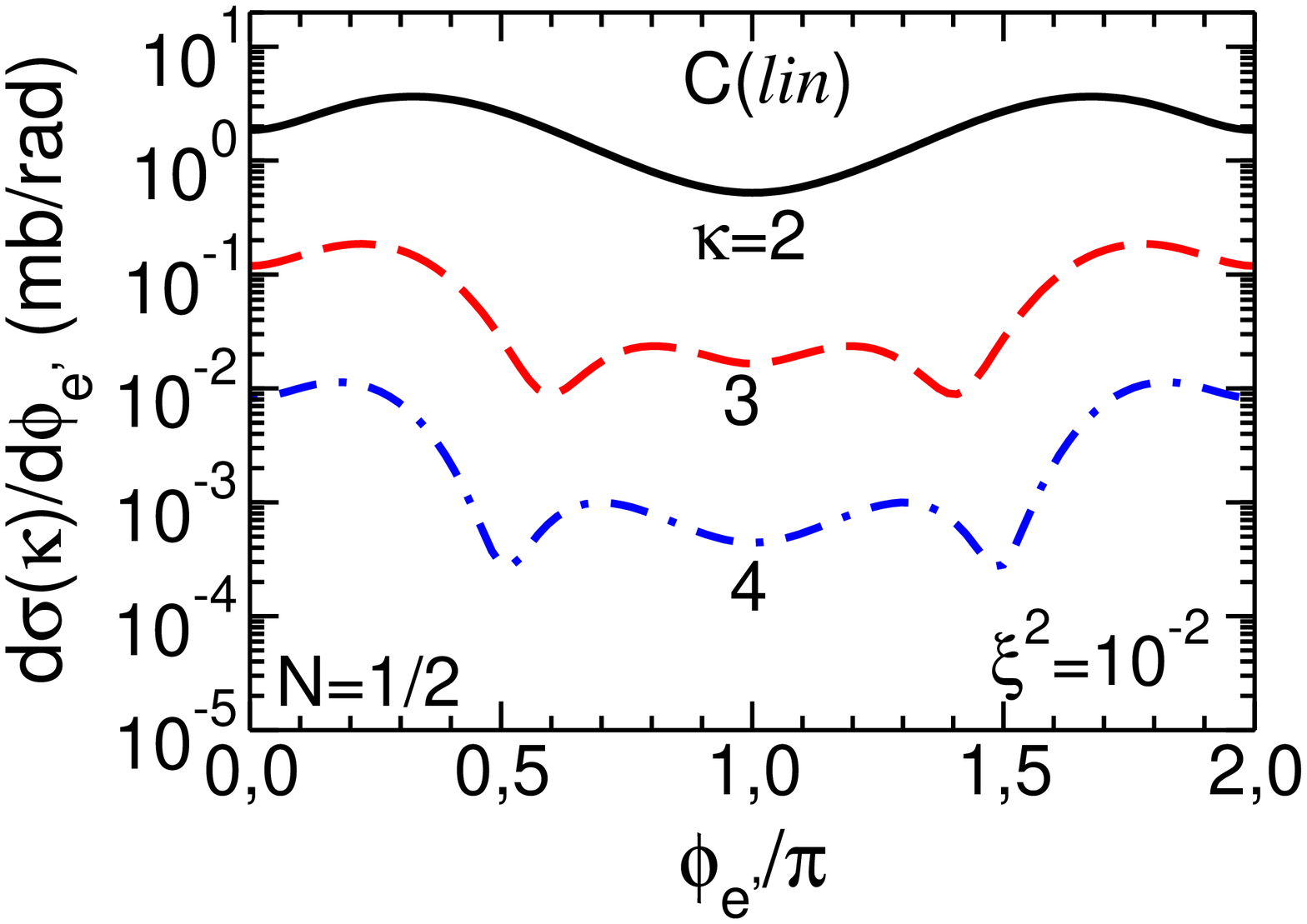}\hfill
 \includegraphics[width=0.48\columnwidth]{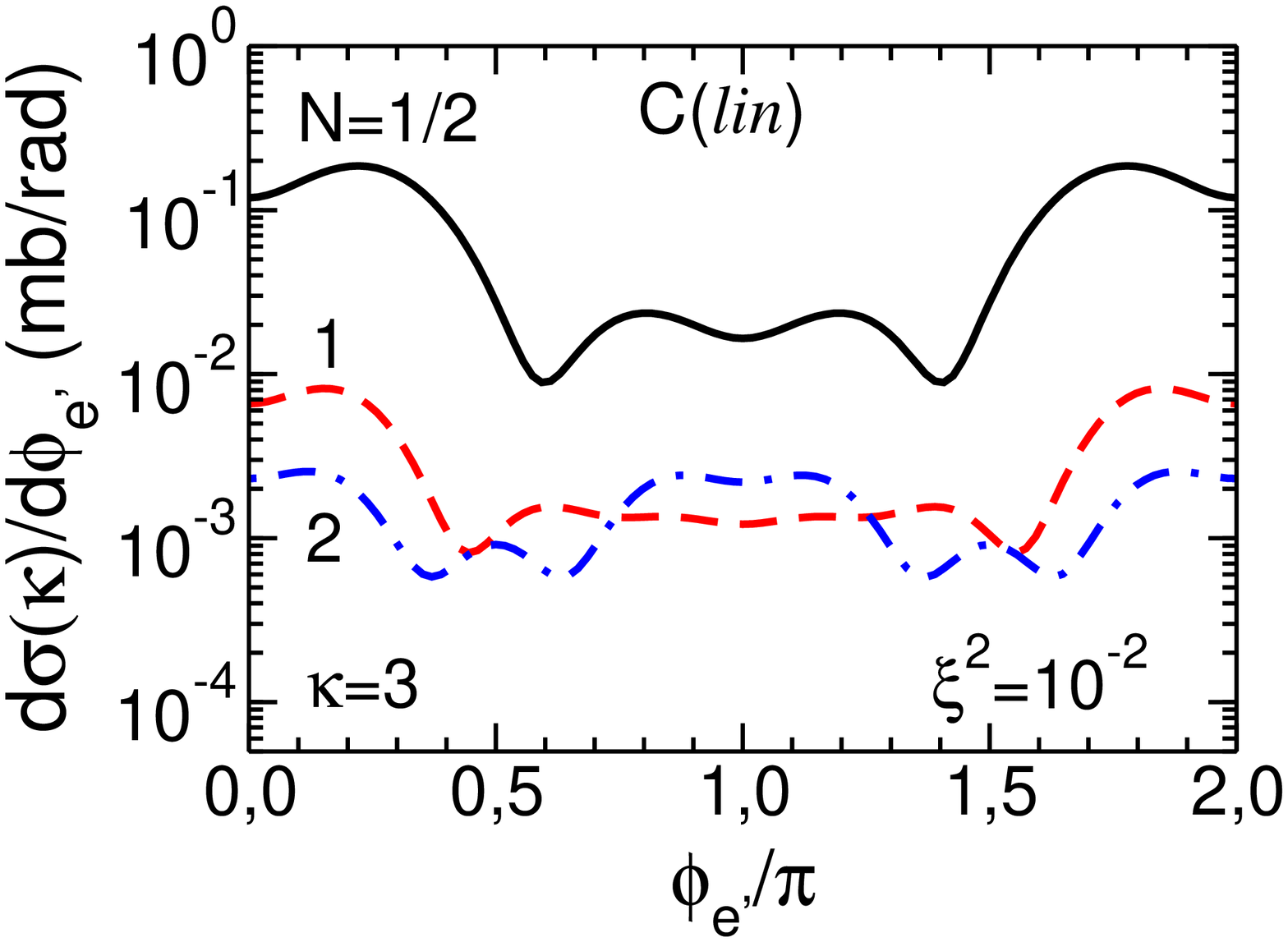}
 \caption{\small{(Color online)
The differential cross sections ${d\sigma^{(lin)} (\kappa)}/{d\phi_{e'}}$
at $\xi^2=10^{-2}$.
The left and right panels are for calculations at $N=1/2$ and different
values of $\kappa$ and for different values $N$  at $\kappa=3$, respectively.
 \label{Fig:014}}}
 \end{figure}

 The differential distributions
 for linear polarization and in FPA
 with $\CEP=0$ are exhibited in Fig.~\ref{Fig:014}.
 %(for circularly polarized beam see~\cite{CEPTitov}).
 The left and right panels correspond to calculations at different
 values of the parameter $\kappa$  for $\xi^2=10^{-2}$ and $N=1/2$,
 and to different values of the pulse duration at the same value
$\xi^2$ and $\kappa=3$, respectively.
 Again, one can see a non-monotonic shape and a multi-bump
 structure of the distributions,
 which is a consequence of the interference of the two terms in the partial
 probabilities of Eq.~(\ref{III6}). The shapes are different from that in
 the case of a circularly polarized pulse~\cite{CEPTitov}.

 \begin{figure}[th]
 \includegraphics[width=0.48\columnwidth]{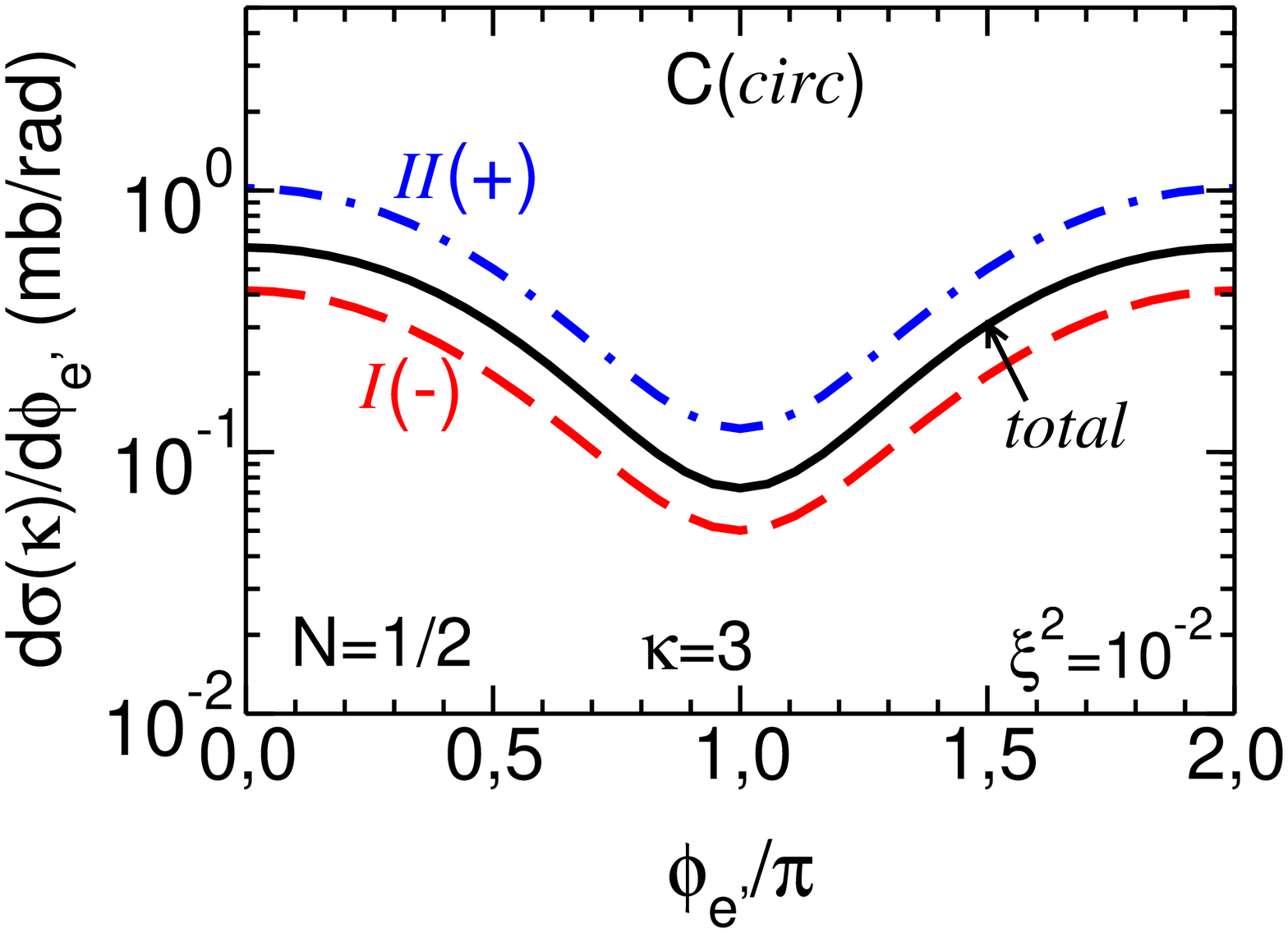}\hfill
 \includegraphics[width=0.48\columnwidth]{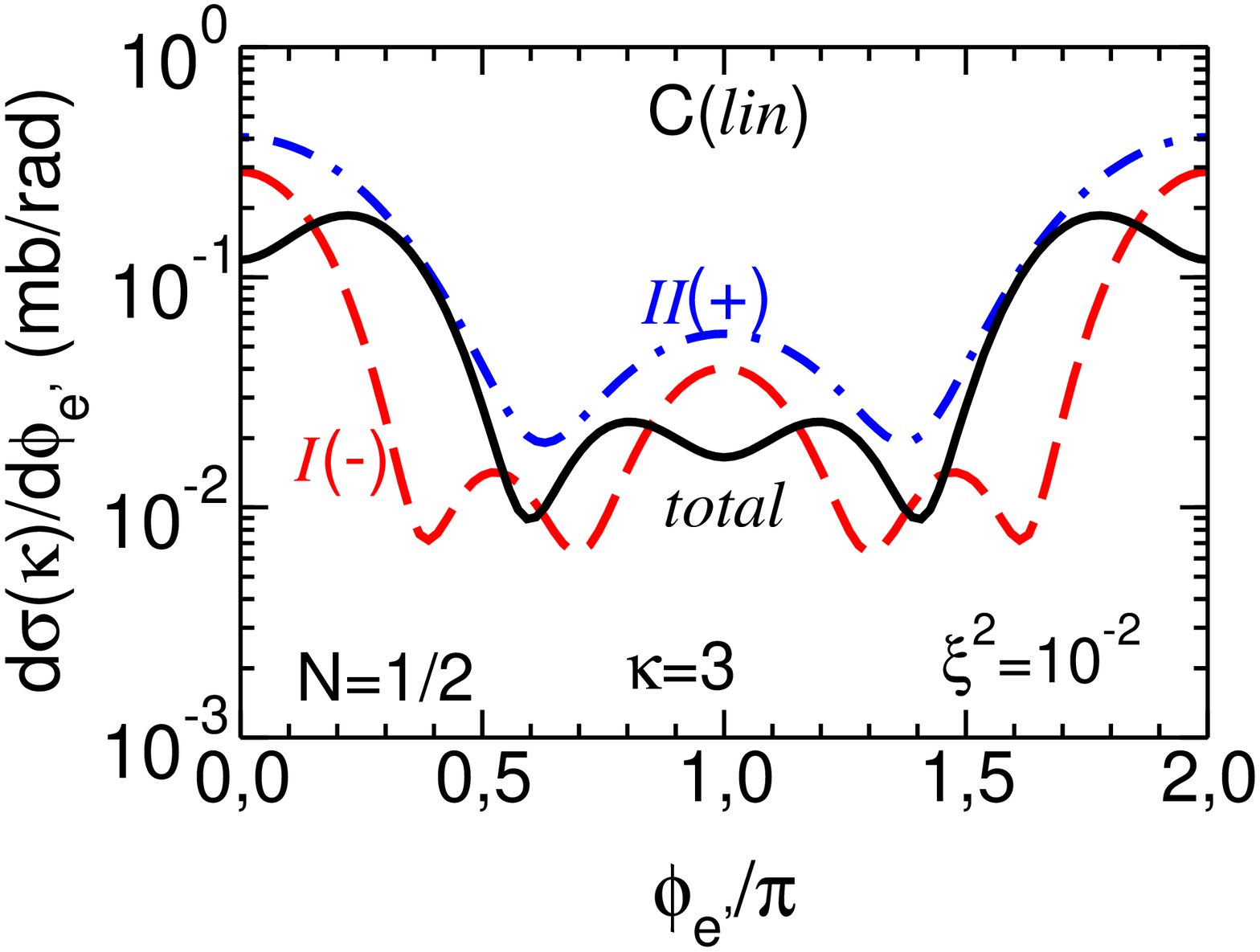}
 \caption{\small{(Color online)
 The differential cross sections ${d\sigma (\kappa) }/{d\phi_{e'}}$
 at $N=1/2$, $\xi^2=10^{-2}$ and $\kappa=3$.
 The left and right panels are for $circ$ and $lin$ pulse polarizations,
 respectively. The dashed and dot-dashed curves labeled by I(-) and II(+) are for
 the first (negative) and second terms in Eqs.~(\ref{III6}), ~(\ref{III7}),
 respectively. The thick solid curve is for the coherent sum of both contributions.
 \label{Fig:0144}}}
 \end{figure}

 In fact, for both circular and linear polarizations, the total probability
 is a coherent sum of negative and positive terms in the partial distributions
 in Eqs.~(\ref{III6}) and (\ref{III7}). But
 the azimuthal angle dependencies of these two terms are either similar or quite different
 for $circ$ or $lin$ pulse polarizations, respectively,
 which is illustrated in the left and right panels of Fig.~\ref{Fig:0144}.
 As a result, the azimuthal angle distributions have either one bump
 at $\phi_{e'}=0\,(2\pi)$ or a multi-bump structure for $circ$ or $lin$ polarizations,
 respectively.

\begin{figure}[th]
 \includegraphics[width=0.48\columnwidth]{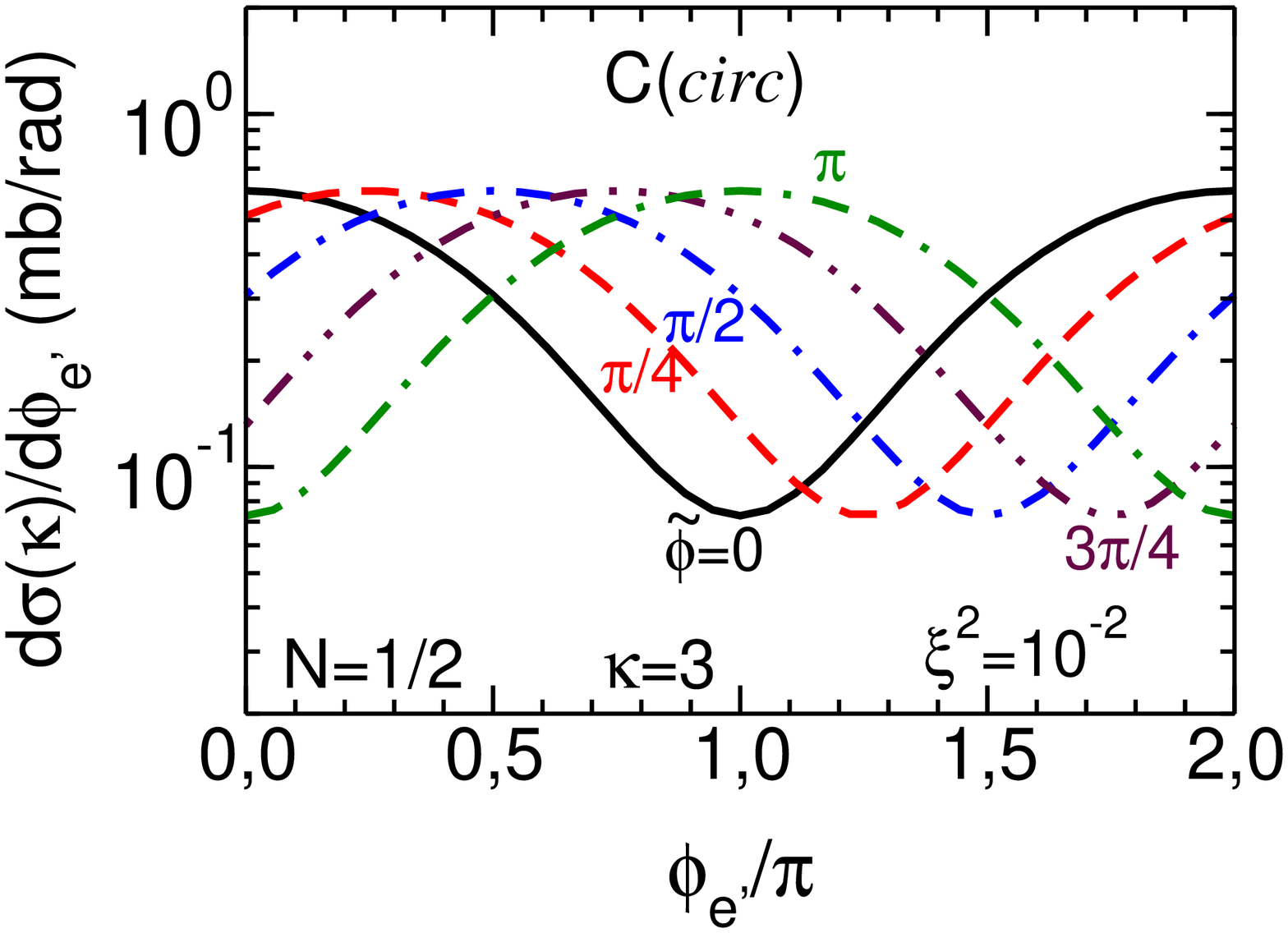}\hfill
 \includegraphics[width=0.48\columnwidth]{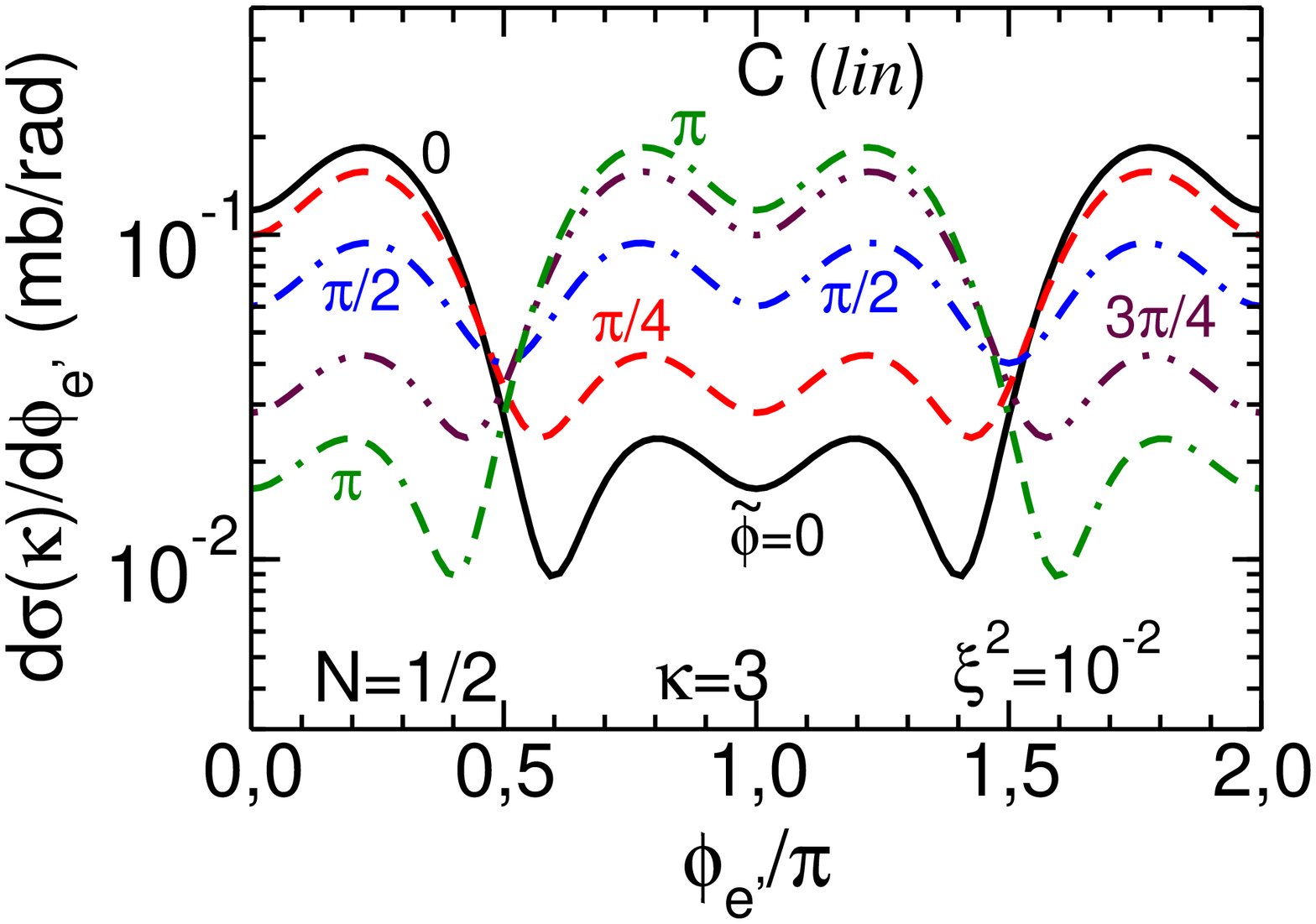}
 ~\\
 \includegraphics[width=0.48\columnwidth]{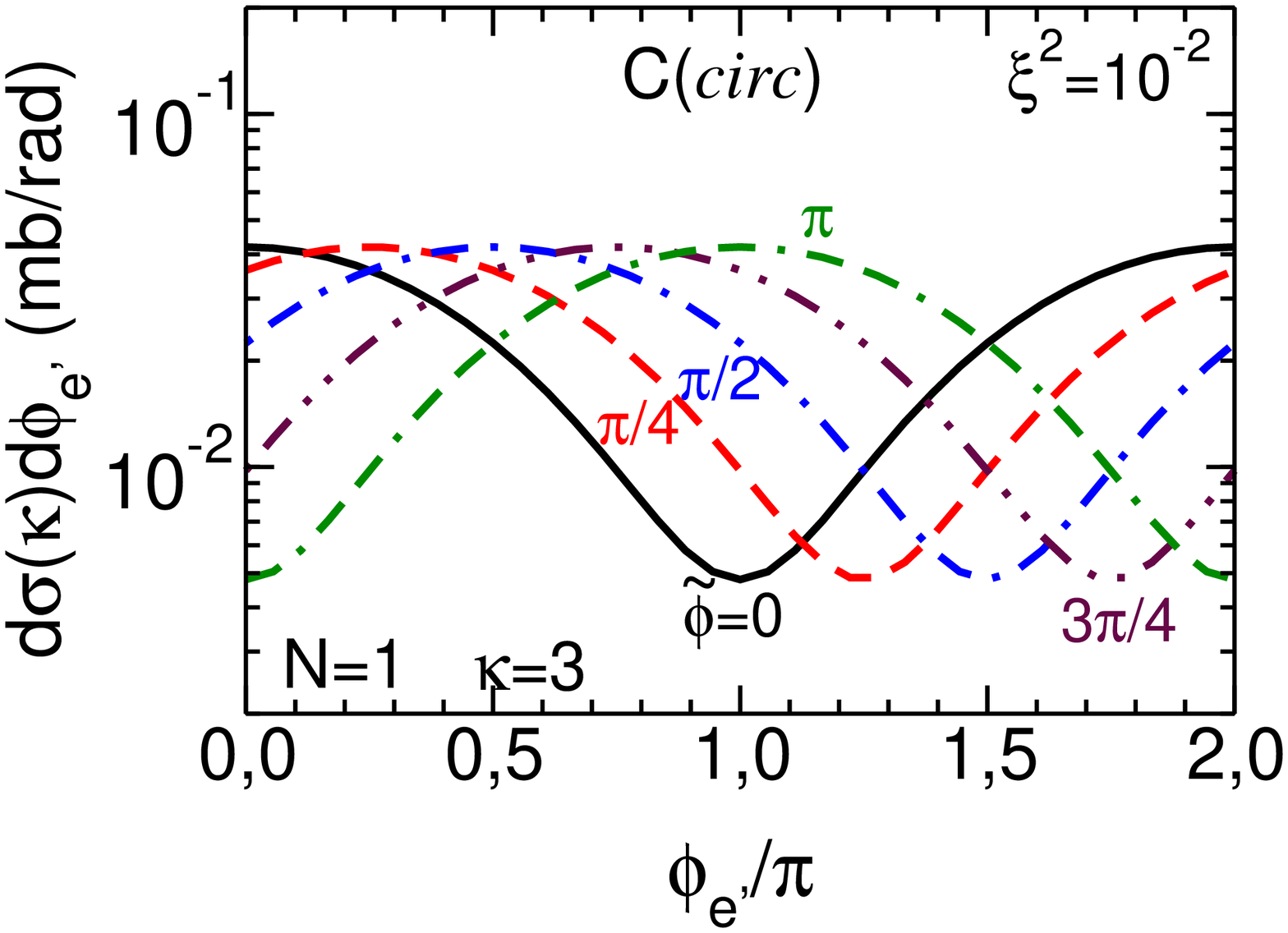}\hfill
 \includegraphics[width=0.48\columnwidth]{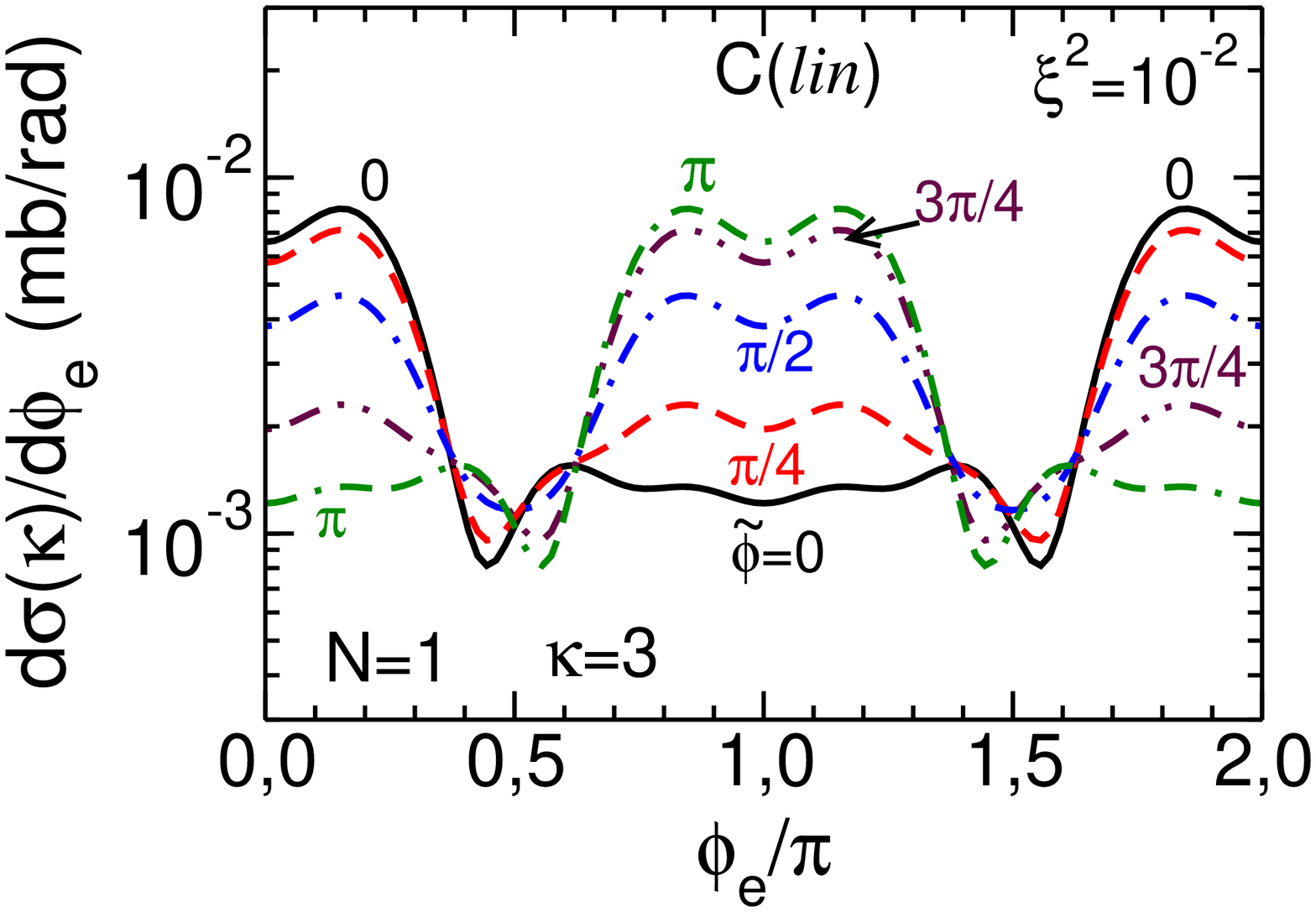}
 ~\\
 \includegraphics[width=0.48\columnwidth]{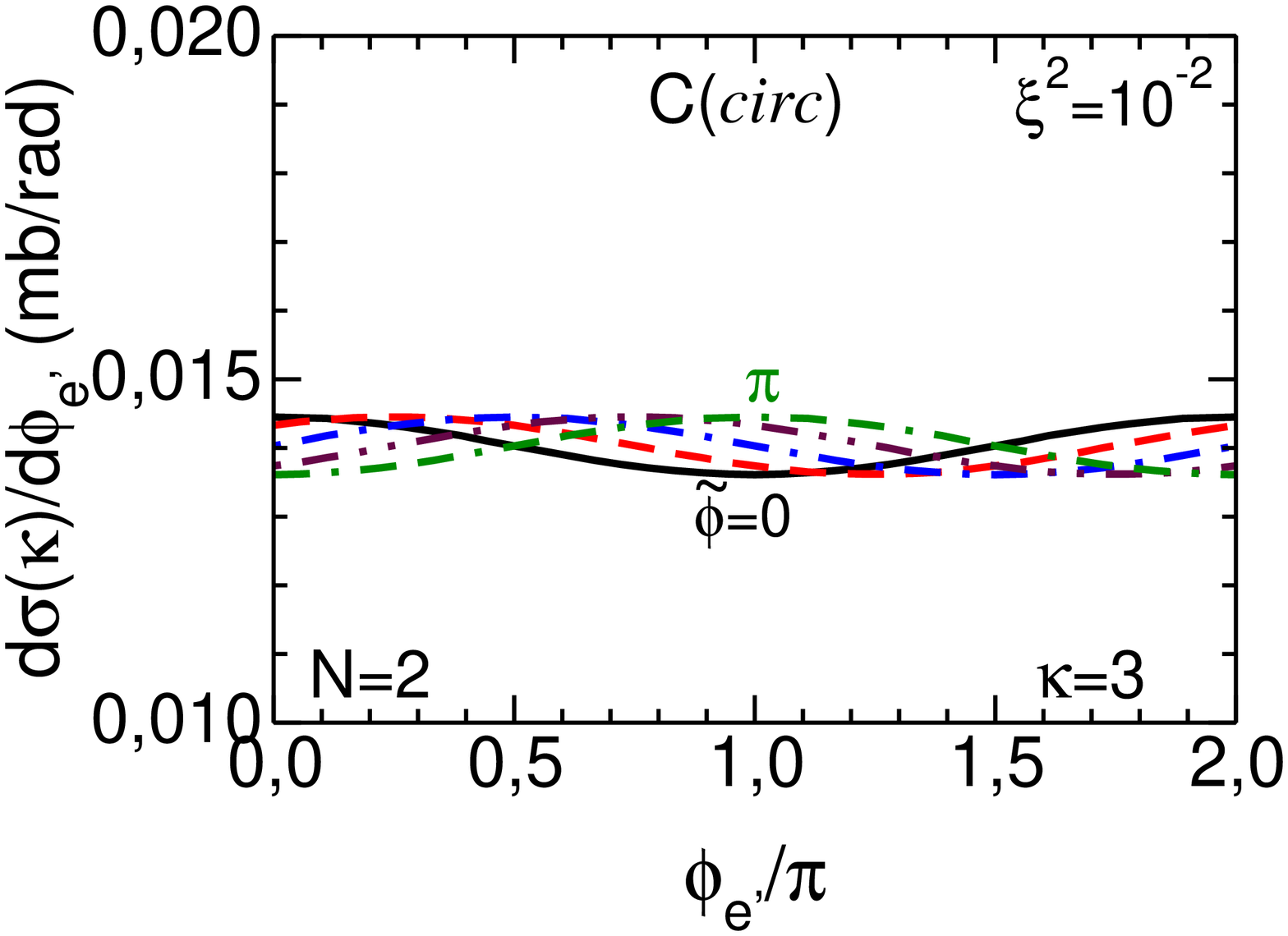}\hfill
 \includegraphics[width=0.48\columnwidth]{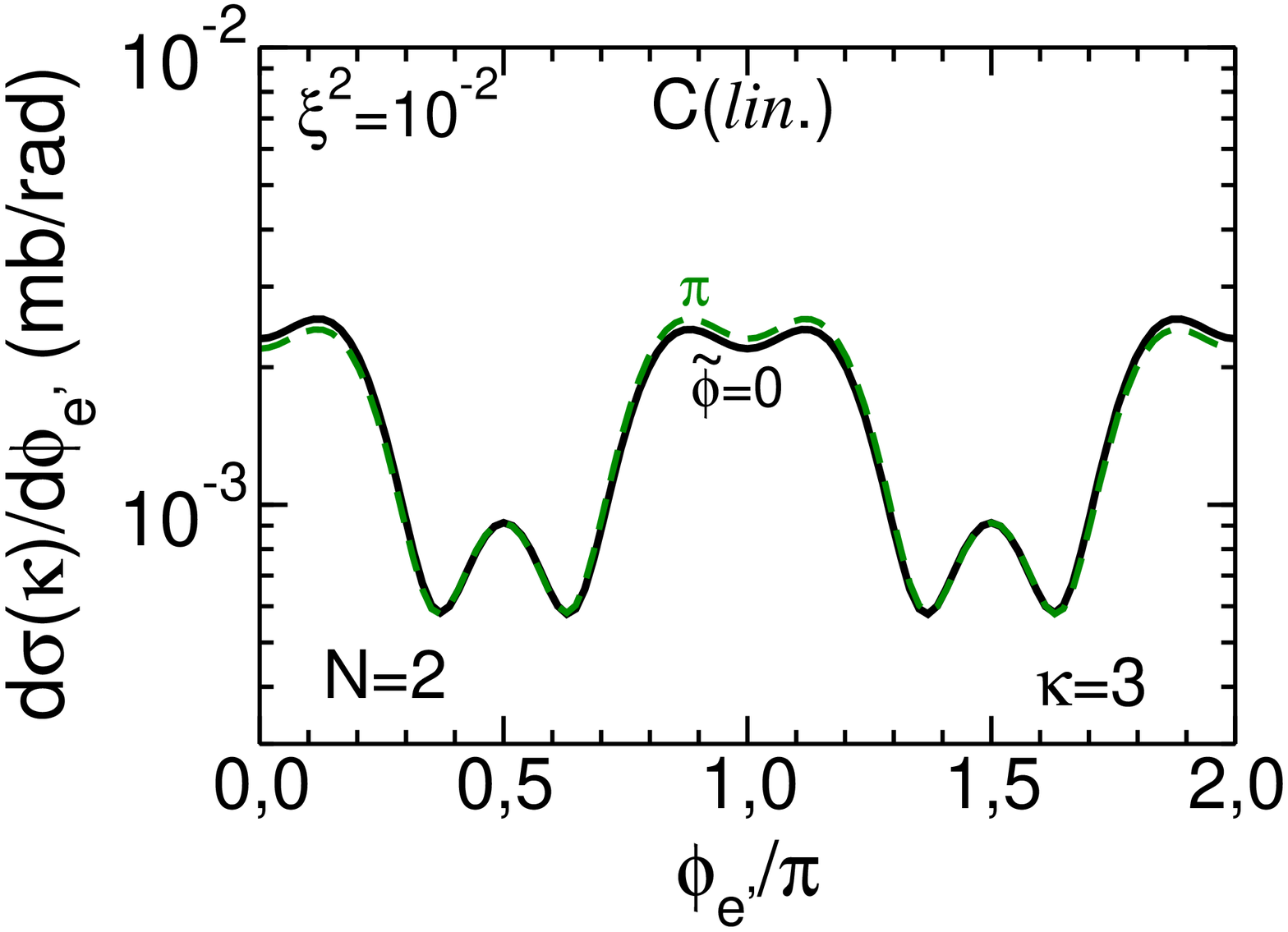}
 \caption{\small{(Color online)
 The differential cross sections ${d\sigma (\kappa)}/{d\phi_{e'}}$
 as a function of $\phi_{e'}$ for different values of $\tilde\phi\equiv\CEP$
 for $\kappa=3$ and $\xi^2=10^{-2}$.
The left and right panels are for  circular and
linear polarizations, respectively.
The top, middle and bottom panels are for $N=1/2$, 1 and 2,
respectively.
 \label{Fig:015}}}
 \end{figure}

Manifestations of the interplay of the azimuthal angle of the outgoing electron $\phi_{e'}$
and the carrier envelope phase $\CEP$ for the  non-linear
Compton scattering in the azimuthal angle distribution in the case of finite pulses
are shown in Fig.~\ref{Fig:015}. The results for circular and linear polarizations
are exhibited in the left and right panels, respectively.
The calculations are for fixed values $\kappa=3$ and $\xi^2=10^{-2}$ and
for different pulse durations with $N=1/2$, 1 and 2 exhibited in the
top, middle and bottom panels, respectively.

Similar to the non-linear Breit-Wheeler process,
in the case of circular polarization, the shapes of the distributions
are smooth curves with maxima and minima at the points
${\phi_{e'}}_{\rm max}=\CEP\pm2\pi$ and
${\phi_{e'}}_{\rm min}={\phi_{e'}}_{\rm max}\pm\pi$.
The explanation for this behavior is the same as in the case of the
non-linear Breit-Wheeler process discussed in Sect.~III.
The shapes of the distributions are similar for
different pulse durations ranging from $N=1/2$ to $N=2$. However,
the relative amplitude of the oscillations,
$d\sigma({\phi_{e'}}_{\rm max})/d\sigma({\phi_{e'}}_{\rm min})$,
changes from about 23 at $N=1/2$ to about 1.06 at $N=2$.
This means that for $N\gtrsim 2$ the azimuthal angle distributions
can be considered as isotropic.

For the linear polarization, the situation is quite different.
As pointed out above, in the case of $\CEP=0$, one gets the results
already discussed above.
The angular distributions $d\sigma_0 (\kappa) / d \phi_{e'})$
have maxima at  $\phi_e'=\pi/4,\,3\pi/4$, $5\pi/4$ and $7\pi/4$.
For sub-cycle and short pulses the height of the first (second)
bump increases when $\CEP$ decreases (increases). That is because
the phase factor in the basis functions
$\widetilde A_m(\ell)$ in Eq.~(\ref{II4})
is determined by the highly oscillating function
$\int dx \exp[-i(\ell x-{\cal P}^{(lin)})]$
depending on the product $c=\cos\phi_{e'}\cos\CEP$, thus
enhancing the cross section for positive values of $c$.
A closer inspection leads to the
bump positions $\phi_{e'}=\pi/4,\,\CEP=0$,
$\phi_{e'}=3\pi/4,\,\CEP =\pi$
and so on. At $\CEP=\pi/2$, the height of the bumps does not depend on
the bump position.
Note that the factor $c$ does not determine directly the bump
positions, which depend also on the interplay
of the two terms in the partial probability~(\ref{III6}).

For relatively long pulse with $N\gtrsim2$,
similarly to the case  of circular polarization,
the angular distribution does not depend on
$\CEP$,  but in contrast to the circular polarization
now it is not isotropic and it becomes close to the
IPA result (cf. Fig.~\ref{Fig:013}).

\section{Summary}

In summary we have performed a simultaneous analysis of  two essentially
non-linear QED processes in circularly and linearly polarized short and intensive
e.m.\ (laser) pulses: (i) non-linear Breit-Wheeler $\ee$ pair creation
in the interaction of a probe photon with such a pulse and
(ii) the photon emission or the non-linear Compton scattering
when an initial electron interacts with
the short and intensive e.m.\ (laser) pulse.
Both processes are analyzed in the multi-photon region,
 where several photons participate at the same time.
In the case of non-linear Breit-Wheeler $\ee$ pair emission, the multi-photon
region is determined uniquely by the threshold variable
$\zeta>1$ or $s < s_{\rm thr}=4m^2$.
For the non-linear Compton scattering, we use the partially integrated
(i.e.\ truncated) cross section,
where the integration starts from the dynamical parameter $\kappa>1$
which selects the multi-photon events and
is an analog of the variable $\zeta$ in the non-linear Breit-Wheeler process.

Our analysis shows a step like
dependence of the total cross sections of $\zeta$ ($\kappa$)
for the non-linear Breit-Wheeler (Compton) process
in the case of relatively
long pulses with the number of oscillations in a pulse $N\gtrsim 2$, similar to the
prediction for the infinitely long pulse.
In the case of a sub-cycle pulse ($N=1/2$), the cross sections  exhibit
an exponential dependence
$\exp[- b_\zeta \zeta]$ ($\exp[- b_\kappa\kappa]$).
The slopes $b_\zeta$, $b_\kappa$ depend on the field intensity and
the pulse duration.

The azimuthal angle distributions are very sensitive to the choice of
parameters and, in particular,
depend on the carrier envelope phase $\CEP$.
In addition to the processes occurring in the circularly polarized
e.m.\ pulses considered earlier \cite{CEPTitov}, the case of linear
polarization leads to a qualitative modification of the azimuthal
distributions of outgoing electrons. These distributions are
non-monotonic functions %of the azimuthal angle of outgoing electrons
with peculiar maxima and minima.
In most cases we gave qualitative explanations
of their positions which are confirmed by the exact numerical calculations.
Their positions, heights, and depths are determined by the
structure of the phase factor  ${\cal P}^{(lin)}$ of the basis functions
$\widetilde A_m$, which in turn
depend on the dynamic variables $\xi^2,\,\zeta,\,\kappa$, and
the carrier envelope phase $\CEP$
and the pulse width ($\Delta$) as well.
In the case of non-linear Compton scattering, the angular distributions are
determined by a nontrivial destructive interference of
the terms in the partial probability $w^{(lin)}(\ell)$.
This result depends on the pulse duration:
for pulses with $N\gtrsim 2$,
the dependence of the azimuthal angle distributions on the CEP disappears.

The effects of the $\CEP$ imprint on azimuthal angle
distributions in the considered processes are comparable
with predictions of CEP phenomena in other processes involving
ultrashort laser pulses~\cite{CEP_1,CEP_2,BrabecKrausz}, such as
CEP manifestations in the polar angular distributions of positrons
in the BW processes~\cite{KrajewskaPRA86},
electron tunneling in a coupled double-quantum-dot system~\cite{CDQDS},
and strong-field ionization~\cite{SFI}.
Our results in Sect.~III.B.2 may be compared with that for non-linear
Compton scattering of recent Refs.~\cite{Li2018,Li2018lqs}, where
the polar angular distribution of outgoing electrons
is analyzed, and it is found that
the distributions have some sharp peaks, the positions
and heights of which depend on $\CEP$. Thus, the height of the peaks is
modified by a factor $\approx 1.8$  when  $\CEP$ changes from
0 to $\pi$. This result can be compared with our prediction
of interference effects in azimuthal angle distributions
 exhibited in Fig.~\ref{Fig:015} (upper panels) where,
 e.g.\ for $N=1/2$, 1 and $\phi_{e'}=0$, the cross section is modified
 by an order of magnitude for the $\CEP$ varying from 0 to $\pi$.
% or effect of $\CEP$ in Fig.~\ref{Fig:015} is much greater.
 However, we  have to stress that our result and the results
 of~\cite{Li2018,Li2018lqs} for relatively long pulses are obtained for quite
 different initial conditions $\gamma\gg\xi$ vs.\ $\gamma \ll\xi$,
 respectively, where $\gamma$ is the Lorentz factor of the incoming electron.
 Therefore, the chosen observables and the specific initial conditions are
 very important for a $\CEP$ determination.

 Our theoretical predictions for circularly
 and linearly polarized laser pulses may be used as a unique
 and powerful input for the design of forthcoming experiments
 in the near future
 and corresponding experimental studies of different
 aspects of the multi-photon dynamics related to non-linear QED processes
 and, in particular, for a $\CEP$ determination.

\section*{Acknowledgments}

The authors gratefully acknowledge the collaboration with D. Seipt, T. Nousch, T. Heinzl,
and useful discussions with A. Ilderton, K. Krajewska, M. Marklund, C. M\"uller, and R.
Sch\"utzhold. A. Ringwald is thanked for explanations w.r.t.\ LUXE. The work is supported
by R. Sauerbrey and T. E. Cowan w.r.t.\ the study of fundamental QED processes for HIBEF.

\section*{Contributions}

All authors have contributed equally to the publication, being variously
involved in the conceptual outline, software development and
numerical evaluations.

\vspace*{0.5cm}
\appendix

\section{\small Multi-photon regime in IPA\label{A}}

\subsection{Breit-Wheeler process \label{A.1}}

 The cross sections of the non-linear BW process in IPA
as a function of $s=4m^2/\zeta$
 are represented by an infinite sum of harmonics labeled by $n$
(cf.~\cite{Greiner})
\begin{eqnarray}
\sigma^{(i)} (s)&=&
\frac{\alpha^2\,\zeta}{4m^2\xi^2 N^{(i)}_0}\nonumber\\
&\times &
\sum\limits_{n=n_{\rm min}}^{\infty}\,
\int\limits_0^{2\pi}d\phi_e\,
\int\limits_{1}^{u_n}
\frac{du}{u^{3/2}\sqrt{u-1}}
\,w_n^{(i)}
\label{II14}
\end{eqnarray}
with $N^{(lin)}_0 = 1/2$ and $N^{(circ)}_0 = 1$,
and the partial harmonics $w^{(i)}$
determined by Eqs.~(\ref{II99}) and (\ref{II13}), respectively.
The minimum number $n$ is determined as the
integer part (Int) of $\zeta=4m_*^2/s$: in detail,
$n_{\rm min}={\rm Int}(\zeta)$ if ${\rm Int}(\zeta)=\zeta$
and $n_{\rm min}={\rm Int}(\zeta)+1$ if ${\rm Int}(\zeta)<\zeta$.
So, for example, for the above-threshold process with $\zeta<1$,
one has $n_{\rm min}=1$.

\subsection{Compton scattering\label{A.2}}

The total cross sections in IPA are expressed in standard notation
as an infinite sum of harmonics labeled by $n$ \cite{LL}
\begin{eqnarray}
 {\sigma^{(i)}}
&=&\frac{\alpha^2}{\xi^2(p\cdot k)N_{0}^{(i)}}\nonumber\\
&\times&\sum\limits_{n=1}^{\infty}
 \,
  \int\limits_{0}^{2\pi}d\phi_{e'}
 \int\limits_{0}^{u_n}\frac{du}{(1+u)^2}
 \,w^{(i)}_n~,\label{III14}
\end{eqnarray}
where $N_{0}^{(i)}$  is defined below Eq.~(\ref{II14}), and
$u_n=2n(p\cdot k)/m^2_{*}$.
The physical meaning of the quantity $n$ is as follows.
It is the number of photons of the laser
beam participating in the process with the formation of a photon with frequency $\omega'$.
The product $n\omega$ is the energy fraction of the background field
involved into the process.

The partial probabilities (harmonics) in Eq.~(\ref{III14}) for $lin$
and $circ$ polarizations read
 \begin{eqnarray}
 \frac12\,w^{(lin)}_n = -A_0^2
 + \xi^2\left(1+\frac{u^2}{2(1+u)}\right)\left( A_1^2
 - A_0A_2\right)
 \label{III8}
 \end{eqnarray}
 and
\begin{eqnarray}
w^{(circ)}_n&=&
-2 J^2_n + \xi^2\left(1+\frac{u^2}{2(1+u)}\right)\nonumber\\
&\times&\left(J^2_{n-1}+ J^2_{n+1}-2J^2_n\right)~,
\label{III9}
\end{eqnarray}
where the notations for the basis functions
are the same as in Sect.~III
(see Eqs.~(\ref{II99}) and (\ref{II13}) for details).
Naturally, in the case of IPA one has to use
kinematics, conservation laws and dynamical variables
$u,\,z$  for dressed fermions~\cite{Ritus-79}.

Selecting
multi-photon processes with $n>1$ from the infinite sum (\ref{III14})
may be done similarly to that given in Sect.~IV.
First, we express the cross sections (\ref{III14}) in an equivalent
but slightly different form
\begin{eqnarray}
 \sigma^{(i)}=
 \sum\limits_{n=1}^{\infty}
 \int\limits_{-1}^{1}d\cos\theta'
 \frac{d\sigma^{(i)}_{n}}{d\cos\theta'}~,
 \label{III114}
\end{eqnarray}
where ${d{\sigma_n^{(i)}}}/{d\cos\theta'}$
is the differential cross section for
each separate harmonic
 \begin{eqnarray}
 \frac{d{\sigma_n^{(i)}}}{d\cos\theta'}
=\frac{\alpha^2}{\xi^2(p\cdot k)N_{0}^{(i)}}\,F_n
\int\limits_{0}^{2\pi}d\phi_{e'}
 \,w^{(i)}_n~
 \label{III144}
\end{eqnarray}
with $F_n={\omega'_n}^2/n(p\cdot k)$.
The frequency of an outgoing photon emitted at polar angle $\theta'$
reads for the $n$th harmonic in IPA
 \begin{eqnarray}
 \omega'_{n}=\frac{n\,\omega (E+|\mathbf {p}|)}
 {E + |\mathbf {p}| \cos\theta'
 + \omega(n+\frac{\xi^2m^2}{2\lambda(k\cdot p)})(1-\cos\theta') }~,
 \label{III12}
\end{eqnarray}
 where $\lambda=1$ or 2 for $circ$ or $lin$ polarizations, respectively.

  \begin{figure}[th]
 \includegraphics[width=0.48\columnwidth]{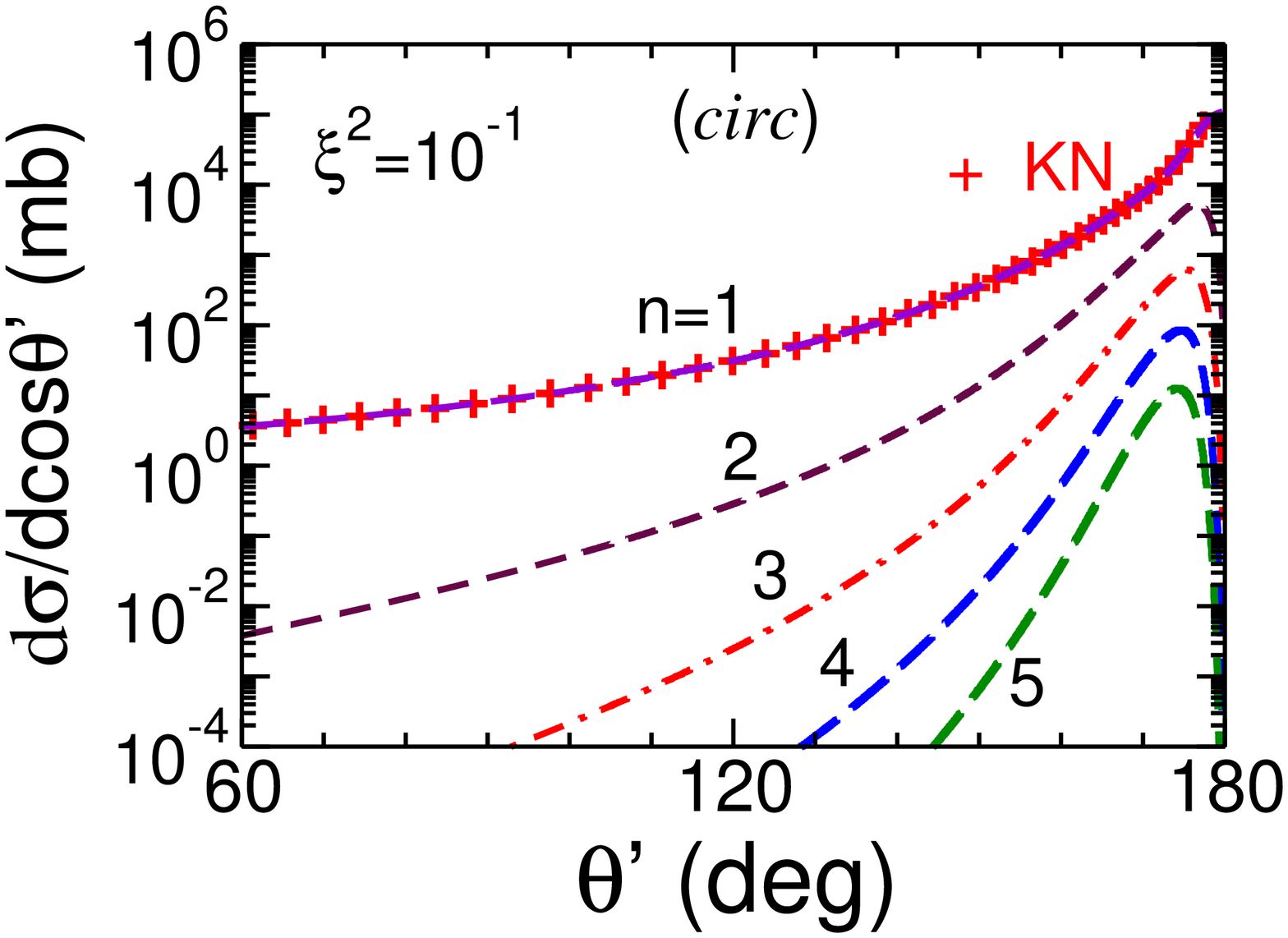}\hfill
 \includegraphics[width=0.48\columnwidth]{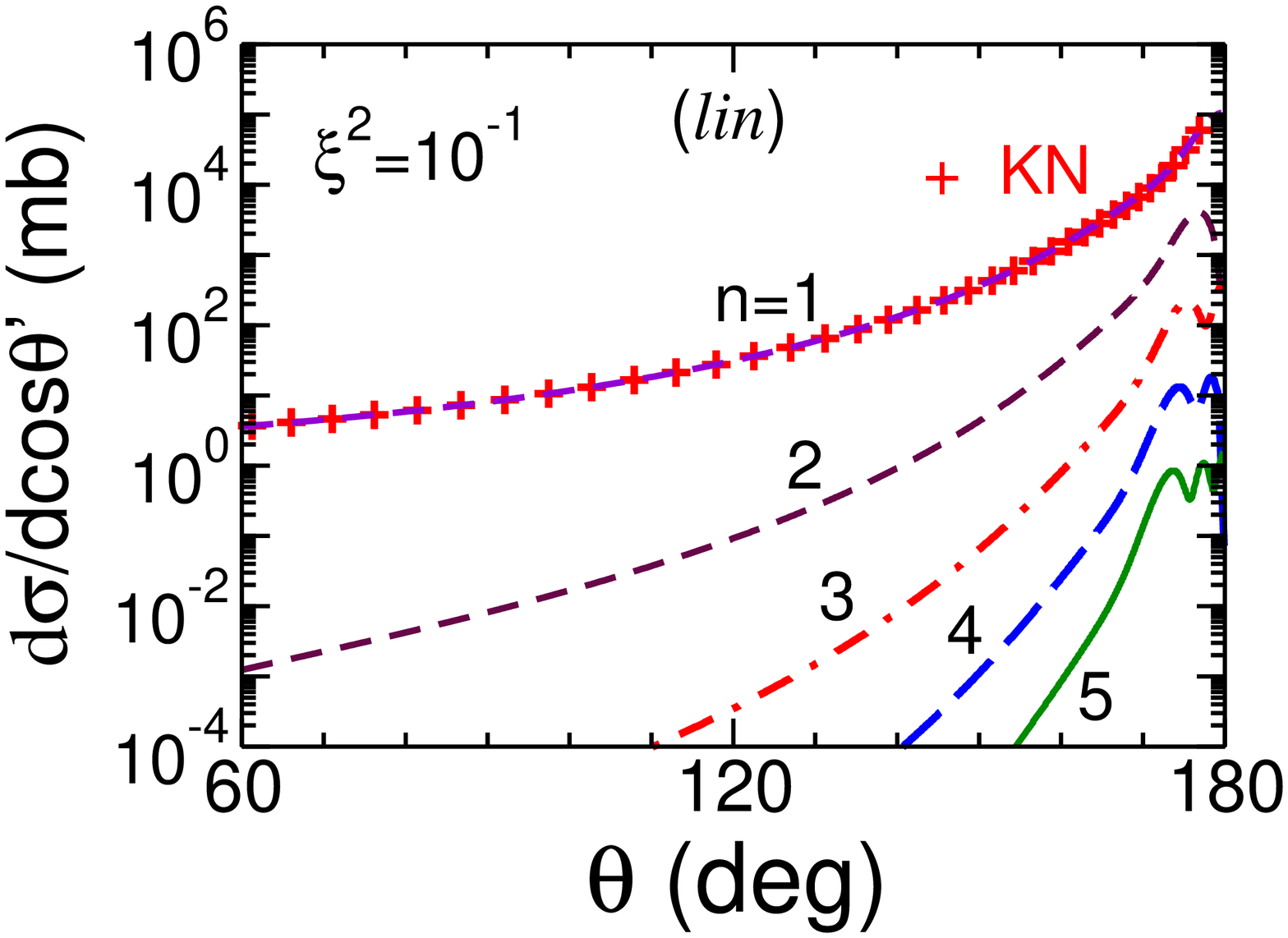}\\
 \includegraphics[width=0.48\columnwidth]{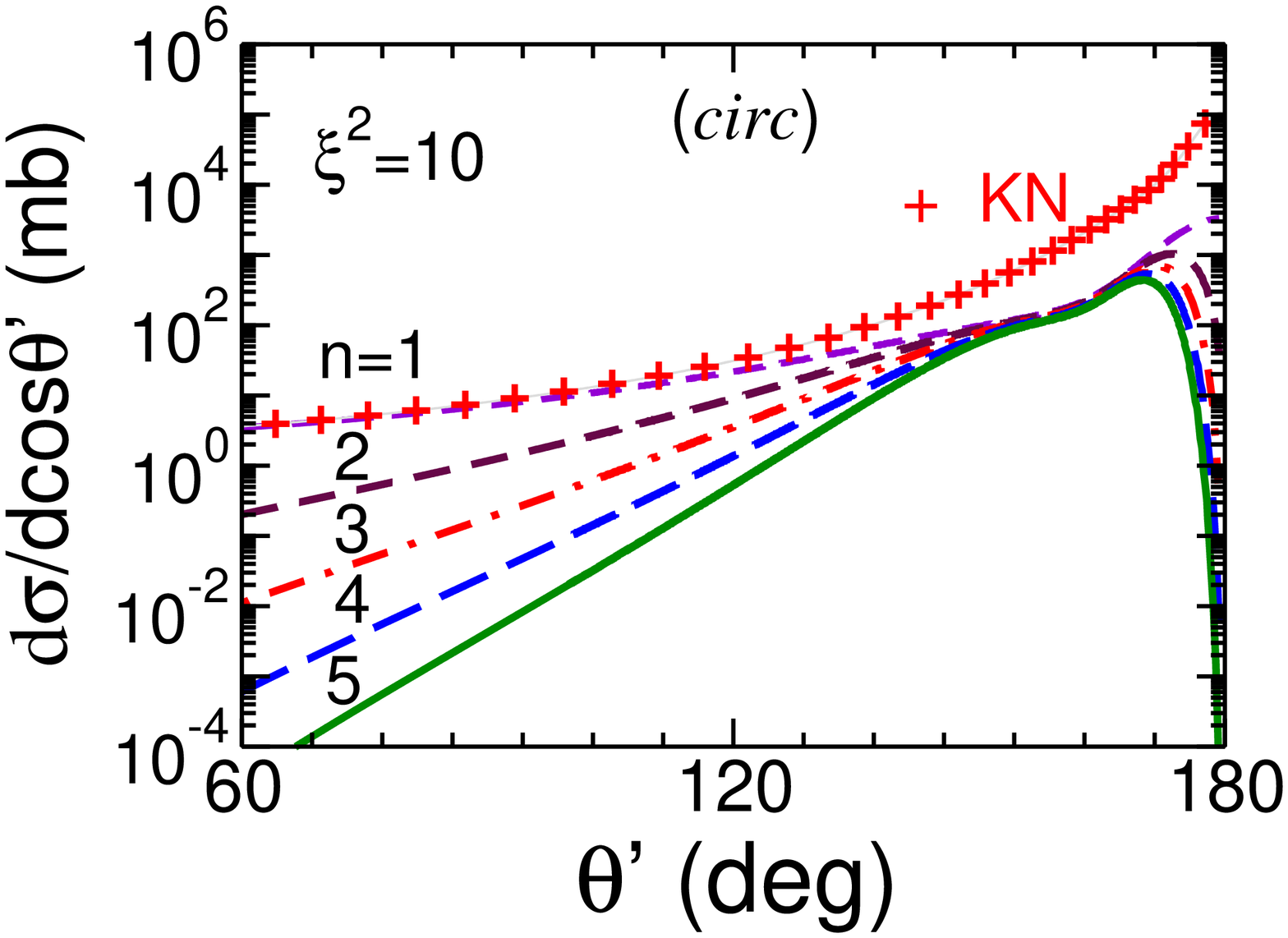}\hfill
 \includegraphics[width=0.48\columnwidth]{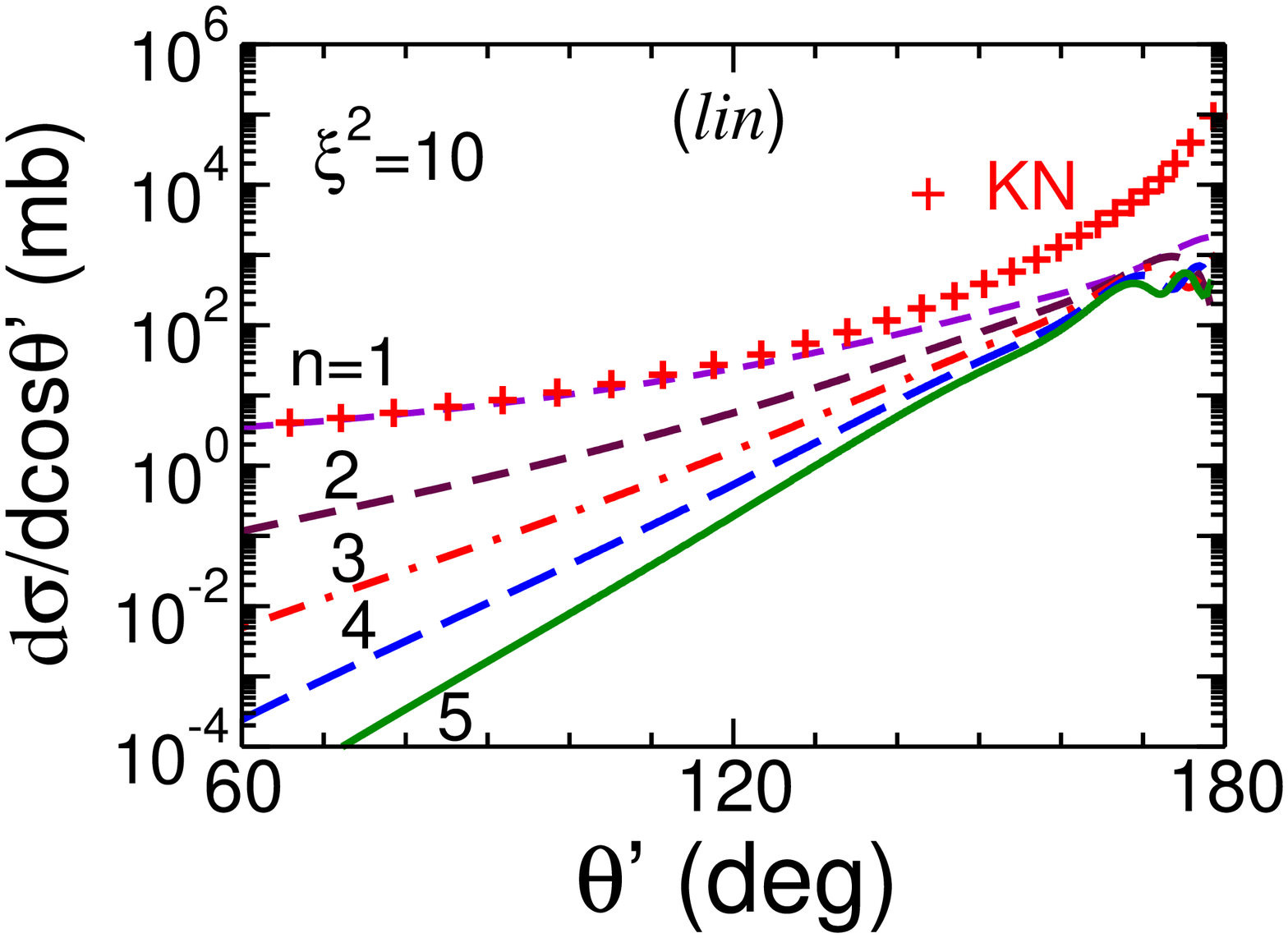}\\

 \caption{\small{(Color online)
The differential cross sections for the individual harmonics (cf. Eq.~(\ref{III144}))
with $n=1 \cdots 5$ (according to legend) and $\xi^2=10^{-1}$ (top panels)
and 10 (bottom panels).
The crosses are for the Klein-Nishina cross section.
The left and right panels are for
$circ$ and $lin$ polarizations, respectively.
The electron energy and the frequency of the photon beam (in lab.\ system)
are 4~MeV and 1.55 eV, respectively.
Note the dead cone for harmonics $n > 1$, i.e.\ no on-axis backscattering,
for $cir$ polarization, while for $lin$ polarization, the odd-number
harmonics are backscattered on-axis.
 \label{Fig:001}}}
 \end{figure}

The differential cross sections  (\ref{III144}) for the individual harmonics
with $n=1 \cdots 5$ and $\xi^2=10^{-1}$ are shown
% differential cross sections for the individual harmonics
 in Fig.~\ref{Fig:001}. The left and right panels are for
 circularly and linearly polarized beams, respectively.
 The initial electron energy and the frequency of the photon beam (in lab.\ system)
 are 4~MeV and 1.55 eV, respectively.
 The crosses are for the Klein-Nishina result~\cite{LL}.
 One can see that the harmonic $n=1$ at $\xi^2\leq 0.1$ fairly well reproduces
 the Klein-Nishina cross section for $\gamma + e \to \gamma' +e'$.
% which, however is beyond scope of our present consideration.
 In the case of multi-photon processes with $n>1$, the calculation
 predicts a strong bump in the backward hemisphere which seems  to be
 preferable to study the multi-photon processes
 (cf.\ our discussion in Sect.~IV.A).
Closer inspection of the cross sections for linear polarization
 in the vicinity of $\theta'=\pi$ shows that they are finite for
 odd harmonics and equal to zero for even harmonics (the so called
 "dead cone behavior", cf.~\cite{Harvey}). For circular polarization,
 the cross sections are equal to zero at $\theta'=\pi$
 for all higher harmonics with $n>1$.

 In order to select essentially multi-photon interactions
 the summation in (\ref{III14}) should start from
 $n=\hat{n}$ equal to  $\omega'/\omega'_1$, which coincides with the integer
 part of the ratio $\kappa=\omega'/\omega'_1$ in the case of FPA.

%%%%%%%

\end{document}